\documentclass[12pt]{article}

\usepackage{bm,amsmath,amssymb,latexsym,graphicx,euscript,multirow,color,url,verbatim}
\usepackage{epsf}
\usepackage[sort&compress,numbers]{natbib}
\usepackage{hyperref}
\usepackage{enumerate}

\allowdisplaybreaks

\newcommand{\capt}[1]{\caption{\small #1}}

\long\def\symbolfootnote[#1]#2{\begingroup%
\def\thefootnote{\fnsymbol{footnote}}\footnote[#1]{#2}\endgroup}

\newcommand{\be}{\begin{equation}}
\newcommand{\ee}{\end{equation}}

\newcommand{\bea}{\begin{eqnarray}}
\newcommand{\eea}{\end{eqnarray}}
\newcommand{\mat}{\begin{pmatrix}}
\newcommand{\rix}{\end{pmatrix}}

\renewcommand{\l}{\left}
\renewcommand{\r}{\right}

\renewcommand{\bar}{\overline}
\renewcommand{\slash}[1]{#1\!\!\!/}

\newcommand{\vv}{\mathbf}

\newcommand{\go}{{\tilde g}}
\newcommand{\gr}{{\tilde G}}

\newcommand{\st}{{\tilde t}}

\begin{document}

\begin{titlepage}

\begin{flushright}
\small{RUNHETC-2011-19}\\
\small{YITP-SB-11-36}\\
\end{flushright}

\vspace{0.5cm}
\begin{center}
\Large\bf
The Status of GMSB After 1/fb at the LHC
\end{center}

\vspace{0.2cm}
\begin{center}
{\sc Yevgeny Kats$^a$\symbolfootnote[1]{kats@physics.rutgers.edu}, Patrick Meade$^b$\symbolfootnote[2]{patrick.meade@stonybrook.edu}, Matthew Reece$^c$\symbolfootnote[3]{mreece@physics.harvard.edu} and David Shih$^a$\symbolfootnote[4]{dshih@physics.rutgers.edu}}\\
\vspace{0.6cm}
\small \textit{$^a$New High Energy Theory Center\\
Rutgers University, Piscataway, NJ 08854, USA\\
\medskip
$^b$ C. N. Yang Institute for Theoretical Physics,\\ Stony Brook University, Stony Brook, NY 11794\\
\medskip
$^c$ Department of Physics\\
Harvard University, Cambridge, MA 02138}
\end{center}

\vspace{0.5cm}
\begin{abstract}
\vspace{0.2cm}
\noindent

We thoroughly investigate the current status of supersymmetry in light of the latest searches at the LHC, using General Gauge Mediation (GGM) as a well-motivated signature generator that leads to many different simplified models. We consider {\it all} possible promptly-decaying NLSPs in GGM, and by carefully reinterpreting the existing LHC searches, we derive limits on both colored and electroweak SUSY production.  Overall, the coverage of GGM parameter space is quite good, but much discovery potential still remains even at 7 TeV.  We identify several regions of parameter space where the current searches are the weakest, typically in models with electroweak production, third generation sfermions or squeezed spectra, and we suggest how ATLAS and CMS might modify their search strategies given the understanding of GMSB at 1/fb. In particular, we propose the use of leptonic $M_{T2}$ to suppress $t{\bar t}$ backgrounds. Because we express our results in terms of simplified models, they have broader applicability beyond the GGM framework, and give a global view of the current LHC reach. Our results on 3rd generation squark NLSPs in particular can be viewed as setting direct limits on naturalness.

\end{abstract}
\vfil

\end{titlepage}

\tableofcontents

\section{Introduction}

This has been a very exciting year for particle physics. The LHC has been performing beyond expectations. At the time of this writing, each experiment has recorded approximately 5/fb of data at 7 TeV center of mass energy. Dozens of searches for new physics have been published by CMS and ATLAS analyzing $\sim 1$/fb of data. Unfortunately, all have yielded null results so far. As we enter the second full year of 7 TeV running, it is a good time to pause and survey  the current status of LHC searches for new physics. The purpose of this paper will be to carry out this survey for a particularly well-motivated theoretical scenario: gauge mediated supersymmetry breaking (GMSB; for a review and original references, see~\cite{Giudice:1998bp}). The searches we will survey are listed in Table~\ref{tab:searches}. These searches have typically expressed their results as constraints on the CMSSM and various ``simplified models" with neutralino LSP (though notable exceptions like~\cite{ATL-PHYS-SLIDE-2011-523,Aad:2011kz,CMS-PAS-SUS-11-009,Chatrchyan:2011ah} set limits on simplified parameter spaces in GMSB). In this paper, we will carefully reinterpret these results in terms of GMSB.

\begin{table}[!t]
\begin{center}
\begin{tabular}{|c|c|c|c|}\hline
Analysis & Collaboration & Luminosity (fb$^{-1}$) & Ref\\
\hline
\hline
jets+MET & ATLAS & 1 & \cite{Aad:2011ib}  \\
 & CMS & 1.1 & \cite{CMS-PAS-SUS-11-004} \\
{\it with} $\alpha_T$ & CMS & 1.1 & \cite{CMS-PAS-SUS-11-003} \\
\hline
6-8 jets+MET & ATLAS & 1.34  & \cite{ATL-PHYS-SLIDE-2011-529}  \\
\hline
$b$-jets+MET & ATLAS & 0.833 & \cite{ATLAS-CONF-2011-098}  \\
             & CMS   & 1.1   & \cite{CMS-PAS-SUS-11-006} \\
\hline
\hline
SS dileptons+jets+MET
& CMS & 0.98 & \cite{CMS-PAS-SUS-11-010}  \\
\hline
OS dileptons +jets+ MET
& CMS & 0.98 & \cite{CMS-PAS-SUS-11-011} \\
\hline
lepton+jets+MET & ATLAS & 1.04 & \cite{Collaboration:2011iu} \\
& CMS & 1.1 & \cite{CMS-PAS-SUS-11-015}   \\
\hline
lepton+$b$-jets+MET & ATLAS & 1.03 & \cite{ATLAS-CONF-2011-130} \\
\hline
$Z(\ell^+\ell^-)$+jets+MET & CMS & 0.98 & \cite{CMS-PAS-SUS-11-017} \\
\hline
$t\bar t$+MET & ATLAS & 1.04 & \cite{Collaboration:2011wc} \\
\hline
\hline
$\gamma\gamma+$MET & ATLAS & 1.07 & \cite{ATL-PHYS-SLIDE-2011-523}  \\
$\gamma\gamma+$jet+MET  & CMS & 1.1 & \cite{CMS-PAS-SUS-11-009}  \\
\hline
$\gamma+$jets+MET & CMS & 1.1 & \cite{CMS-PAS-SUS-11-009}  \\
\hline
$\gamma$+$\ell$+MET & CMS & 0.035 & \cite{Chatrchyan:2011ah}\\
\hline
\end{tabular}
\end{center}
\capt{A summary of the most recent LHC searches with $\gtrsim 1$/fb relevant to GMSB. Included in this table is also the CMS $\gamma$+$\ell$+MET search with 35/pb, since this is so far the only search in this final state. Not included here are: CMS all-hadronic searches with $M_{T2}$~\cite{CMS-PAS-SUS-11-005} and Razor~\cite{CMSRazor} (overlapping with standard jets+MET); ATLAS search for multileptons+MET~\cite{ATLAS-CONF-2011-039} (not updated to 1/fb yet); 2/fb CMS searches for multileptons~\cite{CMSmultileptons}, which were released as this paper was nearing completion; and 1/fb ATLAS searches for OS dileptons and SS dileptons \cite{ATLASleptons}, described in a talk while this paper was in preparation.
}
\label{tab:searches}
\end{table}

We will make use of the general framework for GMSB known as General Gauge Mediation (GGM) ~\cite{Meade:2008wd}.  The advantage of such a framework is that it allows for a theoretically well-grounded, yet model-independent exploration of GMSB phenomenology. The entire physical parameter space for GGM was mapped out with a perturbative messenger model in~\cite{Carpenter:2008wi,Buican:2008ws}. (In defining GGM, we will not discuss the Higgs sector in detail, but simply assume that $\mu$ and $\tan\beta$ can be set freely.) A number of papers have studied both the Tevatron bounds and LHC projections~\cite{Meade:2009qv, Meade:2010ji, Ruderman:2010kj, Ruderman:2011vv, Kats:2011it, Carpenter:2008he, Rajaraman:2009ga,DeSimone:2009ws,Mason:2009qh,Katz:2009qx, Abel:2009ve, Kribs:2009yh,Katz:2010xg, Abel:2010vb, Thalapillil:2010ek, Jaeckel:2011ma}, elaborating on and greatly extending the scope of the pioneering Tevatron-era work on gauge mediated phenomenology~\cite{Dimopoulos:1996vz,Dimopoulos:1996yq, Ambrosanio:1997rv, Culbertson:2000am, MatchevThomas, BMTW}.

\begin{figure}[!t]
\begin{center}
\includegraphics[width=4in]{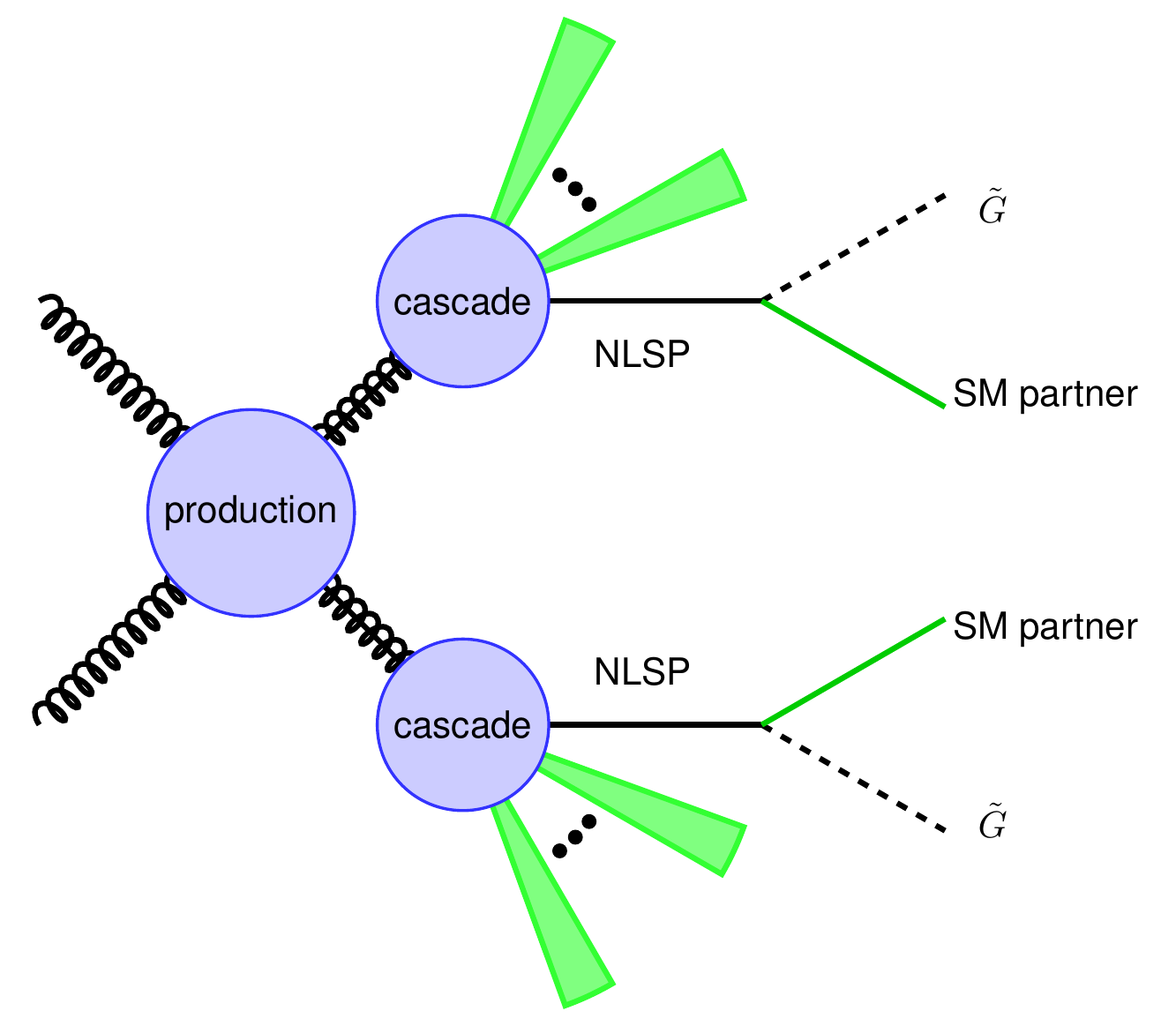}
\end{center}
\capt{Schematic Feynman diagram for a GMSB event. The typical production will be of colored superpartners, e.g., gluinos. Their cascade decays will produce jets and possibly other particles (depicted here as green wedges), and will end in the NLSP. The NLSP will always decay to its SM partner plus an invisible gravitino.}
\label{fig:gmsb}
\end{figure}

A typical GMSB topology is shown in Fig.\ \ref{fig:gmsb}. This figure illustrates a number of important features of GMSB. First, the gravitino is always the LSP. Second, the identity of the next-to-lightest-superpartner (NLSP) dictates much of the phenomenology, because it appears at the bottom of every cascade decay and always decays to its SM partner and the gravitino. (We assume $R$-parity throughout.) Correspondingly, we will partition the parameter space of GMSB primarily via the NLSP identity.

An important feature of the GGM framework is that it allows almost any superpartner to be the NLSP.   We will thoroughly investigate {\it all} NLSP types: neutralino (bino, wino, $Z$-rich higgsino, $h$-rich higgsino), chargino, right-handed slepton, sneutrino, gluino, squark, stop, and sbottom. As we will see, by studying the signatures that arise from every NLSP type in GGM, we will naturally be led to consider most, if not all, of the current LHC searches. In Table~\ref{tab-nlsp-types} we have listed the final states relevant for the various NLSP types.  The table serves as a useful guide for understanding which analyses might be useful for each NLSP type and facilitates a more general application of our results to other models with similar final states. As the table illustrates, GGM is a very effective ``signature generator": it provides a nice unifying framework through which to view the myriad results at the LHC.

\begin{table}[!t]
\begin{center}
\begin{tabular}{|c|c|}\hline
NLSP type & Relevant final states (+MET)\\
\hline
bino & $\gamma\gamma$, $\gamma$+jets\\
\hline
wino & $\gamma\ell$, $\gamma\gamma$, $\gamma+$jets, $\ell+$jets, jets\\
\hline
$Z$-rich higgsino & $Z(\ell^+\ell^-)+$jets, $Z(\ell^+\ell^-)Z(\ell'^+\ell'^-)$,  SS dileptons, jets\\
\hline
$h$-rich higgsino & $b$-jets, SS dileptons, jets\\
\hline
chargino & SS dileptons, OS dileptons, $\ell+$jets, jets\\
\hline
slepton & multileptons, SS dileptons, OS dileptons, $\ell$+jets, jets\\
\hline
squark/gluino & jets\\
\hline
stop & SS dileptons, OS dileptons, $b$-jets, $\ell+$jets, $\ell+b$-jets, $t\bar t$, jets\\
\hline
sbottom & $b$-jets, jets\\
\hline
\end{tabular}
\end{center}
\capt{An overview of GGM phenomenology and the relevant final states. Certain final states in this table are relevant not because of the NLSP decay, but because of the transition from the production channel to the NLSP in the simplified spectra that we will consider.}
\label{tab-nlsp-types}
\end{table}

\begin{figure}[!t]
\begin{center}
\includegraphics[width=1\textwidth]{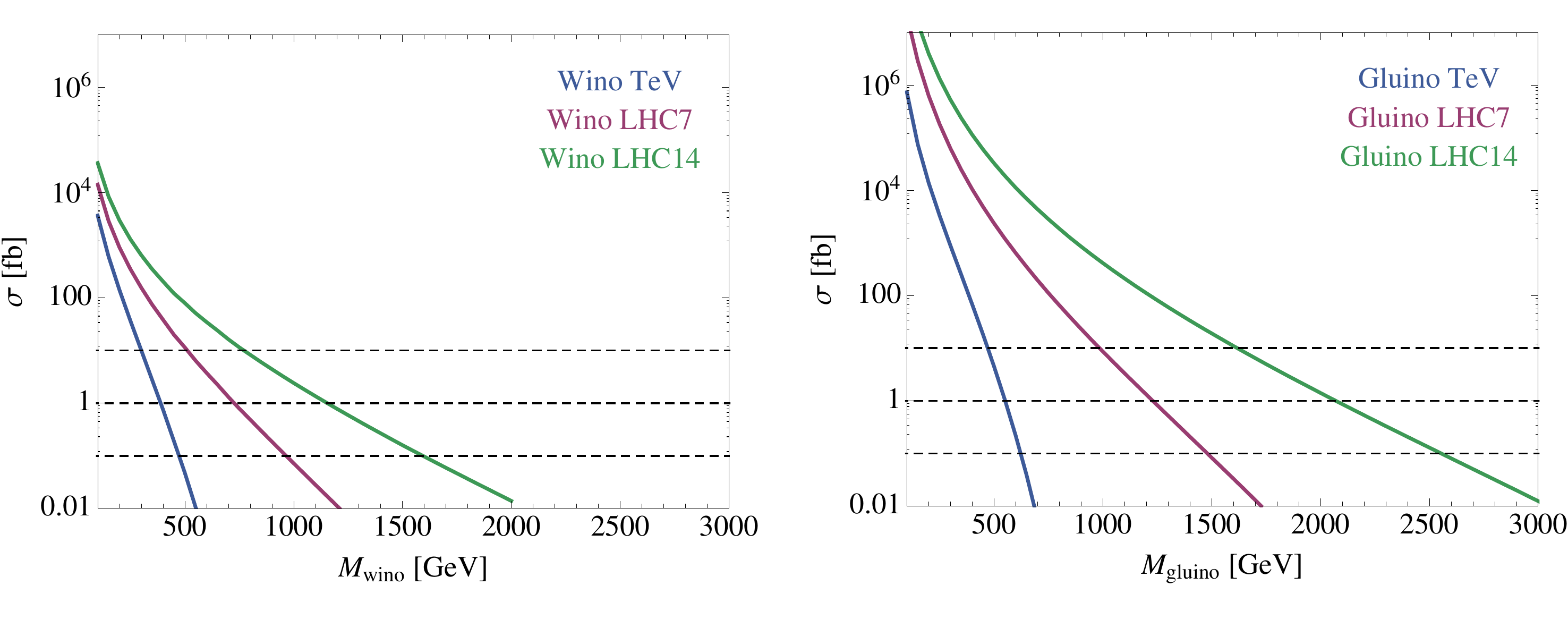}
\end{center}
\capt{NLO production cross sections for wino pairs (left) and gluino pairs (right).  The dashed lines indicate 10~fb, 1~fb and 0.1~fb, while the blue, red and green curves correspond to Tevatron, 7~TeV LHC, and 14~TeV LHC. The 10~fb rate roughly corresponds to the kinematic reach of the current 1/fb LHC searches. The 1~fb rate corresponds to the kinematic limit for the Tevatron and the 7~TeV LHC, both of which will collect ${\mathcal O}(10\,\,{\rm fb}^{-1})$ of data in their complete runs. Finally, the 0.1 fb rate corresponds to the kinematic limit for the 14~TeV LHC, which is expected to collect ${\mathcal O}(100\,\,{\rm fb}^{-1})$ in total.}
\label{fig-xsecs}
\end{figure}

In addition to the NLSP type, the SUSY production mechanism is important for specifying the relevant phenomenology.  Here we could consider either production of colored superpartners (as shown in Fig.\ \ref{fig:gmsb}), or electroweak superpartners. There are very important differences between the two at the LHC. To illustrate this and other points, we have shown in fig.\ \ref{fig-xsecs}  the NLO cross sections (computed using {\sc Prospino}~2.1~\cite{prospino}) for wino production (left) and gluino production (right). The former proceeds via electroweak gauge bosons while the latter goes through gluons; all other SUSY particles are decoupled.

Fig. \ref{fig-xsecs} allows us to understand the ``kinematic limits" at the Tevatron and LHC for electroweak and colored SUSY production. These are approximate idealized limits where an experiment throughout its lifetime would collect $\mathcal{O}(10)$ events before any analysis cuts. (In practice much higher rates will be needed for most experimental analyses, which means the actual reach in mass will often be lower.) We see that the LHC has a huge advantage over the Tevatron for colored production, and much less advantage for electroweak production. In particular, the 7~TeV LHC with 1/fb of data can probe winos up to just $\sim 500$~GeV, which is only slightly beyond the $\sim 400$~GeV reach of the Tevatron. Thus for early LHC running, it makes sense to concentrate on constraining colored production where the corresponding numbers are $\sim 1000$~GeV vs. $\sim 600$~GeV.  Seeing electroweak production will typically require more data (and will also be more difficult in terms of separating signal from background). With ${\cal O}(10/\mbox{fb})$, the 7~TeV LHC can probe up to $\sim 700$~GeV in wino mass, and $\sim 1200$ GeV in gluino mass. Finally, with ${\cal O}(100/\mbox{fb})$, the LHC at 14~TeV will probe wino masses up to $\sim 1500$~GeV, and gluino masses up to $\sim 2500$~GeV.

The 7 TeV numbers of $\sim 700$~GeV for wino production and $\sim 1200$~GeV for gluino production set a yardstick with which to measure the current progress. Among the simplified models that we will benchmark are parameter spaces consisting of either gluino or squark mass vs.\ NLSP mass (with everything else decoupled), and gluino vs.\ squark mass (with fixed NLSP mass and everything else decoupled). We will also consider additional simplified models with electroweak production. These options will allow us to consider the effects of different jet multiplicities and kinematics in determining the limits. We will find that for NLSPs with the cleanest final states (bino NLSP with $\gamma\gamma+$MET; slepton co-NLSPs with same-sign dileptons+MET), the limits on gluino mass are nearly 1000 GeV. So already with 1/fb we are very close to the kinematic limit for 7 TeV LHC in these scenarios. Most of the discovery potential at 7 TeV has already been used up here. For more complicated cases (squeezed spectra, multiple final states, third generation), the limits on the gluino mass are much weaker, ranging typically from 600-800 GeV. So there is considerable room for growth and improvement here. Finally, we find that the only existing LHC searches that constrain electroweak production are the ATLAS and CMS $\gamma\gamma$+MET searches, which constrain winos decaying to bino NLSPs. There is a large amount of growth possible in probing electroweak production of new particles.

There is already a large literature (too large to review here) interpreting LHC results as SUSY limits, so it is worthwhile to make some remarks on our motivation and how our work fits into that broader context. Most of the existing work studies spectra involving {\em all} the MSSM particles, often from a top-down point of view (such as the CMSSM) or in high-dimensional parameter spaces. These models have an abundance of possible production modes and decays, and it is difficult to isolate the physics that goes into setting limits.

We believe that, at this point, a study in terms of simplified spectra is sorely needed, and substantially different from studies of the full MSSM. In the absence of any discovery, our main goal in studying LHC limits on supersymmetry is twofold: first, to obtain a global picture of what types of new physics the LHC has already excluded; second, to reveal under-constrained regions of parameter space. For the first goal, we believe that simplified spectra allow us to present a more comprehensible picture, in terms of clear two-dimensional limit plots, than any study of the full MSSM can achieve. By looking at many different such parameter spaces, a global picture of the LHC's progress can be obtained, expressed in terms of well-understood parameters like physical masses and cross sections rather than high-scale inputs. The second goal---probing for weak points in the current LHC searches---is perhaps the most important, because only by isolating the places where the LHC is not doing well can we take the next step of designing better searches to cover these regions, so that new physics there can be discovered (we hope). Simplified spectra are ideal for this, as they capture precisely the information experimentalists will actually use: physical masses, production rates, and branching ratios. In taking this approach, our outlook is similar to that of other recent work that advocates for studying new physics in terms of physical masses, rates, and branching ratios~\cite{Meade:2006dw,Meade:2009qv, Meade:2010ji, Ruderman:2010kj, Ruderman:2011vv, Kats:2011it, ArkaniHamed:2007fw, Alwall:2008ve,Dube:2008kf, Alwall:2008ag,Alves:2011wf}, and complementary to other work in progress~\cite{chickenfoot}.

In Sections~\ref{sec-results-neutralino}-\ref{sec-results-colored}, we will describe in much more detail the simplified parameter spaces for each NLSP type. We will then exhibit the best current limits on these simplified parameter spaces from the LHC searches in table~\ref{tab:searches}.
For each search in table~\ref{tab:searches}, we have coded up its acceptance and selection cuts using a crude detector simulation, and then validated our code by reproducing the published 95\% exclusion plots (usually in the CMSSM parameter space or a simplified model with neutralino LSP). Generally, we have found that we are in excellent agreement with the official exclusion contours derived by ATLAS and CMS. Details of our simulation and limit-setting procedure are contained in Appendix~\ref{app:simulation}, and in Appendix~\ref{app:exp-requests} we discuss the details that would be useful to include in experimental publications to allow theorists to re-analyze the results reliably. In Section~\ref{sec:leptonicmt2} we will discuss the variable leptonic $M_{T2}$, which can be very useful for searches with $t{\bar t}$ or $WW$ backgrounds. Up to and including that section, we assume that the decaying NLSP is prompt. In GMSB, the coupling to the goldstino is suppressed by the SUSY breaking scale $\sqrt F$ and the NLSP can be long-lived, so in Section~\ref{sec-long-lived} we relax the assumption of prompt NLSPs and comment on searches for long-lived particles. We conclude in Section~\ref{sec-summary} with a summary of our results and the prospects for the near future.

\section{Limits on Neutralino and Chargino NLSPs}
\label{sec-results-neutralino}

We begin the presentation of our results with neutralino and chargino NLSPs. In general, the neutralinos are mixtures of the bino ($\tilde B$), the neutral wino ($\tilde W^0$) and the neutral higgsinos ($\tilde H^0_u$ and $\tilde H^0_d$), while charginos are mixtures of the charged winos ($\tilde W^\pm$) and higgsinos ($\tilde H_u^+$ and $\tilde H_d^-$). For simplicity, we will ignore the possibility of mixing and analyze the collider signals of each gauge eigenstate separately. See~\cite{Meade:2009qv, Ruderman:2011vv} for a recent review of many of the results we quote below.

We will study simplified spectra for these scenarios which include the NLSP, together with squarks and gluinos for colored production, as in \cite{Ruderman:2011vv}. For bino NLSPs, which are never directly produced, we will also investigate a simplified spectrum for electroweak production by including winos. For spectra with squarks, we will assume all the flavors to be approximately degenerate except for right-handed up-type squarks which are decoupled: $m_U \gg m_D = m_Q$. We decouple them in order to be able to satisfy the sum rules of GGM~\cite{Meade:2008wd}. We assume all other SUSY particles, sfermions and gauginos, are decoupled unless otherwise explicitly stated.

\subsection{Bino NLSP}

\begin{table}[!h]
\begin{center}
\begin{tabular}{|c|c|c|}
\hline
particle & mass & relevant decays \\
\hline
$\tilde{g}$ & $M_{\rm gluino}$ & $\tilde{g}\rightarrow j\tilde q^{(*)}$ \\ \hline
$\tilde{q}$ & $M_{\rm squark}$ & $\tilde{q}\rightarrow j\tilde g\,\, {\rm or}\,\,j\chi^0_1$ \\ \hline
$\chi^0_1$ & $M_{\rm bino}$ &  $\chi^0_1 \rightarrow (\gamma\,\,\mathrm{or}\,\, Z)\,\tilde{G}$\\ \hline
\end{tabular}
\end{center}
\capt{Simplified parameter space for bino NLSP and colored production. Here and below, all other sparticles not explicitly shown are assumed to be decoupled, unless otherwise stated. Promising final states here include $\gamma\gamma+$MET and $\gamma+$jets+MET.}
\label{tab-bino}
\end{table}

\begin{table}[!h]
\begin{center}
\begin{tabular}{|c|c|c|}
\hline
particle & mass & relevant decays \\
\hline
$\chi^0_2$ & $M_{\rm wino}$ & $\chi^0_2\rightarrow h^{(*)}\chi^0_1$, $Z^{(*)}\chi^0_1$ \\ \hline
$\chi^\pm_1$ & $M_{\rm wino}$ & $\chi^\pm_1\rightarrow W^{\pm(*)}\chi^0_1$ \\ \hline
$\chi^0_1$ & $M_{\rm bino}$ &  $\chi^0_1 \rightarrow (\gamma\,\,\mathrm{or}\,\, Z)\,\tilde{G}$\\ \hline
\end{tabular}
\end{center}
\capt{Simplified parameter space for wino decaying to bino NLSP.}
\label{tab-wino-bino}
\end{table}

A bino NLSP decays dominantly to $\gamma+\tilde G$ (with BR at least $\cos^2\theta_W$), and the rest of the time to $Z+\tilde G$. Because the bino NLSP cannot be directly produced, we consider two different simplified parameter spaces, one with colored production and the other with electroweak production. These are described  in Tables \ref{tab-bino} and \ref{tab-wino-bino}.

\begin{figure}[!t]
\begin{center}
\includegraphics[width=1\textwidth]{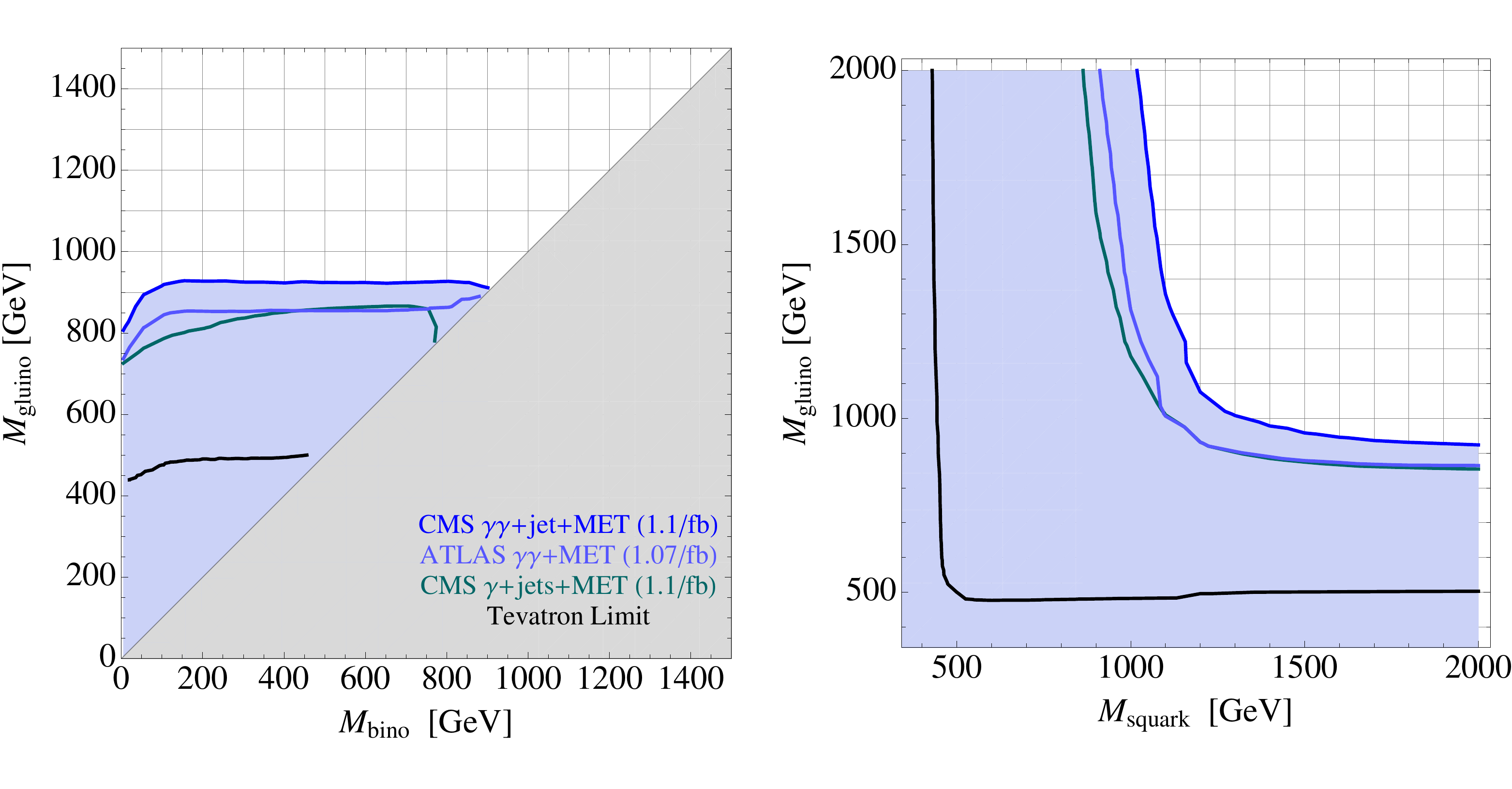}
\end{center}
\capt{Current best limits on the simplified parameter space for bino NLSP described in Table \ref{tab-bino}, together with the Tevatron limit estimated in  \cite{Ruderman:2011vv}. Left: squark masses are decoupled. Right: NLSP mass is fixed at 375 GeV.}
\label{fig-bino-limits}
\end{figure}

Bino NLSP is one scenario where both ATLAS and CMS are already expressing their results in terms of GGM parameter spaces (the same as those in Table~\ref{tab-bino} in fact). Nevertheless, we have attempted to reproduce the official limits using our own code, partly in order to validate our simulations. Shown in Fig.~\ref{fig-bino-limits} are the limits we obtain on the simplified parameter spaces for bino NLSP. The $\gamma\gamma$(+jet)+MET limits agree well with those shown in~\cite{CMS-PAS-SUS-11-009,ATL-PHYS-SLIDE-2011-523} for gluino production.\footnote{CMS also shows the limits for squark production, and here our agreement is less good. We find a much stronger limit on $M_{\rm squark}$ than CMS does, in the limit of decoupled gluino. We believe the problem may stem from the fact that {\sc Prospino} quotes two NLO cross sections: one with the K factor  computed using the average value for the squark mass, and the other with the squark masses kept non-degenerate. It would seem that CMS has used the former cross section from {\sc Prospino}, which would be very inaccurate for this simplified spectrum where $m_U$ has been decoupled in order to comply with the GGM sum rules. } We see that these searches set extremely strong bounds, as expected. The bound is slightly above $900$ GeV for gluino masses with decoupled squarks, and similarly for GGM-degenerate squark masses with decoupled gluinos. The improvement over the Tevatron limit (as estimated in \cite{Ruderman:2011vv} using the results of \cite{Abazov:2010us}) is strikingly good. Note that we have plotted the limit down to vanishing bino masses. The limit degrades somewhat as $M_{\rm bino}\to 0$, because there the bino will sometimes carry very little energy and the resulting MET and photon $p_T$ will be small.

Shown also is the limit from the CMS search for single photon + $\geq 3$ jets + MET. This is nearly as good as the diphoton searches, showing the power of having even a single high $p_T$ photon in cutting down on SM background. The standard jets+MET searches are less sensitive and are not shown here. This is partly due to the fact that these searches effectively {\it veto} on photons (this happens in the event cleaning procedure, which in order to safeguard against detector level noise contributions disqualifies events that contain jets with high electromagnetic component)~\cite{CMSsecretperson,ATLASsecretperson}.

\begin{figure}[!h]
\begin{center}
\includegraphics[width=0.47\textwidth]{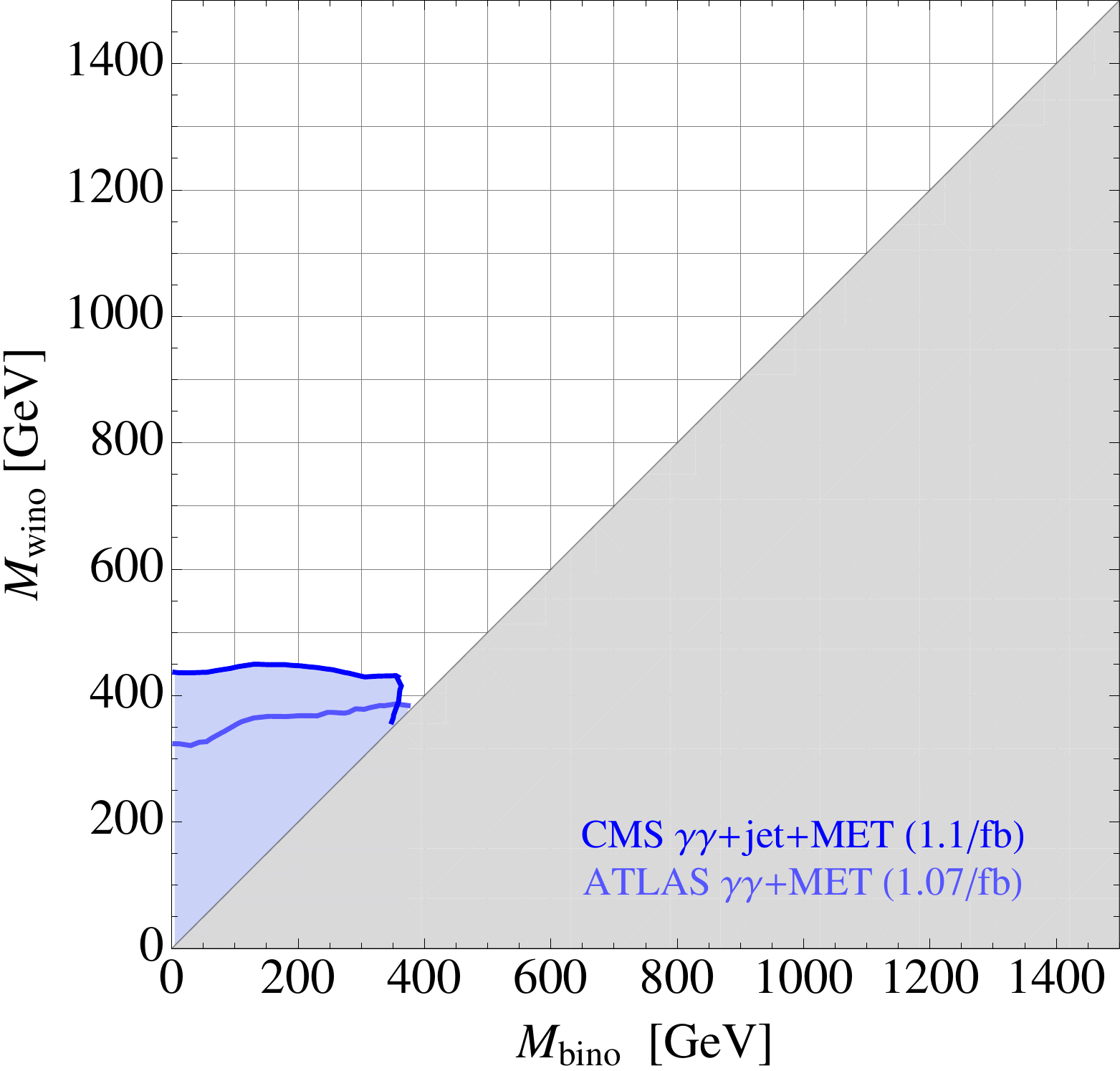}
\end{center}
\capt{Current best limits on the simplified parameter space for wino production and bino NLSP described in Table \ref{tab-wino-bino}.}
\label{fig-wino-bino-limits}
\end{figure}

We have also simulated electroweak production of winos that decay to a bino NLSP, as described in Table~\ref{tab-wino-bino}. The results are presented in Figure~\ref{fig-wino-bino-limits}. We find that the ATLAS and CMS searches for $\gamma\gamma$(+jet)+MET both set strong limits on the wino mass, with the latter giving a slightly better constraint. This is the only case we have simulated in which an electroweak production process is currently constrained by the LHC.  In fact, here  is one of the first instances of the LHC {\it outdoing} the Tevatron in constraining electroweak production. The latest D$\varnothing$ search~\cite{Abazov:2010us} sets a limit of 330 GeV on the wino mass along a ``Minimal Gauge Mediation" model line in which $\sim70\%$ of the cross section was from wino production (and the rest from sleptons). Decoupling the sleptons (as we have here) only further weakens the D$\varnothing$ limit on the wino mass.

\subsection{Wino co-NLSPs}

When the lightest neutralino is mostly wino-like, the charged wino will be nearly degenerate in mass, $\Delta m \sim m_Z^4/\mu^3$, due to an accidental cancellation \cite{Martin:1993ft}. So it will typically be a co-NLSP, in the sense that it will also decay to the gravitino \cite{Meade:2009qv}. The neutral wino will produce a $Z+\tilde G$ or $\gamma+\tilde G$ with branching fractions $\cos^2\theta_W$ and $\sin^2\theta_W$, respectively, for $m_{\tilde W^0} \gg m_Z$, while the charged wino will produce a $W+\tilde G$. The simplified model for the wino co-NLSPs is summarized in Table \ref{tab-wino}.

Although the branching fraction to $Z$ is dominant for winos, decays to photons are non-negligible (especially for lighter winos), so searches with photons can still be relevant. Searches with leptonically decaying $Z$'s may also have some sensitivity, despite the small leptonic branching ratio of the $Z$. Leptonic decays of $W$'s from the charged winos add the novel possibility of a signature combining a photon and a lepton, which has notably been searched for by CMS with 35/pb in the first ever dedicated search for wino co-NLSPs~\cite{Chatrchyan:2011ah}. Finally, jets + MET searches are relevant as well, with jets coming from gluino and squark decays, and also from decays of $W$ and $Z$ bosons.

\begin{table}[!t]
\begin{center}
\begin{tabular}{|c|c|c|}
\hline
particle & mass & relevant decays \\
\hline
$\tilde{g}$ & $M_{\rm gluino}$ & $\tilde{g}\rightarrow j\tilde q^{(*)}$ \\ \hline
$\tilde{q}$ & $M_{\rm squark}$ & $\tilde{q}\rightarrow j\tilde g,\,\,j\chi^0_1\,\,\mathrm{or}\,j\chi^\pm_1$ \\ \hline
$\chi^\pm_1$ & $M_{\rm wino}+\Delta m$ &  $\chi^\pm_1 \rightarrow W^\pm\, \tilde{G}$\\ \hline
$\chi^0_1$ & $M_{\rm wino}$ &  $\chi^0_1 \rightarrow (Z\,\,\mathrm{or}\, \,\gamma)\, \tilde{G}$\\ \hline
\end{tabular}
\end{center}
\capt{Simplified parameter space for wino co-NLSPs. Promising final states include $\gamma$+$\ell$+MET, $\gamma+$jets+MET, $\ell+$jets+MET, and jets+MET.}
\label{tab-wino}
\end{table}

\begin{figure}[!t]
\begin{center}
\includegraphics[width=1\textwidth]{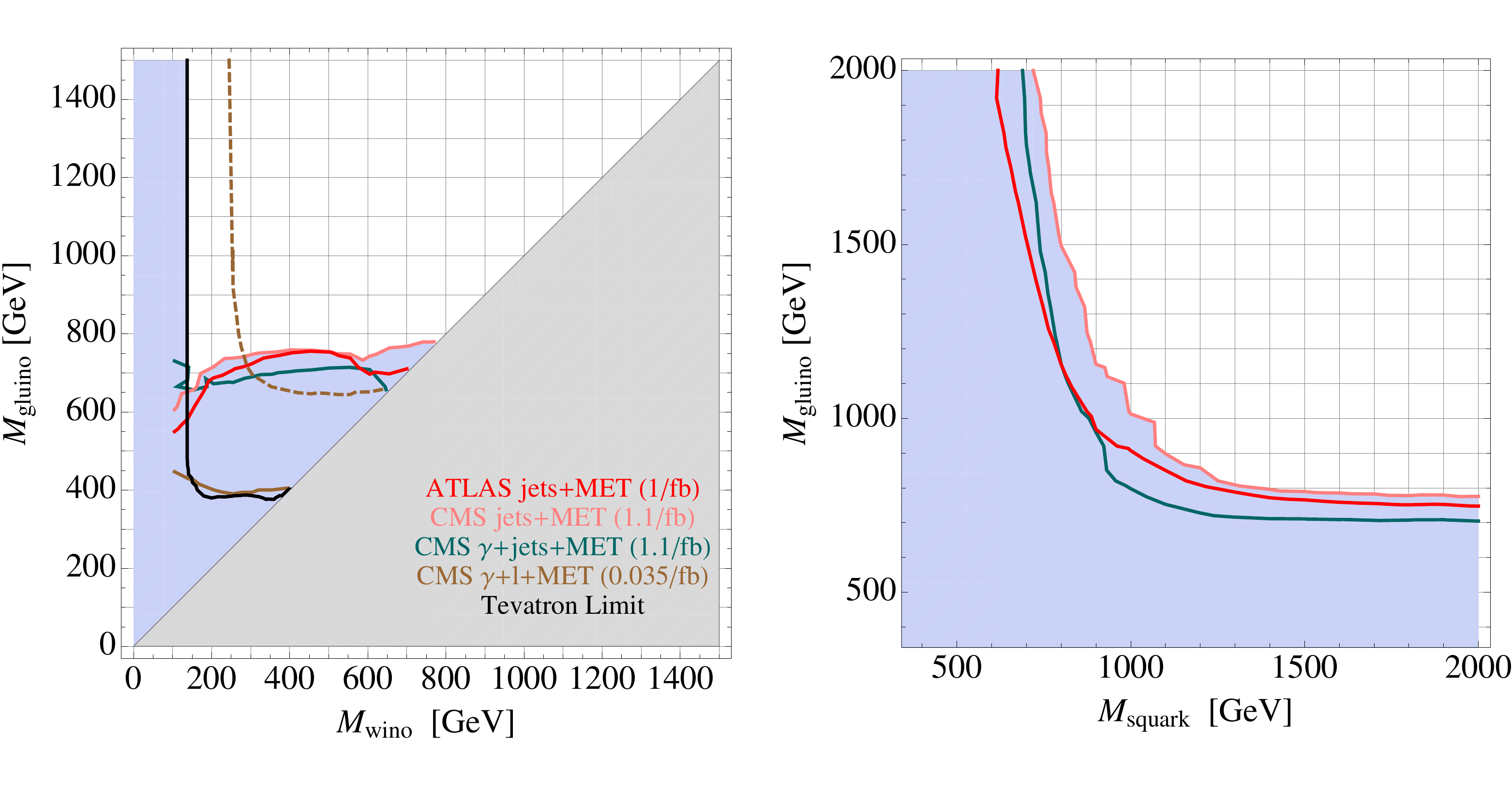}
\end{center}
\capt{Current best limits on the simplified parameter space for wino co-NLSPs described in Table \ref{tab-wino}, together with the Tevatron limit estimated in \cite{Ruderman:2011vv}. Left: squark masses are decoupled. Right: NLSP mass is fixed at 375 GeV.}
\label{fig-wino-limits}
\end{figure}

Shown in Fig.\ \ref{fig-wino-limits}\ are the best limits on the wino co-NLSPs scenario. We see that currently, the jets+MET searches and the $\gamma$+jets+MET search are most sensitive to this scenario. The lepton+jets+MET searches are significantly behind, so they are not shown. Note that the dedicated CMS $\gamma$+$\ell$+MET search for wino co-NLSPs hasn't been updated past 35/pb, so it's not meaningful to directly compare its limits with the 1/fb searches. Still, it is interesting to consider what $\gamma$+$\ell$+MET could do if extrapolated to 1/fb. We have attempted to estimate this in a somewhat optimistic way, by assuming that the backgrounds remain where they are (nearly zero), but that the signal efficiency also remains where it is. In practice of course, to keep the backgrounds low the cuts would need to be hardened, which might degrade signal efficiency. In any event, in our optimistic projection, we see that the updated CMS $\ell$+$\gamma$+MET search (dashed line) could have sensitivity to direct wino production. This would join production of winos decaying to binos as the only LHC constraints so far on direct electroweak SUSY production in our simplified GMSB scenarios. This is a strong motivation for continuing to do the search for wino co-NLSPs in this final state.

\subsection{$Z$-rich Higgsino NLSP}
\label{sec:Zrich}

\begin{table}[!h]
\begin{center}
\begin{tabular}{|c|c|c|}
\hline
particle & mass & relevant decays \\
\hline
$\tilde{g}$ & $M_{\rm gluino}$ & $\tilde{g}\rightarrow j\tilde q,\,\,j\chi^0_{1,2},\,\,t{\bar t}\chi^0_{1,2},\,\,t{\bar b}\chi^-_1\,\,{\rm or}\,\,b{\bar t}\chi^+_1$ \\ \hline
$\tilde{q}$ & $M_{\rm squark}$ & $\tilde{q}\rightarrow j\tilde g,\,\, j\chi^0_{1,2}\,\,\mathrm{or}\,\,j\chi^\pm_1$ \\ \hline
$\chi^0_2$ & $M_{\rm higgsino}+\Delta m$ &  $\chi^0_2 \rightarrow Z^* \chi^0_1\,\,\mathrm{or}\,\,W^{\pm*}\chi_1^\mp$\\ \hline
$\chi^\pm_1$ & $M_{\rm higgsino}+\Delta m'$ &  $\chi^\pm_1 \rightarrow W^{\pm*} \chi^0_1$\\ \hline
$\chi^0_1$ & $M_{\rm higgsino}$ &  $\chi^0_1 \rightarrow Z\, \tilde{G}$\\ \hline
\end{tabular}
\end{center}
\capt{Simplified parameter space for $Z$-rich higgsino NLSPs. Promising final states include $Z(\ell\ell)$+jets+MET, $Z(\ell\ell)Z(\ell'\ell')$+jets+MET, jets+MET, and SS dileptons+MET.}
\label{tab-Zhiggsino}
\end{table}

The NLSP can also be a neutral higgsino, which typically produces a $Z$ or a higgs $h$. Since the branching ratio is model dependent, we will focus on the two extreme cases in which the higgsino decays only to $Z+\tilde G$ or only to $h+\tilde G$, starting in this subsection with the $Z$ case.

The mass splittings between the three higgsino mass eigenstates (two neutral and one charged) are $ \Delta m,\,\Delta m'\sim m_Z^2/M_{1,2}$, so they typically will not be co-NLSPs. This will be the assumption in our scenarios. The simplified model for $Z$-rich higgsino NLSPs is given in Table \ref{tab-Zhiggsino}. Here and in the $h$-rich higgsino simplified models, we have fixed $\tan\beta = 2$ as in \cite{Ruderman:2011vv}, which leads to a fraction of gluino decays to $t{\bar t}\chi^0$ when phase space is available. At larger $\tan\beta$, decays to $b{\bar b}\chi^0$ would be more common, but then we would be in the mixed $Z/h$ higgsino NLSP case, as discussed in \cite{Meade:2009qv}. The gluino decays to $t{\bar b}\chi_1^-$ and to $g\chi^0$ (through a loop) are always present regardless.

\begin{figure}[!t]
\begin{center}
\includegraphics[width=1\textwidth]{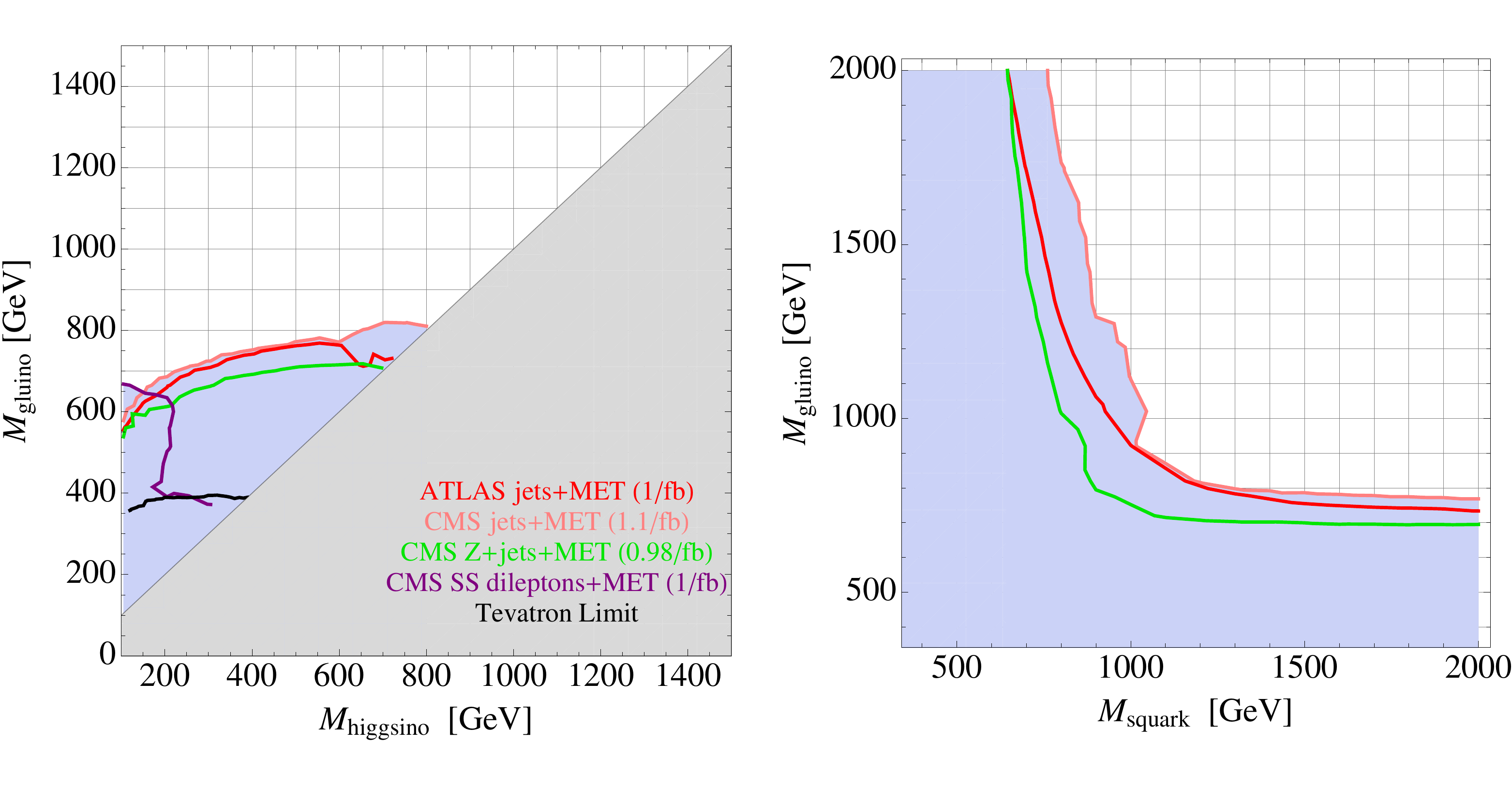}
\end{center}
\capt{Current best limits on the simplified parameter space for $Z$-rich higgsino NLSP described in Table \ref{tab-Zhiggsino}, together with the Tevatron limit estimated in \cite{Ruderman:2011vv}. Left: squark masses are decoupled. Right: NLSP mass is fixed at 375 GeV.}
\label{fig-zhiggsino-limits}
\end{figure}

The $Z$-rich higgsino case can be searched for using leptonically decaying $Z$'s, a possibility that was explored in \cite{Meade:2009qv, Ruderman:2011vv}. Fig.\ \ref{fig-zhiggsino-limits} illustrates the current limits on $Z$-rich higgsino NLSPs. We see that the CMS $Z(\ell\ell)$+jets+MET search sets a slightly worse limit than the all-purpose jets+MET searches.
The reason for this appears to be that jets+MET set the best limit when the requirements on hadronic activity are the highest: specifically, the 4jHM subanalysis for ATLAS and the high $H_T$ subanalysis for CMS. The CMS $Z(\ell\ell)$+jets+MET analysis meanwhile has only a single, much less stringent jet selection. So we would suggest redoing $Z(\ell\ell)$+jets+MET along the lines of the ATLAS 4jHM and/or CMS high $H_T$ jets+MET selections, as one simple way of optimizing for this scenario.

At large gluino-higgsino splitting, there is also a bound from same-sign (SS) dileptons from decay chains with top quarks. At low higgsino masses, this compensates somewhat for the degradation in the jets+MET limits due to the fact that the MET is being squeezed out (since the higgsino is decaying to the massive $Z$). The effect of this squeezing is especially significant, if one takes into account the fact that the cross section for direct higgsino production is growing as the higgsino mass decreases.

We see that at 1/fb, none of the analyses can see direct higgsino production. Combined with the lack of the Tevatron limit, there is still no direct limit on the higgsino NLSP mass from either hadron collider. In~\cite{Ruderman:2011vv}, it was estimated that direct higgsino production could be seen with $Z$+jets+MET or $ZZ$+MET searches with 5/fb. In Section~\ref{sec:leptonicmt2}, we will suggest a new improvement for the $Z$+jets+MET search which should allow the sensitivity to direct higgsino production to be greatly improved.

\subsection{$h$-rich Higgsino NLSP}

\begin{table}[!h]
\begin{center}
\begin{tabular}{|c|c|c|}
\hline
particle & mass & relevant decays \\
\hline
$\tilde{g}$ & $M_{\rm gluino}$ & $\tilde{g}\rightarrow j\tilde q,\,\,j\chi^0_{1,2},\,\,t{\bar t}\chi^0_{1,2},\,\,t{\bar b}\chi^-_1\,\,{\rm or}\,\,b{\bar t}\chi^+_1$ \\ \hline
$\tilde{q}$ & $M_{\rm squark}$ & $\tilde{q}\rightarrow j\tilde g,\,\, j\chi^0_{1,2}\,\,\mathrm{or}\,\,j\chi^\pm_1$ \\ \hline
$\chi^0_2$ & $M_{\rm higgsino}+\Delta m$ &  $\chi^0_2 \rightarrow Z^* \chi^0_1\,\mathrm{or}\,W^{\pm*}\chi^\mp_1$\\ \hline
$\chi^\pm_1$ & $M_{\rm higgsino}+\Delta m'$ &  $\chi^\pm_1 \rightarrow W^{\pm*} \chi^0_1$\\ \hline
$\chi^0_1$ & $M_{\rm higgsino}$ &  $\chi^0_1 \rightarrow h\, \tilde{G}$\\ \hline
\end{tabular}
\end{center}
\capt{Simplified parameter space for $h$-rich higgsino NLSPs. Promising final states include $b$-jets+MET, jets+MET and SS dileptons+MET.}
\label{tab-hhiggsino}
\end{table}

\begin{figure}[!t]
\begin{center}
\includegraphics[width=1\textwidth]{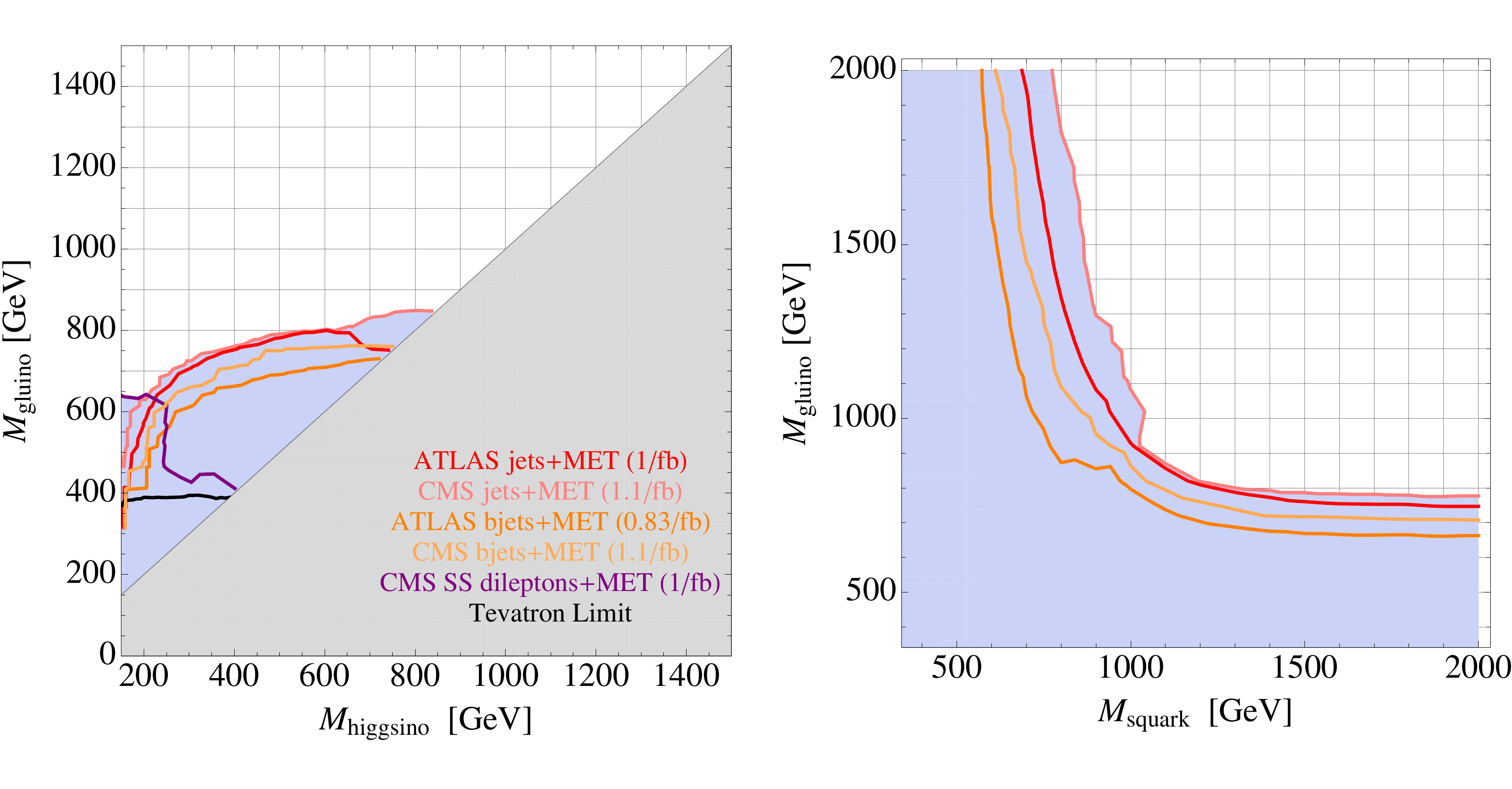}
\end{center}
\capt{Current best limits on the simplified parameter space for $h$-rich higgsino NLSP described in Table \ref{tab-hhiggsino}, together with the Tevatron limit estimated in \cite{Ruderman:2011vv}. Left: squark masses are decoupled. Right: NLSP mass is fixed at 375 GeV.}
\label{fig-hhiggsino-limits}
\end{figure}

For the $h$-rich higgsino case, we will assume the higgs mass to be $m_h = 120$~GeV, so the dominant decay mode is $h \to b\bar b$. In this case, jets + MET searches, especially those that require multiple $b$ jets, can be useful. Our simplified parameter space for $h$-rich higgsino NLSP is shown in Table \ref{tab-hhiggsino}. Again, $\tan\beta = 2$ as in~\cite{Ruderman:2011vv}.

Shown in Fig.\ \ref{fig-hhiggsino-limits} are the best limits on $h$-rich higgsino NLSPs. We see that the standard jets+MET searches are slightly edging out the more specialized $b$-jets+MET searches from ATLAS and CMS. We believe the reason for this is the same as for the $Z$+jets+MET search and $Z$-rich higgsino NLSP. Once again, the standard jets+MET searches set the best limits with high-hadronic-activity categories that the $b$-jets+MET searches lack. So we would again suggest adding a higher-hadronic-activity category to the ATLAS and CMS $b$-jets+MET searches, in order to optimize for this scenario.

Once again, there is a bound from SS dileptons at large gluino-higgsino splitting,  from decay chains with top quarks. This is again complementary to the jets+MET searches which degrade significantly at low higgsino mass due to the MET being squeezed out. Note that the degradation is more pronounced here than in the $Z$-rich case. This is because there the $Z$ could decay to neutrinos, giving a source of MET even in the squeezed region.

We see that at 1/fb, none of the analyses can see direct $h$-higgsino production. This might require an optimized analysis with softer cuts and more data. Requiring more $b$-tags, studying mass distributions of jet pairs, or considering boosted higgses might be interesting strategies. Some further discussion of searches that might be useful for $h$-higgsinos appears in~\cite{Meade:2009qv,Ruderman:2011vv,Kribs:2009yh}.

\subsection{Chargino NLSP}
\label{sec-results-chargino}

\begin{table}[!h]
\begin{center}
\begin{tabular}{|c|c|c|}
\hline
particle & mass & relevant decays \\
\hline
$\tilde{g}$ & $M_{\rm gluino}$ & $\tilde{g}\rightarrow j\tilde q,\,\,j\chi^0_{1,2},\,\,t{\bar t}\chi^0_{1,2},\,\,t{\bar b}\chi^-_1\,\,{\rm or}\,\,b{\bar t}\chi^+_1$ \\ \hline
$\tilde{q}$ & $M_{\rm squark}$ & $\tilde{q}\rightarrow j\tilde g,\,\, j\chi^0_{1,2}\,\,\mathrm{or}\,\,j\chi^\pm_1$ \\ \hline
$\chi^0_{1,2}$ & $M_{\rm chargino}+\Delta m_{1,2}$ &  $\chi^0_{1,2} \rightarrow W^{\pm\ast}\chi^\mp_1$\\ \hline
$\chi^\pm_1$ & $M_{\rm chargino}$ &  $\chi^\pm_1 \rightarrow W^{\pm} {\tilde G}$\\ \hline
\end{tabular}
\end{center}
\capt{Simplified parameter space for chargino NLSPs (which we take to be mostly higgsino). The bino is also present, with mass $\sim 1.2\times M_{\rm chargino}$, but it participates rarely in decay chains and does not contribute significantly to any final state. Promising final states include SS dileptons+MET, OS dileptons+MET, lepton+jets+MET, and jets+MET.}
\label{tab-chargino}
\end{table}

\begin{figure}[!t]
\begin{center}
\includegraphics[width=\textwidth]{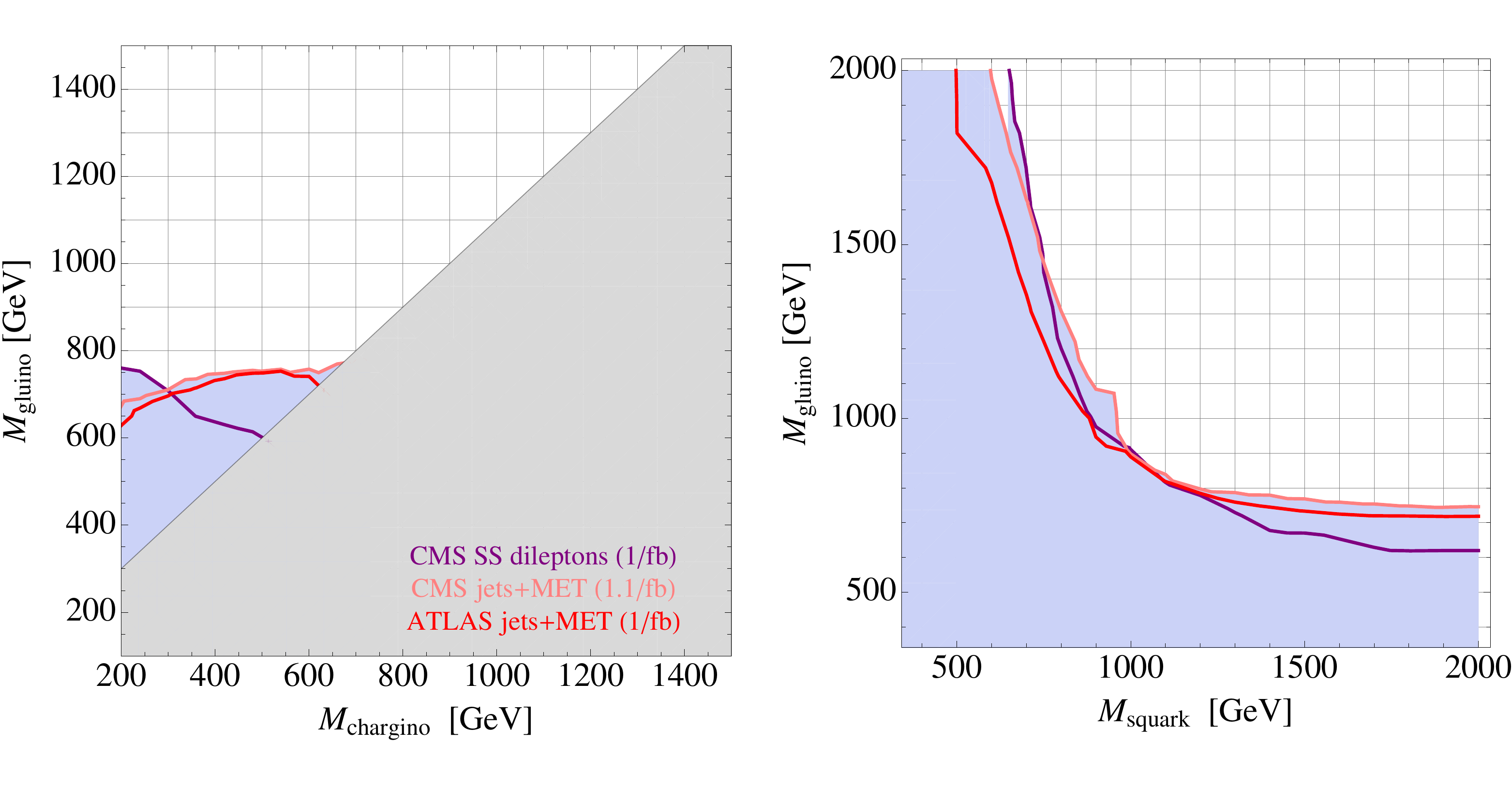}
\end{center}
\capt{Current best limits on the simplified parameter space for chargino NLSP described in Table \ref{tab-chargino}. Left: squark masses are decoupled. Right: NLSP mass is fixed at 375 GeV.}
\label{fig-chargino-limits}
\end{figure}

As discussed in~\cite{Kribs:2008hq}, there is a limited region of parameter space in which a chargino can be the NLSP. Here we study one such slice of parameter space, in which we have light higgsinos and bino with $M_1 = -1.2 \mu$ and $\tan\beta = 2$. (One could also have chargino NLSPs that are winos, but this case would be almost indistinguishable from the wino co-NLSPs case we have analyzed above.) We assume production from gluinos and squarks, as described in table~\ref{tab-chargino}. Bino-higgsino mixings are important for arranging a chargino NLSP, but the mostly-bino state is only rarely involved in the decay chain. Because the bottom of the cascade can be any sign of chargino, there is a strong limit from same-sign dileptons, presented in Figure~\ref{fig-chargino-limits}. (OS dileptons+MET also sets a limit, but it is always weaker than SS dileptons, so we have not shown it here.) There are also limits from jets+MET and lepton+jets+MET, as to be expected. In order to  avoid cluttering the plots, we only show the limits from the former, as the limits from the latter are quite similar.

Again, there does not appear to be any sensitivity to direct chargino NLSP production from any of the 1/fb LHC searches. The signature consists of $W^+W^-+$MET, so it would be difficult to discern under the enormous SM $t\bar t$ and $W^+W^-$ background.  We expect that a variable such as $M_{T2}$ will be very useful for background discrimination in this case, similar to the cases discussed in Section~\ref{sec:leptonicmt2}.

\section{Limits on Slepton co-NLSPs}
\label{sec-results-slepton}

Next we come to another major class of GGM scenarios: those with slepton NLSPs. Either the right-handed (RH) charged sleptons or the left-handed (LH) doublets (containing a charged slepton and a sneutrino) can be at the bottom of the spectrum. In both cases, one needs to specify to what extent the different generations of sleptons are degenerate. The splittings depend on the parameters of the higgs sector, and they can be sufficiently small so that all the flavors are co-NLSPs. We will assume this for simplicity in all of our simplified models.

\subsection{Slepton co-NLSPs}

\begin{table}[!h]
\begin{center}
\begin{tabular}{|c|c|c|}
\hline
particle & mass & relevant decays \\
\hline
$\tilde{g}$ & $M_{\rm gluino}$ & $\tilde{g}\rightarrow jj\chi^0_1$ \\ \hline
$\chi^0_1$ & $M_{\rm bino}$ &  $\chi^0_1 \rightarrow \ell^\pm \tilde{\ell}^\mp_R$\\ \hline
$\tilde{\ell}^\pm_R$  & $M_{\rm slepton}$ &  $\tilde{\ell}^\pm_R \rightarrow \ell^\pm \tilde{G}$\\ \hline
\end{tabular}
\quad
\begin{tabular}{|c|c|c|}
\hline
particle & mass & relevant decays \\
\hline
$\tilde{g}$ & $M_{\rm gluino}$ & $\tilde{g}\rightarrow jj\chi^0_1\,\,\mathrm{or}\,\,jj\chi^\pm_1$ \\ \hline
$\chi^\pm_1$& $M_{\rm wino}$ &  $\chi^\pm_1 \rightarrow  \nu_\tau\tilde{\tau}^\pm_1$ \\ \hline
$\chi^0_1$ & $M_{\rm wino}$ & $\chi^0_1 \rightarrow  \tau^\pm\tilde{\tau}^\mp_1$  \\ \hline
$\tilde\tau_1^\pm$  & $M_{\rm slepton}$ &  $\tilde\tau_1^\pm \rightarrow \tau^\pm \tilde{G}$\\ \hline
\end{tabular}
\end{center}
\capt{Simplified parameter spaces for right-handed slepton co-NLSPs. On the left (right) is the ``flavor democratic" (``tau-rich") scenario.  In the latter case, the other right-handed sleptons are degenerate with $\tilde\tau_1$, but they do not participate in any decay chains. Promising final states include SS dileptons+MET, OS dileptons+MET, multileptons+MET, lepton+jets+MET, and jets+MET. }
\label{tab-slepton}
\end{table}

Simplified parameter spaces for right-handed slepton co-NLSPs were studied recently in~\cite{Ruderman:2010kj}, and here we will mostly follow the same approach.  RH slepton co-NLSPs decay down to their SM partner lepton plus gravitino. Sleptons can be produced directly via electroweak interactions, however the rate and signature are difficult to distinguish from background (which should be mostly $t\bar t$ and $W^+W^-$, for which leptonic $M_{T2}$ might be useful, as discussed in Section~\ref{sec:leptonicmt2}).  More spectacular signatures come from when leptons are produced via the decays of heavier superpartners to sleptons.  In this case additional leptons will be present and same-sign dilepton or multilepton searches would be able to discover such events.

Our simplified parameter spaces for slepton co-NLSPs are shown in Table \ref{tab-slepton}. To include these multilepton signatures we will again assume the initial production to be through a colored state, the gluino.  In order to have no more than 3-body decays, we further assume an intermediate neutralino/chargino state. The mass of the intermediate gaugino state will not be scanned in our results, but will be fixed exactly between the slepton and gluino masses. Following~\cite{Ruderman:2010kj}, we will examine two distinct cases.  First, if there is an intermediate bino, the decays of the bino into the slepton NLSPs will be ``flavor democratic."  Here every event will have four leptons, plus missing energy and jets. The second case will be ``tau-rich" with a wino intermediate state.  As was discussed in~\cite{Ruderman:2010kj}, the wino only couples to right-handed sleptons through mixing, and the mixing can be arranged so that the wino decays exclusively to staus. Then every event will have 2-4 taus, together with missing energy and jets.

\begin{figure}[!t]
\begin{center}
\includegraphics[width=1\textwidth]{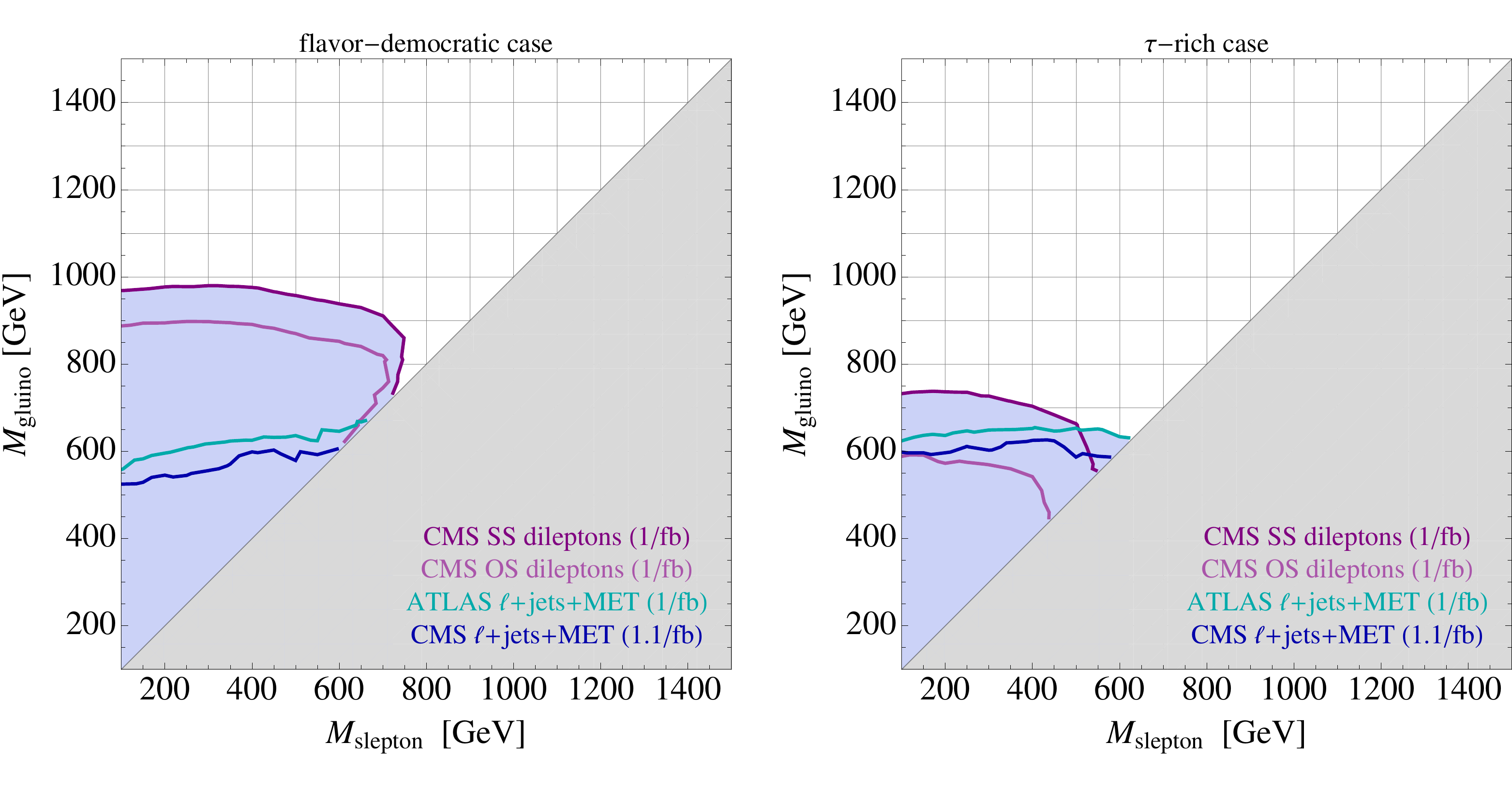}
\end{center}
\capt{Current best limits on the simplified parameter spaces for right-handed slepton co-NLSPs described in Table \ref{tab-slepton}. Left: the ``flavor-democratic" case with intermediate bino. Right: the ``tau-rich" case with intermediate wino. In both cases, the EW-ino mass has been fixed to $(M_{\rm gluino}+M_{\rm slepton})/2$.}
\label{fig-slepton-limits}
\end{figure}

Shown in Fig.\ \ref{fig-slepton-limits} are the current limits on right-handed slepton co-NLSPs. We see that the limits from the CMS same-sign dileptons search are extremely strong for the flavor-democratic case, nearly $M_{\rm gluino}\gtrsim 1$ TeV. This is comparable to the $\gamma\gamma+$MET limit on the bino scenario and is an indication of a super-clean channel. By contrast, the limit in the $\tau$-rich case is much weaker, but still non-negligible, around $M_{\rm gluino}\gtrsim$ 700 GeV. This shows the relative difficulty of constraining $\tau$-rich final states.\footnote{Closer examination of the SS dileptons search reveals an interesting detail. The SS dileptons search has three different analysis categories: high $p_T$, inclusive, and tau. The former two use only $e$ and $\mu$, while the latter requires at least one $\tau$. One can check that in our $\tau$-rich scenario, where every event involves 2-4 taus, the high $p_T$ dilepton selection sets nearly the same limit as the tau selection. So even in this extreme case, including hadronic taus does not lead to much better sensitivity.}

The SS dileptons limit degrades significantly in the $M_{\rm gluino}\to M_{\rm slepton}$ regime. Here, the jets from the gluino decay are being kinematically squeezed out, as are the leptons from the bino/wino decay. The degradation of the limit appears to be mainly due to the former effect, since the SS dileptons search requires at least two jets with $p_T>40$ GeV. We would suggest trying to relax this cut, in order to access this interesting ``squeezed region."

We have also shown the limits from the lepton+jets+MET searches. These veto on additional leptons, but can still have sensitivity to slepton co-NLSPs when the additional leptons are too soft, or are hadronically-decaying taus. Also, the ATLAS lepton+jets+MET search has slightly weaker jet requirements than the SS dileptons search ($p_T>25$ GeV vs. $p_T>40$ GeV for subleading jets). For these reasons, ATLAS lepton+jets+MET turns out to set the best limit in the ``squeezed region" in the tau-rich case.

We have not shown the limits from jets+MET in Fig.\ \ref{fig-slepton-limits}. In the flavor-democratic case, the jets+MET searches have basically no sensitivity, because they veto on leptons. In the $\tau$-rich case, they have better sensitivity, from the hadronic tau decays, and end up setting a slightly worse limit than lepton+jets+MET.

\begin{table}[!h]
\begin{center}
\begin{tabular}{|c|c|c|}
\hline
particle & mass & relevant decays \\
\hline
$\chi^\pm_1$& $M_{\rm wino}$ &  $\chi^\pm_1 \rightarrow  \nu_\tau\tilde{\tau}^\pm_1$ \\ \hline
$\chi^0_1$ & $M_{\rm wino}$ &  $\chi^0_1 \rightarrow \ell^\pm \tilde{\ell}^\mp_R$\\ \hline
$\tilde{\ell}^\pm_R$  & $M_{\rm slepton}$ &  $\tilde{\ell}^\pm_R \rightarrow \ell^\pm \tilde{G}$\\ \hline
\end{tabular}
\end{center}
\capt{Simplified parameter space for wino decaying to slepton co-NLSPs.}
\label{tab-slepton-bino}
\end{table}

Finally, we should briefly mention that we have investigated EW production with slepton co-NLSPs. Analogously to the simplified model in Table~\ref{tab-wino-bino}, we considered wino production, but slepton co-NLSPs instead of bino NLSP. The simplified parameter space for this is summarized in Table~\ref{tab-slepton-bino}, where we arranged for the neutral wino decays to be flavor-democratic as in~\cite{Ruderman:2010kj}. We find no limit on the wino-slepton parameter space from any existing search. In particular, CMS SS dileptons sets no limit because of its jet and $H_T$ requirements. It would be interesting, for both right- and left-handed sleptons, to include the constraints from multilepton searches, especially the 2/fb CMS results that were recently released~\cite{CMSmultileptons}. These would presumably set strong limits on electroweak production.\footnote{Strong limits on the chargino mass were exhibited in~\cite{CMSmultileptons}. However, it is important to keep in mind that these were for spectra with both chargino and left-handed slepton production. So the translation of these limits to our simplified model in Table~\ref{tab-wino-bino} requires more work.}  However, that study required that the transverse impact parameter of leptons be less than $0.02~{\rm cm}$, to reduce heavy-flavor backgrounds, which implies that it is only sensitive to the very lowest possible SUSY breaking scales, beyond which the NLSP lifetime is long enough that this cut is inefficient.

\subsection{Sneutrino co-NLSPs}

\begin{table}[!h]
\begin{center}
\begin{tabular}{|c|c|c|}
\hline
particle & mass & relevant decays \\
\hline
$\tilde{g}$ & $M_{\rm gluino}$ & $\tilde{g}\rightarrow jj\chi^0_1$\\ \hline
$\chi^0_1$ & $M_{\rm bino}$ &  $\chi^0_1 \rightarrow \ell^\pm \tilde{\ell}^\mp_L$, $\nu\tilde\nu$ \\ \hline
$\tilde{\ell}^\pm_L$ & $M_{\rm sneutrino}+\Delta m$ & $\tilde{\ell}_L^\pm \rightarrow W^* \tilde\nu_\ell$\\ \hline
$\tilde\nu_\ell$  & $M_{\rm sneutrino}$ &  $\tilde\nu_\ell \rightarrow \nu_\ell \tilde G$\\\hline
\end{tabular}
\begin{tabular}{|c|c|c|}
\hline
particle & mass & relevant decays \\
\hline
$\tilde{g}$ & $M_{\rm gluino}$ & $\tilde{g}\rightarrow j\chi^0_{1,2},b{\bar t}\chi^+_1,$\\
& & $t{\bar b}\chi^-_1,b{\bar b}\chi^0_{1,2}$ \\ \hline
$\chi^\pm_1$& $M_{\rm higgsino}$ &  $\chi^\pm_1 \rightarrow  \tau^\pm\tilde\nu_\tau$ \\ \hline
$\chi^0_{1,2}$ & $M_{\rm higgsino}$ & $\chi^0_{1,2} \rightarrow  \tau^\pm\tilde{\tau}^\mp_1$  \\ \hline
$\tilde\tau^\pm_1$ & $M_{\rm sneutrino}+\Delta m$ & $\tilde\tau_1^\pm \rightarrow W^* \tilde\nu_\tau$\\ \hline
$\tilde\nu_\tau$  & $M_{\rm sneutrino}$ &  $\tilde\nu_\tau \rightarrow \nu_\tau \tilde G$\\\hline
\end{tabular}
\end{center}
\capt{Simplified parameter spaces for sneutrino co-NLSPs. On the left (right) is the ``flavor democratic" (``tau-rich") scenario.  In the latter case, the other left-handed sleptons are degenerate with $\tilde\tau_1$, $\tilde\nu_\tau$, but they do not participate in any decay chains. The promising final states here are similar to those for right-handed slepton co-NLSPs.}
\label{tab-sneutrino}
\end{table}

\begin{figure}[!h]
\begin{center}
\includegraphics[width=1\textwidth]{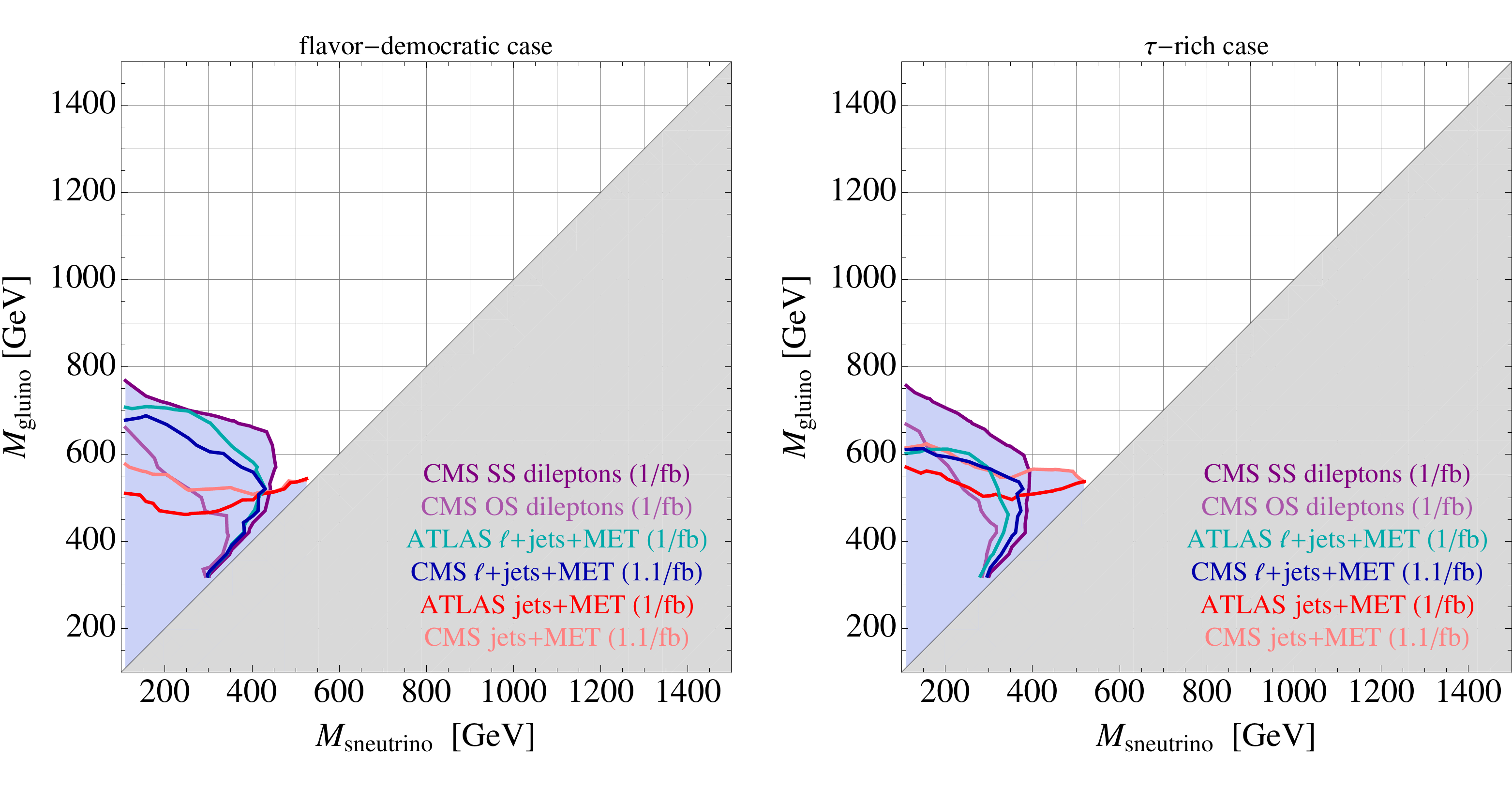}
\end{center}
\capt{Current best limits on the simplified parameter spaces for sneutrino co-NLSPs described in Table~\ref{tab-sneutrino}. Left: the ``flavor-democratic" case with intermediate bino. Right: the ``tau-rich" case with intermediate higgsino. In both cases, the EW-ino mass has been fixed to $(M_{\rm gluino}+M_{\rm sneutrino})/2$.}
\label{fig-sneutrino-limits}
\end{figure}

In the left-handed case, studied extensively in~\cite{Katz:2009qx,Katz:2010xg}, the charged slepton will typically be heavier than the sneutrino by an amount $\Delta m^2 \approx \l|\cos2\beta\r| m_W^2$. So the sneutrinos are the NLSPs, and the charged sleptons will decay to their corresponding sneutrinos plus $W^\ast$. Clearly, the decay of the sneutrino NLSP to neutrino plus gravitino will not itself be visible. Instead, one can hope to see multilepton signatures from decays of heavier states to the charged slepton and from its subsequent decay to the sneutrino. Such signatures are realized in the simplified parameter spaces described in Table \ref{tab-sneutrino}. This follows closely our simplified parameter spaces for RH slepton NLSPs in the previous subsection. In particular, there is a ``flavor democratic" scenario with intermediate bino, and a ``tau rich" scenario, now with intermediate higgsino instead of wino. The higgsino is induced to decay solely to the third generation by taking large $\tan\beta$.

Shown in Fig.\ \ref{fig-sneutrino-limits} are the current limits on left-handed sneutrino NLSPs.  As can be seen from the figure, the LH sneutrino bounds are generally worse than the RH charged slepton bounds. In the flavor-democratic case this happens because the RH scenario always has 4 leptons per event, while in the LH scenario there tend to be fewer and softer leptons (some binos decay invisibly to $\nu\tilde\nu$ and the charged slepton decays through a $W^*$). Similar but slightly more complicated counting applies to the $\tau$-rich case.

\section{Limits on Colored NLSPs}
\label{sec-results-colored}

Finally, we come to the third major class of NLSPs in GGM: colored NLSPs. These range from gluino NLSP and squark co-NLSPs ---which probably furnish the simplest possible ``simplified models" in the MSSM---to the more nontrivial cases of stop and sbottom NLSP.

\subsection{Gluino NLSP or Squark co-NLSPs}

The simplified spectrum for a gluino NLSP or squark co-NLSPs is very simple indeed.  Here we can truly decouple everything else in the MSSM and focus on direct NLSP production. For the squark NLSP case, we will assume that all the generations are roughly degenerate (with the right-handed up-type squarks decoupled, as before) so that all the squarks (except for the $\st_1$ and $\tilde b_2$) are co-NLSPs decaying to a quark and a gravitino. The third generation squarks $\st_1$ and $\tilde b_2$ are split enough from the other squarks by $D$-terms that they always prefer to decay instead to $\tilde b_1+W^\ast$ and $\tilde b_1+Z^\ast$ respectively. Our simplified models for gluino and squark NLSP are summarized in Table \ref{tab-gluino-squark}.

\begin{table}[!t]
\begin{center}
\begin{tabular}{|c|c|c|}\hline
particle & mass & relevant decays \\ \hline
$\tilde{g}$ & $M_{\rm gluino}$ & $\tilde{g}\rightarrow g\tilde{G}$ \\ \hline
\end{tabular}
\quad
\begin{tabular}{|c|c|c|}\hline
particle & mass & relevant decays \\ \hline
$\tilde t_1$ & $M_{\rm squark}+\Delta m$ & $\tilde t_1\to \tilde b_{1,2} W^{+\ast}$\\\hline
$\tilde b_2$ & $M_{\rm squark}+\Delta m'$ & $\tilde b_2\to \tilde b_1 Z^\ast$\\\hline
$\tilde b_1$ & $M_{\rm squark}$ & $\tilde b_1\to b\tilde{G}$\\\hline
$\tilde{q}$ & $M_{\rm squark}$ & $\tilde{q}\rightarrow q\tilde{G}$ \\ \hline
\end{tabular}
\end{center}
\capt{Simplified parameter spaces for gluino NLSP (left) and squark co-NLSPs (right). In the latter case, we take all squarks to be degenerate, except for $m_{u_R}$ which is decoupled in order to satisfy the GGM sum rules, as before. The most promising final state here is clearly jets+MET.}
\label{tab-gluino-squark}
\end{table}

\begin{figure}[!t]
\begin{center}
\includegraphics[width=1\textwidth]{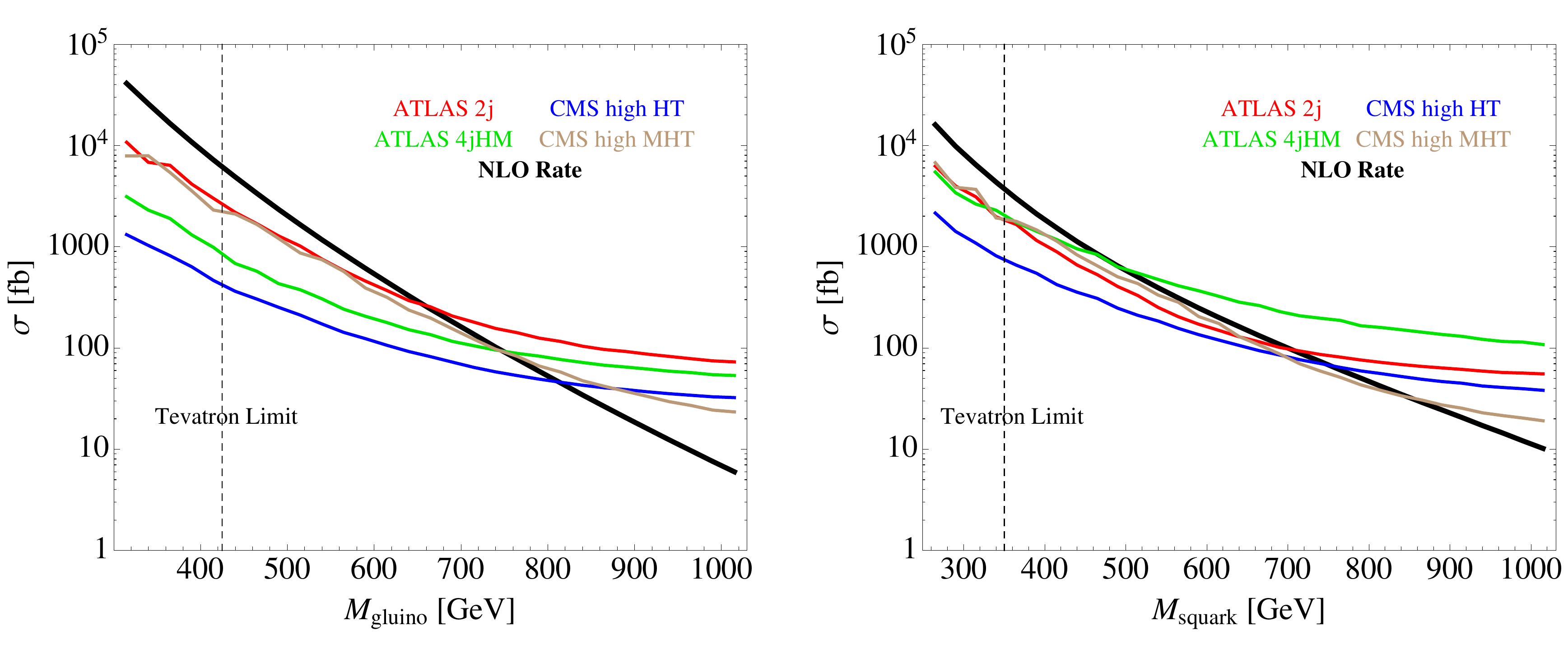}
\end{center}
\capt{Gluino NLSP (left) and squark co-NLSPs (right). The thick black lines indicate the NLO pair production cross sections for gluino and squarks, respectively. The colored lines indicate the limits on $\sigma_{\rm prod}$ from representative jets+MET searches.}
\label{fig-gluino-squark-limits}
\end{figure}

Pair production of either gluino NLSPs or squark co-NLSPs would result in events with (at least) two energetic jets and missing energy from the NLSP decays. Additional high $p_T$ jets are typically generated from ISR and FSR. We find that the high $p_T$ jet multiplicity is significantly higher for gluino NLSP than for squark co-NLSPs, because gluons tend to radiate more than quarks.

Obviously, gluino and squark NLSPs will be highly constrained by jets+MET searches. The current limits are shown in Fig.~\ref{fig-gluino-squark-limits}. For each experiment, we present results from a search region that is best for the gluino NLSP and one that is best for the squark co-NLSPs scenario. Some interesting differences emerge from this comparison. For example, looking at the ATLAS jets+MET limits, we see that the 2j search sets the best limit for squarks, but the 4jHM search sets the best limit for gluinos. This can be traced back to the higher jet multiplicity for gluino NLSPs discussed above.

Despite these minor differences, we see that the best limits on both scenarios are about $M_{\rm NLSP}\gtrsim 800-850$ GeV. These extremely stringent constraints on the bottom of the MSSM spectrum strongly disfavor the promptly-decaying gluino NLSP and degenerate squark co-NLSPs scenarios. Of course, if the squarks are not all degenerate, the limit can be much weaker; if the left-handed squarks are decoupled, so that only the right-handed down-type squarks are light, the limit is only 450 GeV.

\subsection{Third Generation Squark NLSPs}

Given that the point of weak-scale SUSY is to solve the gauge hierarchy problem, and that stop loops contribute most of all to the renormalization of the higgs mass, light stops are very well-motivated components of a natural MSSM spectrum, as is the light sbottom that comes with the left-handed stop. In general, even in a flavor blind mediation scheme such as GGM, the third generation squarks can be split from the first two generations. Already at tree level, the third generation is split due to off-diagonal mass matrix elements proportional to Yukawa couplings, which can be large either due to large $A$-terms or large $\mu$. Further large splittings among generations can arise when renormalization-group running effects are included from the SUSY breaking scale to the weak scale, due to the larger Yukawa couplings of the third generation. In this subsection we investigate the possibility that the third generation squarks are in fact the NLSP.  Having a third generation NLSP presents an interesting opportunity and challenge for the experiments given that $b$-jets can be tagged, and top quarks provide an entirely different experimental signature than all other quarks.

\begin{table}[!b]
\begin{center}
\begin{tabular}{|c|c|c|}
\hline
particle & mass & relevant decays \\
\hline
$\tilde{q}$ & $M_{\rm squarks}$ & $\tilde{q}\rightarrow j\chi^0_1$ \\ \hline
${\tilde b}_2$ & $M_{\rm squarks}$ & ${\tilde b}_2 \to h {\tilde b}_1$ \\ \hline
$\chi^0_1$ & $M_{\rm bino}$ &  $\chi^0_1 \rightarrow b\tilde{b}_1$ \\ \hline
$\tilde{b}_1$   & $M_{\rm sbottom}$ &  $\tilde{b}_1 \rightarrow b \tilde{G}$\\ \hline
\end{tabular}
\quad
\begin{tabular}{|c|c|c|}
\hline
particle & mass & relevant decays \\
\hline
$\tilde{q}$ & $M_{\rm squarks}$ & $\tilde{q}\rightarrow j\chi_1^0,\,\,j\chi_1^\pm$ \\ \hline
${\tilde b}_2$ & $M_{\rm squarks}$ & ${\tilde b}_2 \to h {\tilde b}_1$ \\ \hline
$\chi^0_1,\chi^\pm_1$ & $M_{\rm wino}$  & $\chi^\pm_1 \rightarrow t^{(*)}\tilde{b}_1$ or $\chi^0_1 \rightarrow b\tilde{b}_1$ \\ \hline
$\tilde{b}_1$   & $M_{\rm sbottom}$ &  $\tilde{b}_1 \rightarrow b \tilde{G}$\\ \hline
\end{tabular}
\\\vspace{4mm}
\begin{tabular}{|c|c|c|}
\hline
particle & mass & relevant decays \\
\hline
$\go$ & $M_{\rm gluino}$ & $\go \to  b \tilde{b}_1$ \\ \hline
$\tilde{b}_1$   & $M_{\rm sbottom}$ &  $\tilde{b}_1 \rightarrow b\tilde{G}$\\ \hline
\end{tabular}
\end{center}
\capt{Simplified parameter spaces for sbottom NLSP. In both, we take all the squarks but $\tilde b_1$ to be degenerate, with no regard to the GGM sum rules (in contrast with the rest of the paper). On the top left (squark-bino-sbottom), the signatures consist of $4b+$jets+MET. On the top right (squark-wino-sbottom) they include $4b+$jets+MET, $3b+t^{(\ast)}+$jets+MET, and $2b+2t^{(\ast)}+$jets+MET. Finally on the bottom (gluino-sbottom), the signature is $4b$+MET. This differs from the squark-bino-sbottom by not having the extra jets.}
\label{tab-sbottom}
\end{table}

In gauge mediation, $A$-terms are typically small, but this does not preclude interesting splittings from $\mu$ and running. An interesting question for the case of light third generation squarks is how much they can be split from the first and second generations in GGM. This has never been fully answered and deserves a closer look. However, it is beyond the scope of this paper. Instead, we will treat the stop or sbottom mass as a free parameter which can be arbitrarily light compared to the other squarks, though this is not necessarily realistic in the context of any given model. Even if part of this parameter space proves to be outside the reach of GGM, it could be compatible with other models of low-scale SUSY breaking with light gravitinos. Some interesting examples of model-building along these lines can be found in~\cite{ArkaniHamed:1997fq,Franco:2009wf,Craig:2011yk,Gabella:2007cp}.

In this section, more than others, we highlight squark rather than gluino production, simply because the presence of light third-generation squarks may lead in many (but not all) models to the other generations of squarks nearby. The cross section for several squarks degenerate in mass is large, so these scenarios can be strongly constrained. But in the case in which the other squarks are decoupled and only a single squark NLSP is light, the cross section can be much lower, and the NLSP is still allowed to be very light.

\subsubsection{Sbottom NLSP}

\begin{figure}[!b]
\begin{center}
\includegraphics[width=1\textwidth]{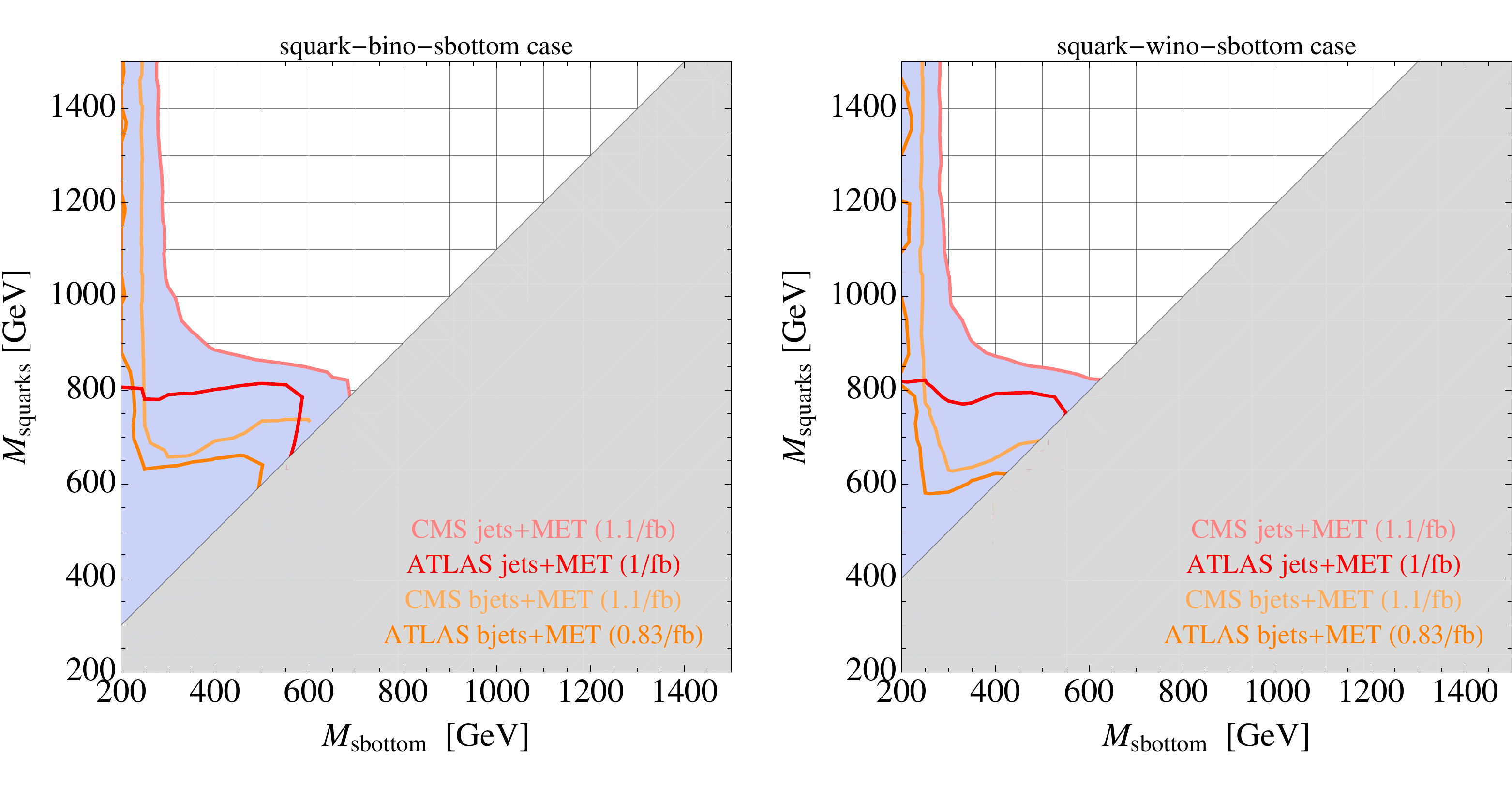}
\end{center}
\capt{Current best limits on the simplified parameter spaces for  sbottom NLSP described in Table \ref{tab-sbottom}. In both cases, the EW-ino mass has been fixed to $(M_{\rm squark}+M_{\rm sbottom})/2$. In the plot at right, the diagonal is positioned to allow on-shell three-body ${\tilde W}^- \to  b W^-{\tilde b}_1$ decays. In the plot at left, it is simply chosen to allow a 50 GeV gap between states.}
\label{fig-sbottom-limits}
\end{figure}

The simplified spectra for the sbottom NLSP are described in Table \ref{tab-sbottom}. We take  the sbottom to be mostly-right-handed (but a small LH component to allow for wino decays). We either allow for gluino production with decoupled squarks, or vice versa.

In the case of squark production, for simplicity, we fix {\it all} the other squarks to be at the same mass, and to be completely independent of $M_{\rm sbottom}$. Note that this means we are relaxing the strict adherence to the GGM sum rules in this subsection (and the next), in contrast with the rest of the paper. As in the case of the simplified models for the sleptons, we include either an intermediate bino or wino at a fixed mass (the average of the sbottom and squark masses), partly to keep the decays simple, and partly to populate different final states. The bino case gives $4b+$jets+MET, while the wino case gives a mixture of this along with $3b+t^{(\ast)}+$jets+MET and $2b+t^{(\ast)}+$jets+MET. So it is interesting to compare the two scenarios.

\begin{figure}[!t]
\begin{center}
\includegraphics[width=.5\textwidth]{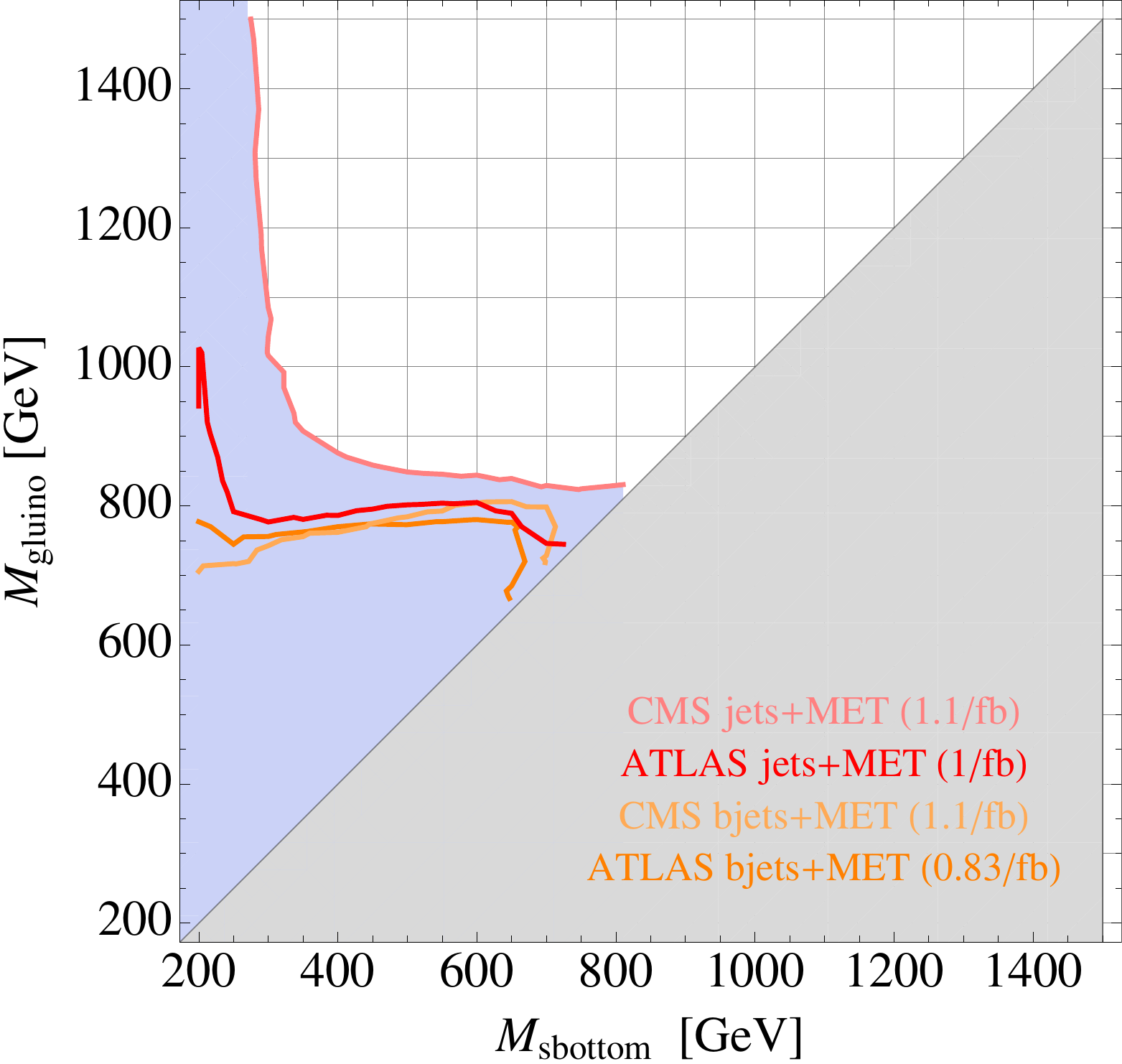}
\end{center}
\capt{Current best limits on the simplified parameter spaces for  sbottom NLSP with gluino production described in Table \ref{tab-sbottom}.}
\label{fig-gluino-sbottom-limits}
\end{figure}

The limits on these three simplified models  are shown in Figs.~\ref{fig-sbottom-limits} and \ref{fig-gluino-sbottom-limits}. Note the vertical branch of the CMS jets+MET limit curve, which shows that it is sensitive to direct production of a single sbottom even when the other squarks (and the gluino) are decoupled. The limit on the sbottom mass is about 280 GeV, comparable to (given the uncertainty of our estimate) the D$\varnothing$ limit of 250 GeV~\cite{Abazov:2010wq}.

\subsubsection{Stop NLSP (direct production)}

\begin{figure}[!t]
\begin{center}
\includegraphics[width=0.7\textwidth]{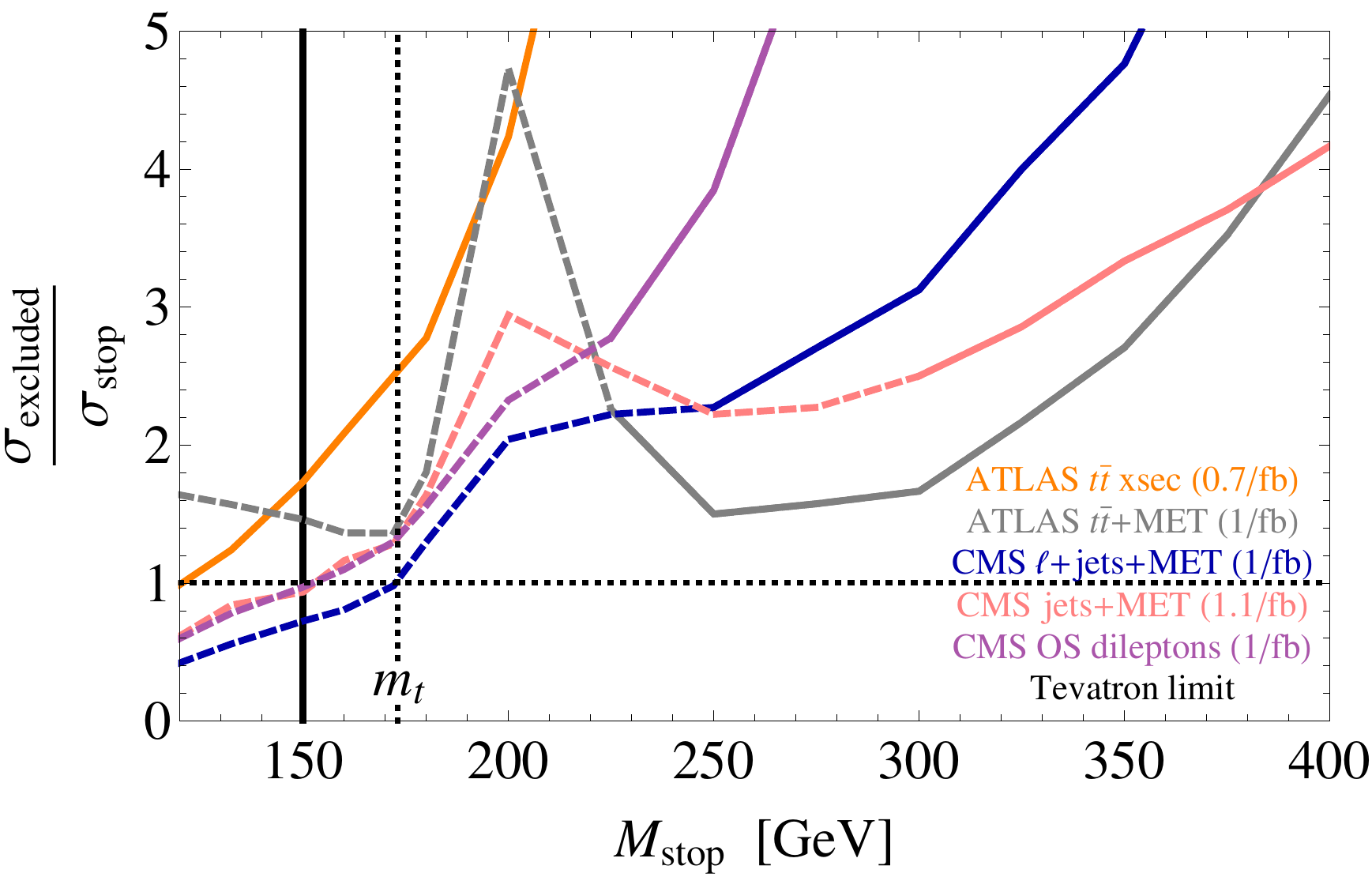}
\end{center}
\capt{Stop NLSP: limits on direct production (excluded cross section divided by the NLO+NLL stop production cross section from~\cite{Beenakker:2010nq}). Along with the best SUSY searches from Table~\ref{tab:searches}, we show the limits from the pre-tag sample of the ATLAS $t\bar t$ cross section measurement in the dilepton channel~\cite{ATLAS-CONF-2011-100} (orange) and the ATLAS search for $t\bar t$ events with large MET~\cite{Collaboration:2011wc} (gray). The curves are dashed in the low mass region where the efficiency of the jet-related and MET-related cuts (but not the leptonic selection) is below 1\%. This is to indicate that our results may not be reliable there, since the signal efficiencies are extremely low. The black line is the Tevatron limit estimated in~\cite{Kats:2011it} using the CDF search~\cite{Aaltonen:2009sf}.}
\label{fig-stop-direct-limits}
\end{figure}

The case of a stop NLSP is special because it may be very light, perhaps even lighter than the top. Stop NLSPs decay as $\st \to W b\gr$ (or $\st \to t \gr$ if $m_\st > m_t$). As a result, it is difficult to see direct production of stop pairs on top of the $t\bar t$ background. In~\cite{Kats:2011it}, simulations of the relevant Tevatron and LHC analyses were used to argue that a stop NLSP considered in isolation could still be as light as 150~GeV. We will re-examine the situation in view of the more recent searches. In the next subsection, we will consider more general scenarios in which the stops are produced from the decays of gluinos or squarks.

An obvious place to look for directly produced light stop NLSPs is in analyses of $t\bar t$ (or $t\bar t$-like) samples since they have similar signatures. In~\cite{Kats:2011it}, it was estimated that two of the $35\mbox{/pb}$ LHC analyses available at that time would be able to probe stops up to about $180$~GeV, if updated to 1/fb. The updated versions of these two analyses are the non-$b$-tagged sample of the ATLAS $t\bar t$ cross section measurement in the dilepton channel~\cite{ATLAS-CONF-2011-100} (0.7/fb) and the ATLAS search for $t\bar t$ events with anomalously large missing energy~\cite{Collaboration:2011wc} (1.04/fb). We find, using the methods of~\cite{Kats:2011it}, that neither sets the expected limit due to tightened analysis cuts. The cross section limits are shown in Fig.~\ref{fig-stop-direct-limits}. Since both analyses use lepton triggers, it may still be possible to repeat them with softer cuts. Overall, Fig.~\ref{fig-stop-direct-limits} indicates that~\cite{Collaboration:2011wc} is a very promising search up to 300 GeV or more. Its weakness near $M_{\rm stop} \sim 200$~GeV stems from the fact that for stops that are only slightly heavier than the top, the gravitinos carry very little energy (unlike in the 3-body decays of the lighter stops or the 2-body decays of the heavier stops) and therefore the cut on the transverse mass $m_T$ eliminates much of the signal.

At the same time, we see in Fig.~\ref{fig-stop-direct-limits} that some of the SUSY searches have become competitive and {\it may} have already surpassed the Tevatron limit by excluding direct production of stops up to approximately the top mass. However, since all these searches have very low efficiencies in that low mass region, the systematic uncertainties on our simulation are likely to be large, so the precise exclusion limits are highly uncertain. The very low efficiencies indicate that in applying these cuts, we are making use of the far tails of the kinematic distributions in the {\it signal}. We expect that these tails depend on radiation in the event, which we are simulating with Pythia. A more careful approach would use matching of matrix elements and parton showers to simulate stop pair production plus jets, which would be an interesting follow-up to pursue. It is also important to note that the CMS lepton+jets+MET search imposes a cut on transverse impact parameter of muons at $0.2~{\rm mm}$. This could be inefficient when NLSPs have $c\tau$ which is similar or larger, so the limit shown is strictly only valid for very low SUSY-breaking scales in GMSB.

Due to the key role played by naturalness in theoretical investigations of supersymmetry, it would be very interesting for any of the analysis groups mentioned above (ATLAS $t\bar t$+MET, CMS jets+MET, CMS lepton+jets+MET, CMS OS dileptons+MET) to release an official limit on light stop NLSPs. It would also be interesting to statistically combine different searches to achieve the strongest possible limit. Our results here amount to only a crude estimate; at best we can say that some of the current SUSY searches should have good sensitivity to ultra-light stops.

\begin{table}[!b]
\begin{center}
\begin{tabular}{|c|c|c|}
\hline
particle & mass & relevant decays \\
\hline
$\tilde{q}$ & $M_{\rm squarks}$ & $\tilde{q}\rightarrow j\chi^0_1$ \\ \hline
${\tilde t}_2$ & $M_{\rm squarks}$ & ${\tilde t}_2 \to h {\tilde t}_1$ \\ \hline
$\chi^0_1$ & $M_{\rm bino}$ &  $\chi^0_1 \rightarrow t^{(\ast)}\tilde{t}_1$ \\ \hline
$\tilde{t}_1$   & $M_{\rm stop}$ &  $\tilde{t}_1 \rightarrow t\tilde{G}$\\ \hline
\end{tabular}
\quad
\begin{tabular}{|c|c|c|}
\hline
particle & mass & relevant decays \\
\hline
$\tilde{q}$ & $M_{\rm squarks}$ & $\tilde{q}\rightarrow j\chi$ \\ \hline
${\tilde t}_2$ & $M_{\rm squarks}$ & ${\tilde t}_2 \to h {\tilde t}_1$ \\ \hline
$\chi^0_1,\chi^\pm_1$ & $M_{\rm wino}$  & $\chi^\pm_1 \rightarrow b\tilde{t}_1$ or $\chi^0_1 \rightarrow t^{(\ast)}\tilde{t}_1$ \\ \hline
$\tilde{t}_1$   & $M_{\rm stop}$ &  $\tilde{t}_1 \rightarrow t\tilde{G}$\\ \hline
\end{tabular}\\\vspace{4mm}
\begin{tabular}{|c|c|c|}
\hline
particle & mass & relevant decays \\
\hline
$\go$ & $M_{\rm gluino}$ & $\go \to t \st_1$ \\ \hline
$\tilde{t}_1$   & $M_{\rm stop}$ &  $\tilde{t}_1 \rightarrow t\tilde{G}$\\ \hline
\end{tabular}

\end{center}
\capt{Simplified parameter spaces for stop NLSP. In the first two cases, we take all the squarks but $\tilde t_1$ to be degenerate, with no regard to the GGM sum rules (in contrast with the rest of the paper). On the top left (squark-bino-stop), the signatures consist of $4t^{(\ast)}$+jets+MET. On the top right (squark-wino-stop) they include this signature, together with $3t^{(\ast)}$+$b$+jets+MET, and $2t^{(\ast)}$+$2b$+jets+MET. Finally, on the bottom (gluino-stop), the signature is $4t^{(*)}$+MET. This differs from the squark-bino-stop by not having the extra jets.}
\label{tab-stop}
\end{table}

It may also be interesting to do a search in the dilepton channel where the $t\bar t$ background can be reduced by using the variable ``leptonic $M_{T2}$'' that we will discuss in the next section. One could also look in the jets+MET channel of the $t\bar t$ sample, as in the recent CDF search~\cite{Aaltonen:2011na}. In this channel, stop events would naturally be enhanced relative to the top since such a final state does not usually arise for the top unless hadronically decaying taus are involved. Given all of these possible strategies, we believe that in the near future the LHC should either discover or rule out a light stop up to several hundred GeV.

\subsubsection{Stop NLSP (from additional colored production)}

\begin{figure}[!t]
\begin{center}
\includegraphics[width=.92\textwidth]{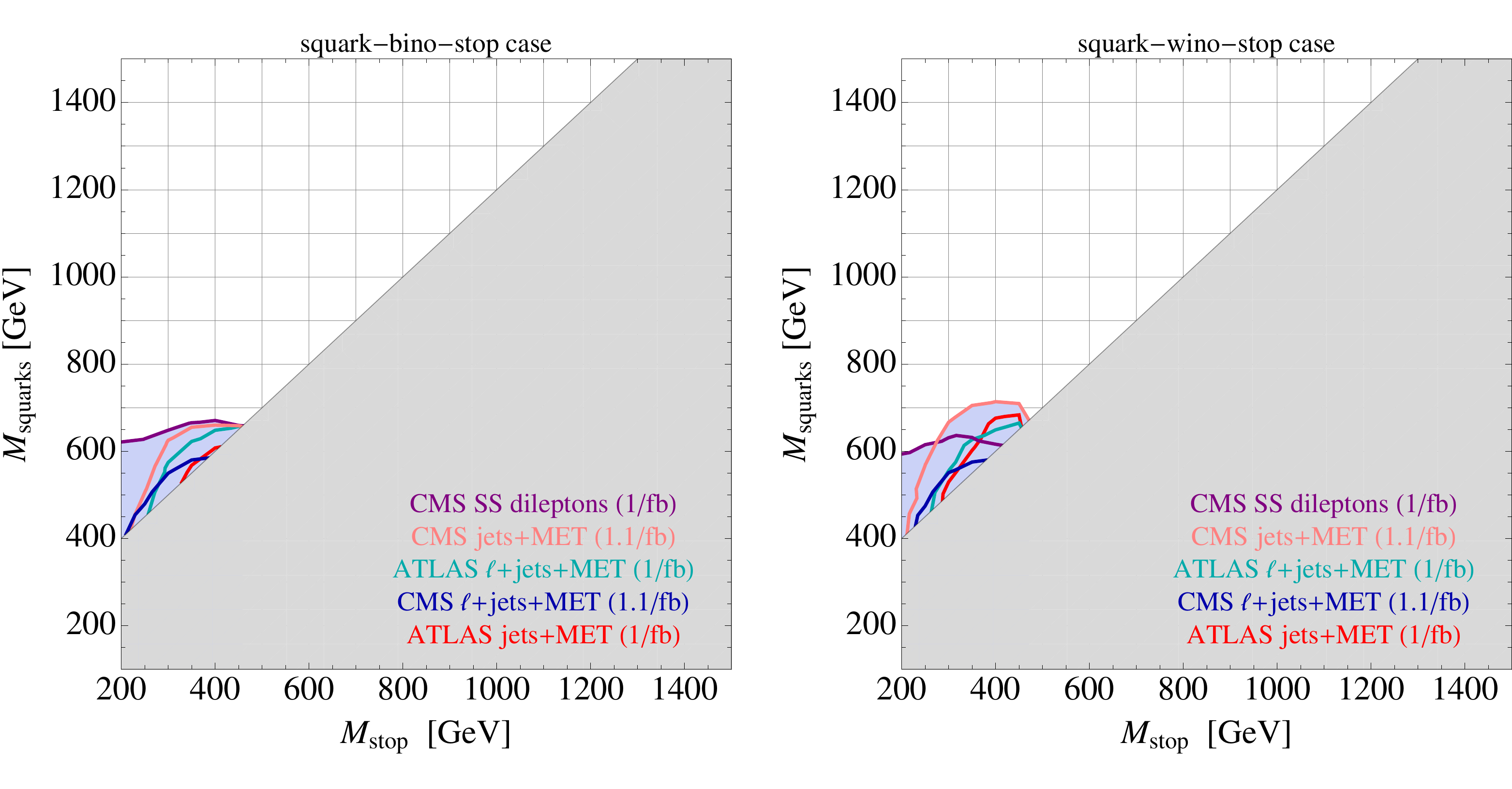}
\end{center}
\capt{Current best limits on the simplified parameter spaces for stop NLSPs produced via squarks, as described in Table~\ref{tab-stop} (top row). In both cases, the EW-ino mass has been fixed to $(M_{\rm squark}+M_{\rm stop})/2$. The shaded region either has $M_{\rm stop} > M_{\rm squarks}$ or has no on-shell 3-body decay ${\tilde \chi}^0 \to {\bar b}W^-{\tilde t}_1$.}
\label{fig-stop-limits}
\end{figure}

To investigate stops that are produced from heavier squarks we choose the same simplified parameter spaces as for the case of the sbottom in the previous subsection, simply replacing the sbottom with the stop. These are summarized in Table~\ref{tab-stop} (first row). The current limits are shown in Figure~\ref{fig-stop-limits}. Note that the bino case performs worse than the wino case. The wino case often involves a decay ${\tilde \chi}^+_1 \to {\bar b}{\tilde t}_1$, whereas the bino case necessarily involves ${\tilde \chi}^0_1 \to t{\tilde t}_1^\ast~{\rm or}~{\bar t}{\tilde t}_1$. This is the general phenomenon that when more top quarks are present, searches in jets and missing $E_T$ perform worse. A search in same-sign dileptons, however, picks up some of the slack.

\begin{figure}[!t]
\begin{center}
\includegraphics[width=0.46\textwidth]{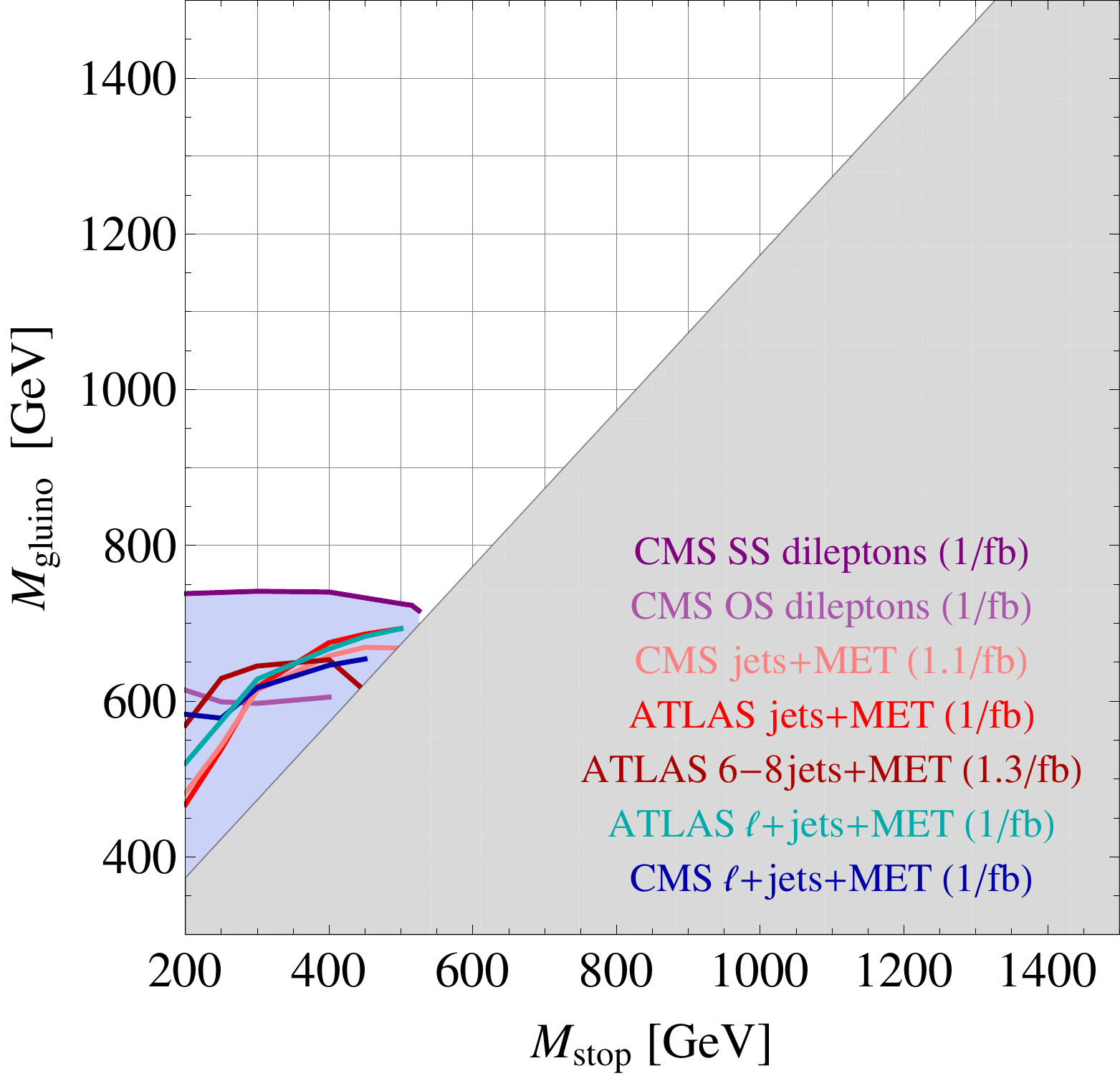}
\end{center}
\capt{Current best limits on the simplified parameter space for stop NLSPs produced from gluinos, described in Table~\ref{tab-stop} (bottom row). The diagonal is positioned to allow on-shell ${\tilde g} \to {\bar t} {\tilde t}_1$.}
\label{fig-stop-go-limits}
\end{figure}

Another simplified model that we considered is the production of stops from gluinos. As described in Table~\ref{tab-stop} (second row), this scenario has multi-top signatures. As a result, SS dilepton searches are most sensitive to it, as shown in Figure~\ref{fig-stop-go-limits}.

\section{Leptonic $M_{T2}$}
\label{sec:leptonicmt2}

In this section, we would like to explore a promising new method for rejecting $t\bar t$ and $W^+W^-$ backgrounds in analyses which search for OS dileptons and missing energy, such as \cite{CMS-PAS-SUS-11-011,CMS-PAS-SUS-11-017}. As we will show, our proposed method  could greatly improve on the current sensitivity to a variety of GGM scenarios, such as $Z$-rich higgsino NLSPs, slepton and sneutrino co-NLSPs, and stop NLSPs. For the sake of concreteness, we will focus our discussion in this section on the first scenario.

To begin, recall from our discussion around Figure~\ref{fig-zhiggsino-limits}, that the limits from both CMS $Z$+jets+MET and the jets+MET searches degrade significantly in the light higgsino region. Moreover, there is currently no limit on direct production of light higgsino NLSPs. Let us examine the reasons for this in more detail.

\begin{figure}[!t]
\begin{center}
\includegraphics[width=0.5\textwidth]{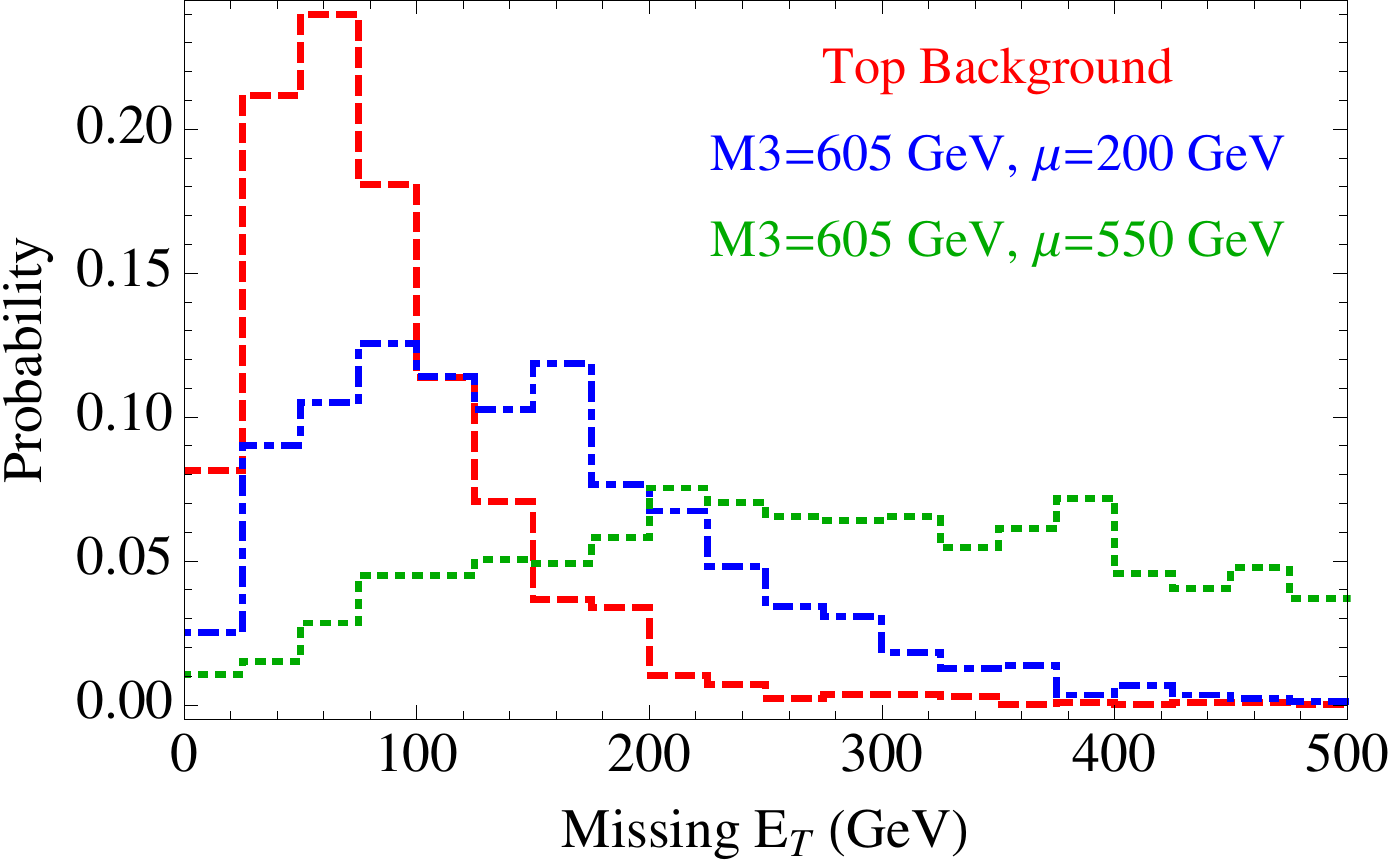}\includegraphics[width=0.5\textwidth]{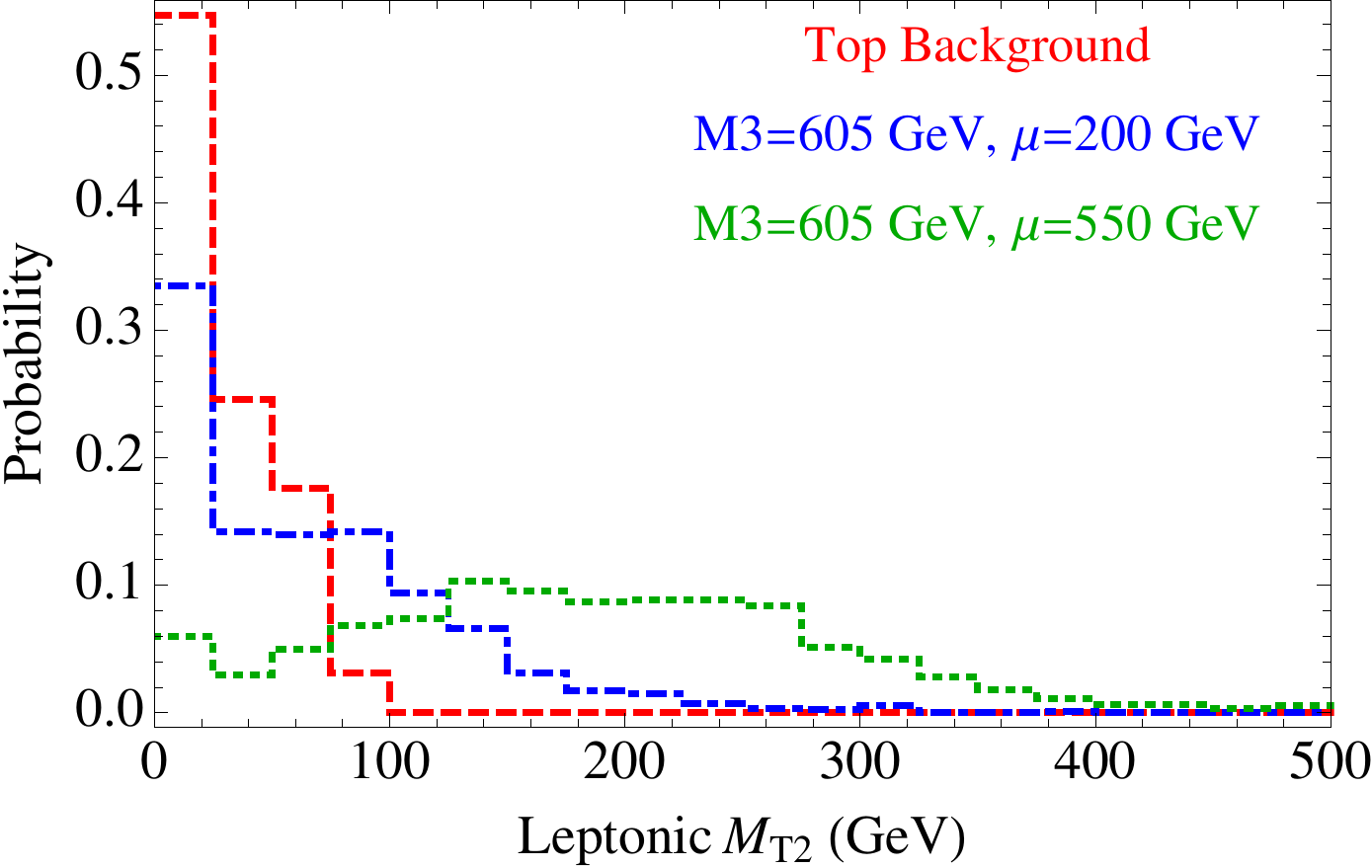}
\end{center}
\capt{$Z$-rich higgsino NLSP: signal and background distributions for missing $E_T$ and leptonic $M_{T2}$. All cuts from the CMS analysis~\cite{CMS-PAS-SUS-11-017} except for the cut on missing $E_T$ have been applied. Two different choices of higgsino mass are shown, illustrating that missing $E_T$ and related variables are suppressed when the higgsino is light.}
\label{fig-HZ-metmt2compare}
\end{figure}

When the higgsino is heavy, near the gluino mass, most of the energy of the gluino goes into the higgsino mass and the result is an energetic $ZZ{\tilde G}{\tilde G}$ final state, for which hard missing $E_T$ cuts perform well. However, when the higgsino is light, much of the gluino energy goes into jets, and the $Z$ bosons and gravitinos end up carrying a smaller fraction of the energy. The dominant background, as shown in~\cite{CMS-PAS-SUS-11-017}, is $t{\bar t}$ production with both tops decaying leptonically, and the mass of the lepton pair accidentally falling near the $Z$ mass. In Figure~\ref{fig-HZ-metmt2compare} at left, we show the missing $E_T$ distributions for two different values of the higgsino mass and for $t{\bar t}$ background, after applying all the cuts of~\cite{CMS-PAS-SUS-11-017} except the 100 or 200~GeV cut on missing $E_T$. It is apparent that at small higgsino masses, a cut on missing $E_T$ at 100 GeV keeps a substantial fraction of both signal and background, while a cut at 200 GeV eliminates most of the $t{\bar t}$ background at the cost of low signal efficiency.

\begin{figure}[!t]
\begin{center}
\includegraphics[width=0.5\textwidth]{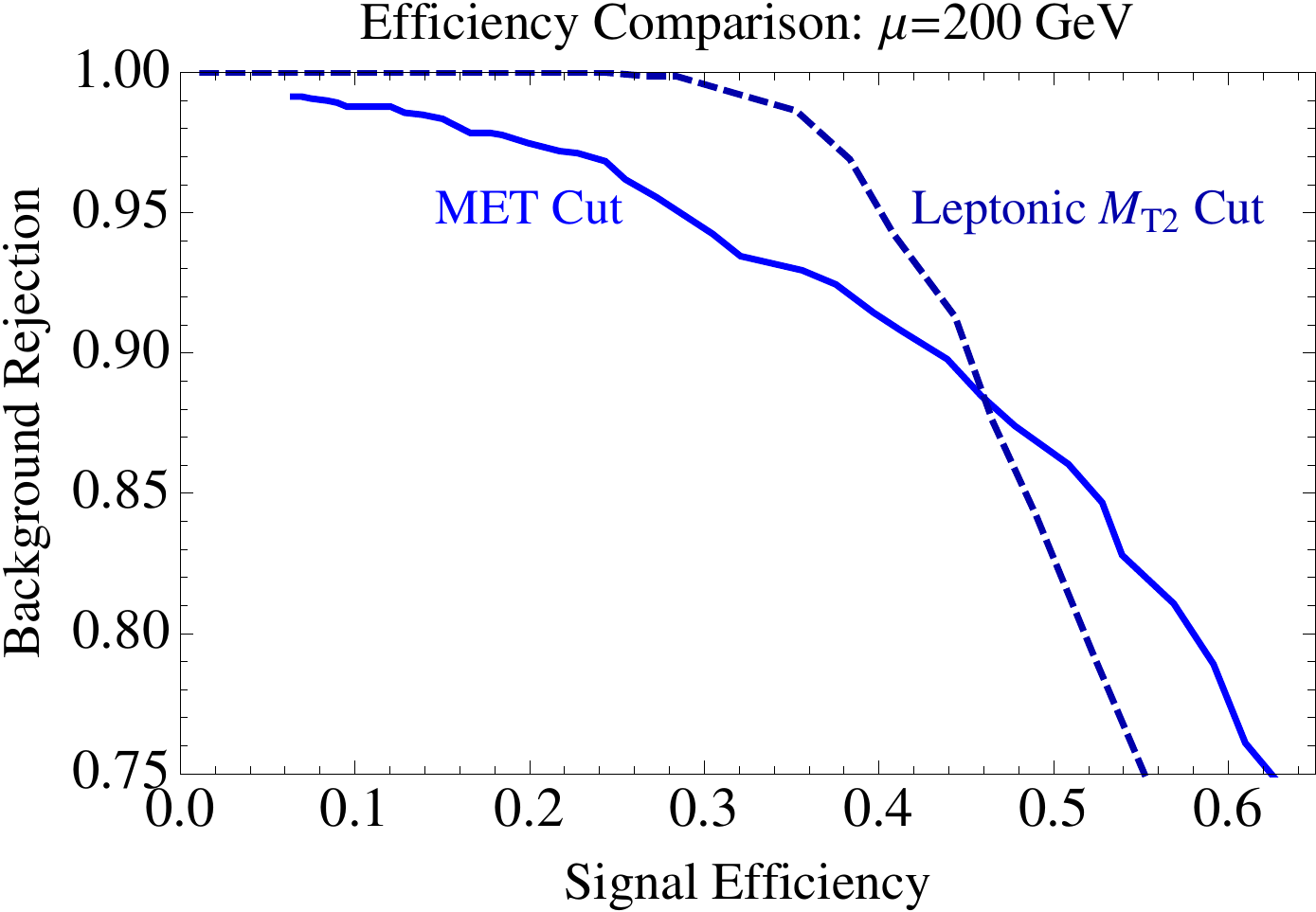}
\end{center}
\capt{Comparison of the efficiency of cutting on missing $E_T$ (solid blue line) and on leptonic $M_{T2}$ (dashed, darker blue line) for signal and the rejection rate (i.e., 1 minus efficiency) for $t{\bar t}$ background. At high background rejection (large coordinates on the vertical axis), the leptonic $M_{T2}$ cut keeps a substantially larger fraction of the signal. The crossover happens at a cut of about 140 GeV on missing $E_T$, with the same efficiency as a cut of about 65 GeV on leptonic $M_{T2}$. We don't show results for heavier higgsinos, because in that case either cut is extremely efficient on signal while strongly suppressing background.}
\label{fig-HZ-effcompare}
\end{figure}

We would like to propose a specific choice of cut, ``leptonic $M_{T2}$," which can more  precisely remove background and probe the region of small higgsino masses. Because the dominant background is $t{\bar t}$, in which both the leptons and the missing transverse energy come from the decay of two $W$ bosons, we can remove the background very precisely. The variable $M_{T2}$ (or ``stransverse mass") is useful, since $M_{T2}$ computed from the two leptons and missing $E_T$ is bounded above by the $W$ mass~\cite{Lester:1999tx}. In particular, because both the leptons and the missing neutrinos are essentially massless, we can use an analytic formula for $M_{T2}$~\cite{Lester:2011nj} rather than a time-consuming iterative method. In the right panel of Figure~\ref{fig-HZ-metmt2compare}, we show the $M_{T2}$ distributions for signal and background. Note that the $t{\bar t}$ background falls off rapidly at about 80 GeV. Previous discussions of the use of $M_{T2}$ to reject backgrounds have generally included jets~\cite{Barr:2009wu}; an $M_{T2}$ variable with jets and leptons could be used with an edge at the top mass, but deciding which jets to use involves a combinatorial problem. Leptonic $M_{T2}$ in dilepton events, by contrast, is a well-defined variable without combinatoric ambiguities, and we expect it will be more precisely measured than similar variables using jets.

A cut on leptonic $M_{T2}$ of near 80 GeV, then, can almost completely eliminate $t{\bar t}$ background while keeping an order-one fraction of the signal. In Figure~\ref{fig-HZ-effcompare}, we compare the signal efficiency and $t{\bar t}$ background rejection rate of cutting on missing $E_T$ and on leptonic $M_{T2}$. It is clear that the leptonic $M_{T2}$ cut is preferable at high background rejection. In particular, cutting on this variable would be very worthwhile with larger amounts of data, and should help to make the electroweak production of higgsino pairs accessible. Note that $M_{T2}$ is only nonzero if there is missing energy in an event. For events in which missing energy is not due to a decay of a pair of identical particles like $W$ bosons, $M_{T2}$ and missing $E_T$ are correlated. Thus, for a background like $Z(\to\ell\ell)$+jets, the $M_{T2}$ cut is not an improvement over a missing $E_T$ cut but we expect it to have comparable efficiency. Thus, we assume that the $Z(\to\ell\ell)$+jets background will remain relatively unimportant, even when cutting on $M_{T2}$ rather than missing $E_T$. Because the missing energy in $Z(\to\ell\ell)$+jets events is fake, it is difficult for us to simulate accurately; thus, this point requires further study.

To reiterate---here we have focused on light higgsino NLSPs, but more generally, leptonic $M_{T2}$ should prove useful for eliminating $WW$ and $t{\bar t}$ backgrounds in any new physics search with dileptons plus missing energy.
For example, it could be used in searches for direct slepton co-NLSP production, or direct stop NLSP pair production with both stops decaying leptonically. We will present a more detailed look at how to improve dilepton searches with $M_{T2}$ elsewhere \cite{lepmt2paper}.

\section{Comments on Long-Lived NLSPs}
\label{sec-long-lived}

So far we were assuming that the decays of the NLSPs are prompt. This may not be the case. Since the couplings to the gravitino are non-renormalizable, it is very natural for the NLSP to be long-lived. For a fundamental SUSY-breaking scale $\sqrt{F}$ of a few hundred TeV or below, lifetimes are generally below about a millimeter. Here we expect identification of most physics objects is unaffected, and so the results presented in previous sections should apply.\footnote{Exceptions include the CMS lepton+jets+MET study~\cite{CMS-PAS-SUS-11-015} and multileptons+MET study~\cite{CMSmultileptons}. The former (latter) rejects muons (all leptons) with transverse impact parameter above 0.2~mm, and thus they have reduced sensitivity to all but the lowest SUSY-breaking scales $\sqrt{F} \approx 10$~TeV.} For larger values of $\sqrt{F}$, lifetimes become long, many aspects of particle identification can change, and new searches should be used.

If the NLSP is neutral and stable on collider scales, the visible signatures reduce to those of high-scale mediation. These are well studied, so we will not consider them here.

On the other hand, if the NLSP is charged or colored, or neutral but decays within the detector, more dedicated searches are necessary. In this section we will survey existing LHC searches for long-lived particles (listed in Table~\ref{tab-gmsb-analyses-long}) and discuss to what extent they, or similar searches, can be used for looking for long-lived NLSPs.

\begin{table}[t!]
\begin{center}
\begin{tabular}{|c|c|c|c|c|c|}\hline
Analysis & Collaboration & Luminosity (fb$^{-1}$) & Ref \\
\hline
\hline
Stopped HSCP & CMS & 0.9 &  \cite{CMS-PAS-EXO-11-020}  \\

\hline
HSCP ($dE/dx$, TOF) & CMS & 1.1 & \cite{CMS-PAS-EXO-11-022} \\

\hline
\hline
Displaced lepton pair & CMS & 1.1-1.2 & \cite{CMS-PAS-EXO-11-004} \\

\hline

Displaced photons & CMS & 2.1 &   \cite{CMSdisplacedphotons} \\

\hline

Disappearing tracks & ATLAS & 1 & \cite{ATLASdisappearingtrack} \\

\hline
Displaced jets + high $p_T$ muon & ATLAS & 0.033 & \cite{Aad:2011zb}\\

\hline
\end{tabular}
\end{center}
\capt{A summary of the most recent LHC searches for  detector-stable or detector-metastable particles relevant to GMSB. HSCP is short for ``heavy stable charged particle"; TOF for ``time-of-flight."}
\label{tab-gmsb-analyses-long}
\end{table}

\subsection{Heavy Stable Charged Particles}

A number of ATLAS and CMS searches look for signatures of heavy stable charged particles (HSCPs) that propagate through the tracker without decaying, or stop in the detector and decay later.  Here we will review the most recent analyses (both CMS) with $\sim$ 1/fb.

The 1.1/fb CMS search~\cite{CMS-PAS-EXO-11-022} looks for high-$p_T$ isolated tracks with large ionization energy loss $dE/dx$. Another analysis presented in the same study requires also the presence of a slow muon candidate which allows loosening of some of the other cuts. For colored NLSPs the slow muon signature is only relevant if their $R$-hadrons do not all neutralize before reaching the muon detectors. The limits resulting from the two approaches are similar.

CMS excludes direct pair production of stable gluinos up to 800--900~GeV, and of stops up to 500--600~GeV (uncertainties are from unknown $R$-hadron physics). They also obtained a limit in a Minimal Gauge Mediation (MGM) scenario with stau NLSP, with production primarily through heavier squarks, gluinos, and winos. This was expressed as a limit on the mass of the stau, but this is only meaningful in the context of MGM where the stau mass is related to the masses of all the other sparticles. Clearly, it would be interesting to understand the limits on staus more model-independently, e.g.\ as a limit on the direct stau production cross section.  It would also be useful to know how the limits from this search apply to other scenarios, such as long-lived sbottoms (constrained by a smaller-statistics ATLAS study of $R$-hadrons~\cite{Aad:2011yf} using 34/pb that obtained a slightly weaker limit on sbottoms than stops), or chargino NLSPs~\cite{Kribs:2008hq} (on which a recent D$\varnothing$ search~\cite{Abazov:2011pf} with 5.2/fb set a limit of 217~GeV on a higgsino-like chargino and 267 GeV on a gaugino-like chargino).

CMS has also searched~\cite{CMS-PAS-EXO-11-020} for long-lived colored particles that form charged $R$-hadrons which ($\sim 10$\% or more of the time) come to rest due to interactions with the detector material and decay much later, often out-of-time with respect to the collisions. The study set limits on pair-produced gluinos and stops with lifetimes between $75$~ns and $10^6$~s. However, the strongest limits from these searches are weaker than the direct search~\cite{CMS-PAS-EXO-11-022}.

Overall, searches for heavy stable charged particles are a relatively mature science, and our main suggestion here is, wherever possible, to present the limits on several reasonable scenarios (stops, sbottoms, gluinos, staus, charginos) rather than a single one. Also, we believe these limits are more meaningful when phrased in terms of physical masses and cross sections rather than in terms of artificial model parameters such as those of MGM.

\subsection{Displaced Decays}

GMSB is a rich source of well-motivated simplified models that can be used to guide experimental searches for displaced decays. These decays fall into two types: decays of neutral particles (with displaced vertex, timing, or pointing signatures) and decays of charged particles (with additional kinked track signatures). In either case, reconstructing the displaced decay can be a powerful tool for precise measurements of new physics~\cite{Kawagoe:2003jv,Meade:2010ji,Park:2011vw}. Kinks and other odd features in tracks are currently under-studied and new physics with large cross sections could be hiding there~\cite{Meade:2011du}. Thus, a more comprehensive approach to studying such displaced decays is very well-motivated.

So far both ATLAS and CMS have released one study of displaced vertices that could be relevant for GGM. The ATLAS search~\cite{Aad:2011zb} looks for vertices that are displaced significantly from the primary vertex, in events containing a high-$p_T$ muon (from any vertex). They set limits on an RPV scenario in which squarks decay to neutralinos decaying as $\tilde\chi_1^0 \to \mu q \bar q'$. Limits are set on particles with lifetimes $0.8\mbox{ mm} \leq c\tau \leq 1\mbox{ m}$. Typical excluded cross sections are ${\cal O}(1-10)$~pb, with some mass dependence. These limits (based on only 33/pb of data) will improve with more statistics since the search is background-free. This kind of search can be useful for GGM scenarios in which a long-lived NLSP can decay into jets (to create the $\geq 4$ tracks needed for identifying the displaced vertex) as long as there is also a high-$p_T$ muon in the event. One example is when two neutralino NLSPs are produced from the decays of heavier particles, and at least one of the neutralinos decays into a hadronically decaying $Z$ or $h$, with the muon either produced from the cascade decay at the primary vertex or in the decay of the second neutralino into a $Z \to \mu^+\mu^-$ \cite{Meade:2010ji}.  Another example could be pair production of stop NLSPs, although a stop NLSP would travel inside an $R$-hadron, and it is unclear to us whether the search will still work as-is in cases where the $R$-hadron is charged and thus has a track going from the primary to the secondary vertex.

The CMS search~\cite{CMS-PAS-EXO-11-004} looks for displaced vertices formed by $e^+e^-$ or $\mu^+\mu^-$ pairs. A search of this kind can be relevant for long-lived neutralino NLSPs that can decay into a $Z$ (the branching ratio can be close to $1$ for a higgsino), as studied in detail in~\cite{Meade:2010ji}. The CMS search sets limits of 10 to 100 fb on $\sigma \times \mbox{BR}$ for lifetimes from 0.1 cm to 100 cm, with mild dependence on choices of masses. However, since the benchmark model used in~\cite{CMS-PAS-EXO-11-004} does not include MET, the search includes the requirement that, in the transverse plane, the total momentum of the leptons must point along the line connecting the primary and secondary vertex. This will not be the case in the GMSB scenarios because some of the momentum is carried by the gravitino. We estimate that the efficiency of this requirement for 250 GeV higgsinos is 50\% in the electron channel (where the cut is $|\Delta\phi| < 0.8$) and 16\%  in the muon channel ($|\Delta\phi| < 0.2$). Overall, we expect that interesting limits can be placed on long-lived neutralinos with this search, especially if the cut on $|\Delta\phi|$ can be relaxed.

In recent talks two more results have been presented. CMS has analyzed displaced decays to photons, in events with one well-identified photon and another that converts in the tracker~\cite{CMSdisplacedphotons}. This applies directly to neutralino NLSPs with $\gamma\gamma{\tilde G}{\tilde G}$ final states. It would be interesting to relax some of the requirements, e.g. to allow an electron instead of a photon, to attempt to include cases where one decay might go to $Z(\to e^+e^-) {\tilde G}$.  The other new result is an ATLAS search for disappearing tracks~\cite{ATLASdisappearingtrack}, motivated by degenerate wino LSPs. From the available information it is unclear to us whether this would constrain signals like a stau decaying in flight, but it would be interesting to see the result interpreted in such GMSB contexts.

These searches are interesting, but much more territory remains to be explored. At the most trivial level, any of our simplified models could be supplemented with an additional parameter $c\tau$ for the NLSP decay, which can range from less than a millimeter to much larger than the detector size. More generally, it is important to keep in mind the important physical parameters that go into determining the sensitivity of a given analysis. These include:

\begin{itemize}

\item {\it Production cross section:} Searches should make sure to specify the production mode precisely. This could be controlled by the NLSP mass itself, in the case of direct production, or by the mass of heavier states (e.g.\ gluinos) which are produced more copiously.

\item {\it NLSP lifetime:} This parameter is obviously is crucial for any search that looks for displaced objects. It is of paramount importance in any such search to show the limit or acceptance as a function of the lifetime.

\item {\it NLSP mass:} This will control the kinematics of the displaced decay products of the NLSP, which can in turn affect greatly the sensitivity of a given search.

\item {\it NLSP boost:} If the NLSP is boosted, e.g.\ due to the decay of a heavier state, then this could also affect the kinematics of the displaced objects. It would be very interesting to see the dependence of the sensitivity on the NLSP boost.

\end{itemize}

Finally, a comment about triggering. This is a major challenge for any inclusive displaced object search. We would just like to point out that in GMSB, displaced objects are always accompanied with {\it significant missing energy}, since the long-lived NLSP is always decaying to the gravitino in addition to a visible object. So MET triggers could prove to be a key ingredient for GMSB-motivated displaced-object searches.

\section{Summary and Outlook}
\label{sec-summary}

In this paper we have surveyed the space of all possible NLSP types in gauge mediation.  For each case, we have analyzed a variety of production mechanisms, including cascade decays from strongly and electroweakly produced particles and direct production.  In some of the cases we varied additional parameters in order to cover the largest possible part of the signature space. Even though our results are cast in the framework of GGM, their description in terms of simplified models provides information about many spectra that appear in other SUSY breaking schemes as well. Thus our work gives a broad general picture about the status of SUSY in light of the 1/fb of data from the LHC.

\begin{figure}[!t]
\begin{center}
\includegraphics[width=0.7\textwidth]{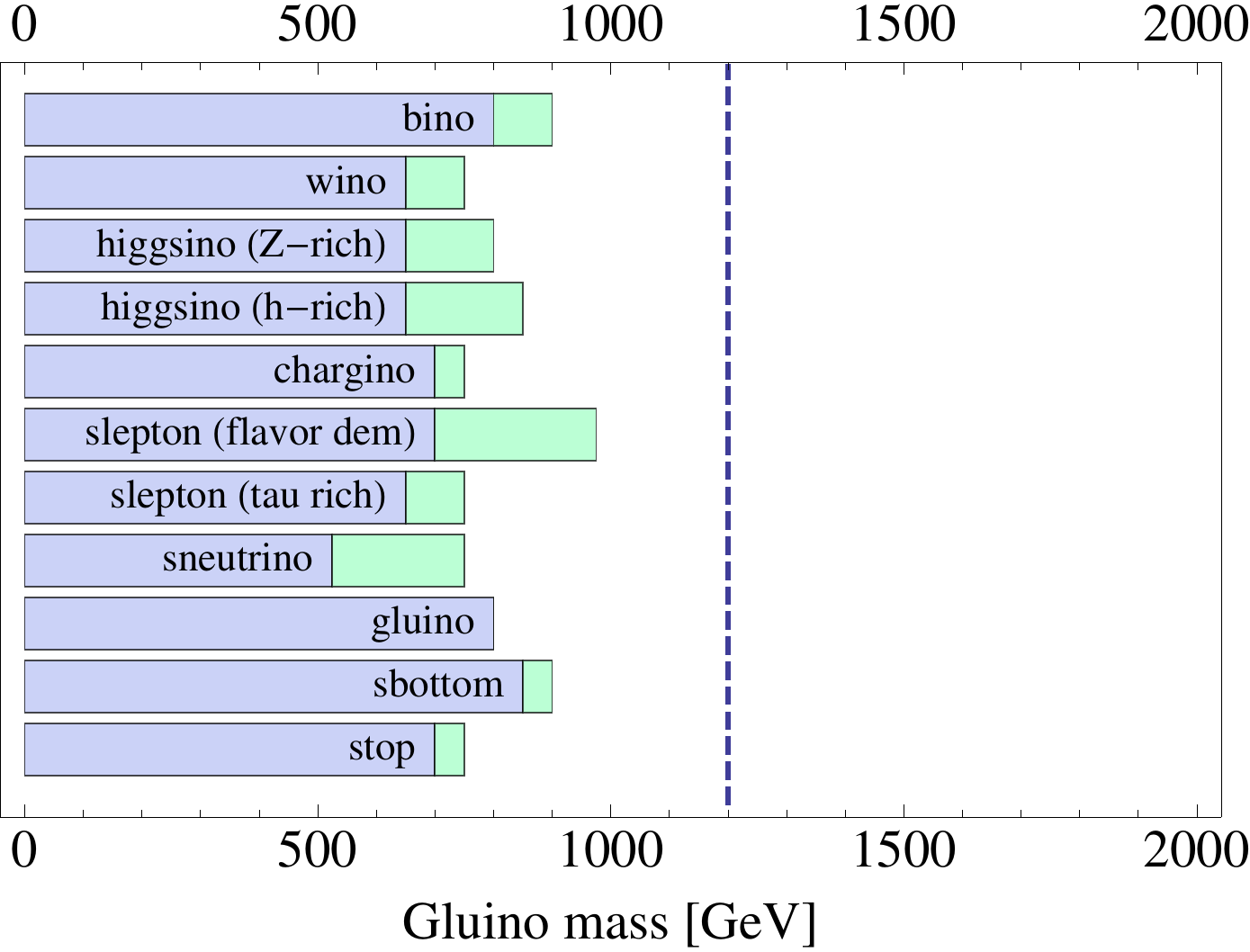}
\end{center}
\capt{A summary of the limits on gluino mass for various NLSPs, in the gluino-NLSP plane (one of the many simplified models considered in this paper). Masses in the blue band are ruled out, while the green band represents the range of possible excluded values as the NLSP mass varies. The dashed vertical line represents the idealized ``kinematic limit" of LHC7, as discussed in the Introduction.}
\label{fig-limits-summary}
\end{figure}

We have found that in some cases the colored superpartners can still be very light, even without any special fine-tuning of the spectrum. For example, the gluino can be as light as $\sim 550$~GeV in the sneutrino co-NLSPs scenario. In some other scenarios the colored particles are much more constrained. The limits on gluinos for all NLSP types are summarized in Figure~\ref{fig-limits-summary}.  Some of the limits are strong and are already close to the  kinematic limit of the 7~TeV LHC (depicted by the dashed vertical line in Figure~\ref{fig-limits-summary}). Still, a large amount of parameter space remains viable at 7 TeV.

Scenarios in which just a single third-generation squark is light have much weaker limits. We have found a limit of $M_{\rm sbottom} \gtrsim 280$~GeV on the direct pair production of sbottom NLSPs. We have also found that multiple SUSY searches are interesting for light stop NLSPs, becoming competitive with the current estimated Tevatron limit of $M_{\rm stop} \gtrsim 150$~GeV. It should be noted though that the cuts used in these searches are rather hard relative to the stop mass, and the light stop events make it into the sample only due to their large cross sections. Because of these tiny efficiencies, the results strongly depend on the tails of distributions which we cannot claim to have simulated reliably. Therefore, we prefer not to quote a limit on the stop NLSP mass, which might be somewhere around $175$~GeV. More dedicated experimental searches for stop NLSPs, or more careful simulations of the existing ones, are sorely needed. We have pointed out that small optimizations of some of the existing analyses, or certain other types of searches, can probe the light stop regime more thoroughly. Clearly, stop NLSPs can still be as light as $\sim 200$~GeV, and we have also seen that in such a scenario all the other squarks can be relatively light as well, near just 600~GeV.

Constraining electroweak production is harder, and overall the LHC is far from reaching its 7 TeV kinematic limit ($\sim 700$ GeV for wino production, as discussed in the Introduction). Still, we found that the $\gamma\gamma$+MET searches set a limit of $M_{\rm wino} \gtrsim 400$~GeV on the production of winos decaying to bino NLSPs. We have also estimated that directly produced wino co-NLSPs can be constrained to $M_{\rm wino} \gtrsim 250$~GeV with the data available today. Another potentially clean scenario of electroweak production is winos decaying to slepton co-NLSPs, which may have good sensitivity in a SS dilepton search without jet requirements, or a multi-lepton search.

Overall, in most cases, we have found that the strongest limits are set by very general searches for jets and missing $E_T$. The exceptions are searches with extremely low backgrounds: $\gamma\gamma$+MET, which is by far the strongest limit on bino NLSPs, and same-sign dileptons, which sets powerful limits on slepton NLSPs and scenarios with multiple top quarks, e.g.\ squarks or gluinos decaying to stop NLSPs. Clearly, such searches are very powerful and play to the LHC's strengths. We have seen though that other specialized searches such as $Z$+jets+MET and $b$-jets+MET could do better if they included harder cuts on hadronic activity similarly to some of the jets+MET search regions.

As the LHC moves from 1/fb studies to 5/fb or more, channels which are clean but have low branching fractions will become increasingly powerful. For instance, currently the best limits on $Z$-rich higgsino NLSPs come from jets + MET and $Z(\ell^+\ell^-)$+jets+MET searches, but with more data, $Z(\ell^+\ell^-)Z(\ell'^+\ell'^-)$+MET becomes a promising channel as well.  Another example, relevant to $b$-rich scenarios, is requiring three or more $b$-tags in an event, which should greatly reduce background at the cost of lowered signal efficiency.

With greater integrated luminosity, more sophisticated search strategies will also become viable. We have argued that a cut on leptonic $M_{T2}$ can enhance the power of searches involving dileptons by removing the largest backgrounds in a more targeted way. Brute force is enough for the LHC to trounce Tevatron limits, but there is still a need for more refined techniques to significantly extend what energy alone has accomplished so far.

For the next round of LHC searches, it may be useful to keep in mind some theoretical expectations. Many models in which the known forces are unified predict that colored superpartners are significantly heavier than uncolored superpartners, which provides motivation for directly aiming searches at electroweak production of gauginos, higgsinos, or sleptons. As we have seen, so far such processes are almost unconstrained. Furthermore, naturalness motivates light stops to cancel divergences in the Higgs mass. By 5/fb the LHC, with the combination of additional statistics and potentially refined search strategies, should be a sensitive probe of light stops. Thus, two of the most well-justified signals from a theoretical point of view---light electroweak superpartners and light stops---will be among the highly relevant new pieces of information that potentially will arrive with 5/fb searches. Finally, we hope that long lifetimes will be explored more systematically in upcoming analyses.  While SUSY was not right around the corner, our results have identified a large region that still has room for interesting discoveries.

\section*{Acknowledgments}
We gratefully acknowledge Warren Andrews, Monica D'Onofrio, Michael Flowerdew, Bill Gary, Yuri Gershtein, Tobias Golling, Eva Halkiadakis, Steve Martin, Dave Mason, Sophio Pataraia, Finn Rebassoo, George Redlinger, Seema Sharma, Anna Sfyrla, Dan Tovey, Iacopo Vivarelli, Frank Wuerthwein, Takashi Yamanaka, and Chris Young for helpful conversations and/or correspondence. We thank Andrey Katz, Michele Papucci, and Josh Ruderman for useful discussions on their related work. MR and DS would like to thank the KITP for hospitality while part of this work was being completed (supported in part by the National Science Foundation under Grant No.~PHY05-51164).  The research of YK is supported in part by DOE grant DE-FG02-96ER40959.   The work of PM is supported in part by NSF CAREER Award NSF-PHY-1056833. MR is supported by the Fundamental Laws Initiative of the Center for the Fundamental Laws of Nature, Harvard University. The research of DS was supported in part by a DOE Early Career Award.

\appendix

\section{Simulation Description}\label{app:simulation}

Superpartner spectra for the GMSB scenarios were generated using {\sc SuSpect~2.41}~\cite{Djouadi:2002ze} and the decay tables using {\sc SDECAY 1.3}~\cite{Muhlleitner:2003vg}. The decays of the NLSPs and co-NLSPs to the gravitino were forced in {\sc Pythia}.

We base the simulation on {\sc Pythia~8.150}~\cite{Sjostrand:2007gs} interfaced with {\sc FastJet}~\cite{Cacciari:2005hq}. Since the implementation of some of the SUSY processes in {\sc Pythia~8} is not yet complete we use the {\sc Pythia~6}--{\sc Pythia~8} interface to import all the SUSY processes from {\sc Pythia~6.4.22}~\cite{Sjostrand:2006za}. For parton distributions we use CTEQ6L1~\cite{Pumplin:2002vw} (obtained via {\sc LHAPDF}~\cite{Whalley:2005nh}).

{\sc Pythia} output is processed with a crude detector simulation code that mimics the object identification procedures of a generic LHC detector. In particular, we take the effective calorimeter coverage to be $|\eta| < 5$. For charged particles, we only accept $p_T > 1$~GeV. We allow the identification of tracks, photons, electrons, muons, hadronic taus and $b$ jets within $|\eta| < 2.5$ (further cuts are applied at the analysis level). We compute the missing energy vector $\slash{\vv{E}}_T$ based on the energy deposits in the calorimeters, as well as the full energies of muons. For a lepton ($e$ or $\mu$) to be isolated, the sum, within $R < 0.3$, of the $p_T$ of all other tracks, plus the transverse energy deposited in the calorimeter, except the contribution of the lepton itself, must be smaller than $15\%$ of the lepton $p_T$. For a photon to be isolated, the summed $p_T$ of tracks, the transverse electromagnetic energy and the transverse hadronic energy within $R < 0.4$ (excluding the photon contribution) should satisfy $\sum p_T < 0.001 p_T^\gamma + 2$~GeV, $E_T^{\rm em} < 0.006 p_T^\gamma + 4.2$~GeV, and $E_T^{\rm had} < 0.0025 p_T^\gamma + 2.2$~GeV. Jets are formed from the calorimeter deposits (except those of the isolated leptons and photons) using the anti-$k_T$ jet algorithm~\cite{Cacciari:2008gp} with cone size $R = 0.4$. Hadronically decaying taus that originate from the hard process (but not from jet fragmentation) are ``reconstructed'' by taking their momentum before the decay and subtracting the momentum of the neutrino. Hadronic tau jets (usually a single jet) are identified as those jets that are within $R < \l(5\mbox{ GeV}/p_T^\tau\r)$ around the reconstructed tau. For $b$ tagging, for each $B$ meson, we find the jet closest in $R$, with $R < 0.4$ and $p_T$ within $\pm 60\%$ of $p_T$ of the $B$ meson. If such a jet exists, we indicate it as a $b$ candidate. Later, in the analysis stage, we apply approximate trigger, object identification and $b$ tagging efficiencies, to the extent that such information is available in the description of the relevant ATLAS or CMS analysis.

Our simulation is significantly faster than PGS~\cite{PGS} because it does not discretize the calorimeters. We also do not simulate the energy resolutions of the various objects. Neither detail is essential for our purposes.

In the calculation of efficiency, each event that starts with two colored superpartners is weighted by a corresponding NLO $K$ factor from {\sc Prospino}~2.1~\cite{prospino}. These are also used in combination with the LO cross sections from {\sc Pythia} for determining the cross section.

To compute limits on our simplified parameter spaces, we have used the ``model-independent upper limits" quoted in the experimental publications, wherever available. When these numbers were not available, we have used the standard $CL_s$ statistic~\cite{CLsrefs}. For searches with multiple subanalyses, we have used the subanalysis which gave the best limit.

\section{Things we like to see in experimental publications}\label{app:exp-requests}

An essential condition for the validity of a work like ours is the ability to reliably simulate the experimental searches using only publicly available information from the experiments. Luckily, many of the recent LHC searches put a special effort into making their results suitable for re-analysis by theorists who want to apply them to different new physics scenarios. But in some cases certain important pieces of information were missing. We would like to list several items that in our opinion are important to be included in papers presenting experimental results:
\begin{itemize}
\item Typical identification efficiencies of the various objects (leptons, photons, $b$-jets), with a rough parameterization of the $p_T$ or $|\eta|$ dependence where relevant.
\item The likely treatment of other types of objects in the event besides those relevant to the particular search. Suppose, for example, that a hard photon is present in an analysis that does not explicitly identify photons. Will it simply be counted as a jet, or fail certain jet quality requirements? If it fails to be a jet, will it consequently be ignored, or will it disqualify the entire event from being included in the sample?
\item The description of other experimental effects whose precise simulation happens to be crucial for the particular analysis (e.g., resolutions of certain quantities).
\item Simulated event yields for several signal points. These are important for validating the theorists' simulation of the analysis.
\item Whenever possible, a model-independent $CL_s$ bound on the rate of signal events passing the cuts. Details of the treatment of systematic errors would be helpful. Even more helpful would be public release of $CL_s$-calculating codes or likelihoods.
\item Precise definitions of variables such as $H_T$ and $m_{\rm eff}$, with all conditions on the number of jets to be summed over, their $p_T$ cuts, and so on.
\end{itemize}
Much of the work in this paper has been in reinterpreting existing LHC searches in terms of GGM parameter spaces for which they were not originally intended. Clearly, it would be very useful to have an official framework like the proposed RECAST~\cite{Cranmer:2010hk} which could automate and standardize this process.


\begin{thebibliography}{99}

\small

\bibitem{Giudice:1998bp}
  G.~F.~Giudice and R.~Rattazzi,
  ``Theories with gauge-mediated supersymmetry breaking,''
  \href{http://dx.doi.org/10.1016/S0370-1573(99)00042-3}{Phys.\ Rept.\  {\bf 322}, 419 (1999)}
  \href{http://arxiv.org/abs/hep-ph/9801271}{\tt arXiv:hep-ph/9801271}.

\bibitem{Aad:2011ib}
{\bf ATLAS} Collaboration, G.~Aad {\em et al.}, ``{Search for squarks and
  gluinos using final states with jets and missing transverse momentum with the
  ATLAS detector in $\sqrt s = 7$~TeV proton-proton collisions},''
  \href{http://arxiv.org/abs/1109.6572}{{\tt arXiv:1109.6572 [hep-ex]}}.

\bibitem{CMS-PAS-SUS-11-004}
{\bf CMS} Collaboration, ``{Search for supersymmetry in all-hadronic events
  with missing energy},'' CMS PAS SUS-11-004,
  \url{http://cdsweb.cern.ch/record/1378478}.

\bibitem{CMS-PAS-SUS-11-003}
{\bf CMS} Collaboration, ``{Search for supersymmetry in all-hadronic events
  with $\alpha_T$},'' CMS PAS SUS-11-003,
  \url{http://cdsweb.cern.ch/record/1370596}.

\bibitem{ATL-PHYS-SLIDE-2011-529}
{\bf ATLAS} Collaboration, ``{Search for new phenomena in final states with large jet multiplicities and missing transverse momentum using $\sqrt s = 7$~TeV $pp$ collisions with the ATLAS detector},''
\href{http://arxiv.org/abs/1110.2299}{{\tt arXiv:1110.2299 [hep-ex]}}.


\bibitem{ATLAS-CONF-2011-098}
{\bf ATLAS} Collaboration, ``{Search for supersymmetry in $pp$ collisions at
  $\sqrt{s} = 7$~TeV in final states with missing transverse momentum, $b$-jets and no leptons with the ATLAS detector},''
  ATLAS-CONF-2011-098,
  \url{http://cdsweb.cern.ch/record/1369212}.

\bibitem{CMS-PAS-SUS-11-006}
{\bf CMS} Collaboration, ``{Search for new physics in events with $b$-quark
  jets and missing transverse energy in proton-proton collisions at 7~TeV},'' CMS PAS SUS-11-006,
  \url{http://cdsweb.cern.ch/record/1390493}.

\bibitem{CMS-PAS-SUS-11-010}
{\bf CMS} Collaboration, ``{Search for new physics with same-sign isolated
  dilepton events with jets and missing energy},'' CMS PAS SUS-11-010,
  \url{http://cdsweb.cern.ch/record/1370064}.


\bibitem{CMS-PAS-SUS-11-011}
{\bf CMS} Collaboration, ``{Search for new physics in events with opposite-sign
  dileptons and missing transverse energy},'' CMS PAS SUS-11-011,
  \url{http://cdsweb.cern.ch/record/1370065}.

\bibitem{Collaboration:2011iu}
{\bf ATLAS} Collaboration, G.~Aad {\em et al.}, ``{Search for supersymmetry in
  final states with jets, missing transverse momentum and one isolated lepton
  in $\sqrt{s} = 7$~TeV $pp$ collisions using $1$~fb$^{-1}$ of ATLAS data},''
  \href{http://arxiv.org/abs/1109.6606}{{\tt arXiv:1109.6606 [hep-ex]}}.

\bibitem{CMS-PAS-SUS-11-015}
{\bf CMS} Collaboration, ``{Search for supersymmetry in $pp$ collisions at
  $\sqrt s = 7$~TeV in events with a single lepton, jets, and missing
  transverse momentum},'' CMS PAS SUS-11-015,
  \url{http://cdsweb.cern.ch/record/1380922}.

\bibitem{ATLAS-CONF-2011-130}
{\bf ATLAS} Collaboration, ``{Search for supersymmetry in $pp$ collisions at
  $\sqrt{s} = 7$~TeV in final states with missing transverse momentum, $b$-jets
  and one lepton with the ATLAS detector},'' ATLAS-CONF-2011-130, \url{http://cdsweb.cern.ch/record/1383833}.

\bibitem{CMS-PAS-SUS-11-017}
{\bf CMS} Collaboration, ``{Search for new physics in events with a $Z$ boson
  and missing energy},'' CMS PAS SUS-11-017,
  \url{http://cdsweb.cern.ch/record/1370066}.


\bibitem{Collaboration:2011wc}
{\bf ATLAS} Collaboration, ``{Search for New Phenomena in $t\bar t$ Events With Large Missing Transverse Momentum in Proton-Proton Collisions at $\sqrt s = 7$~TeV with the ATLAS Detector},'' \href{http://arxiv.org/abs/1109.4725}{{\tt arXiv:1109.4725 [hep-ex]}}.


\bibitem{ATL-PHYS-SLIDE-2011-523}
{\bf ATLAS} Collaboration, ``{Search for SUSY and UED in Final States with
  Photons and Missing Transverse Energy with the ATLAS Detector},''
  ATL-PHYS-SLIDE-2011-523, \url{http://cdsweb.cern.ch/record/1380305}.

\bibitem{CMS-PAS-SUS-11-009}
{\bf CMS} Collaboration, ``{Search for supersymmetry with photons, jets and
  MET},'' CMS PAS SUS-11-009, \url{http://cdsweb.cern.ch/record/1377324}.

\bibitem{Chatrchyan:2011ah}
{\bf CMS} Collaboration, S.~Chatrchyan {\em et al.}, ``{Search for
  supersymmetry in events with a lepton, a photon, and large missing transverse
  energy in $pp$ collisions at $\sqrt{s} = 7$~TeV},''
  \href{http://dx.doi.org/10.1007/JHEP06(2011)093}{{\em JHEP} {\bf 06} (2011)
  093},
\href{http://arxiv.org/abs/1105.3152}{{\tt arXiv:1105.3152 [hep-ex]}}.

\bibitem{CMS-PAS-SUS-11-005}
{\bf CMS} Collaboration, ``{Search for supersymmetry in all-hadronic events
  with $M_{T2}$},'' CMS PAS SUS-11-005,
  \url{http://cdsweb.cern.ch/record/1377032}.

\bibitem{CMSRazor}
{\bf CMS} Collaboration, S.~Chatrchyan {\em et al.}, ``{Inclusive search for
  squarks and gluinos in $pp$ collisions at $\sqrt{s} = 7$~TeV},''
\href{http://arxiv.org/abs/1107.1279}{{\tt arXiv:1107.1279 [hep-ex]}}.

\bibitem{ATLAS-CONF-2011-039}
{\bf ATLAS} Collaboration, ``{SUSY Searches at ATLAS in Multilepton Final
  States with Jets and Missing Transverse Energy},'' ATLAS-CONF-2011-039,
  \url{http://cdsweb.cern.ch/record/1338568}.

\bibitem{CMSmultileptons}
{\bf CMS}~Collaboration, ``Search for Supersymmetry Using Multilepton Signatures in $pp$ Collisions at 7~TeV,''
CMS PAS SUS-11-013, \url{http://cdsweb.cern.ch/record/1393719};
{\bf CMS}~Collaboration, ``Search for Anomalous Production of Multilepton Events and $R$-Parity-Violating Supersymmetry in $\sqrt{s} = 7$~TeV $pp$ Collisions,''
CMS PAS EXO-11-045, \url{http://cdsweb.cern.ch/record/1393758}

\bibitem{ATLASleptons}
T.~M\"{u}ller, ``Searches for Jets and Missing Transverse Energy with Leptons
at ATLAS," \url{https://indico.cern.ch/materialDisplay.py?contribId=18&sessionId=5&materialId=slides&confId=149404}

\bibitem{Aad:2011kz}
{\bf ATLAS} Collaboration, G.~Aad {\em et al.}, ``{Search for Diphoton Events
  with Large Missing Transverse Energy with 36 pb$^{-1}$ of 7 TeV Proton-Proton
  Collision Data with the ATLAS Detector},''
\href{http://arxiv.org/abs/1107.0561}{{\tt arXiv:1107.0561 [hep-ex]}}.

\bibitem{Meade:2008wd}
P.~Meade, N.~Seiberg, and D.~Shih, ``{General Gauge Mediation},''
  \href{http://dx.doi.org/10.1143/PTPS.177.143}{{\em Prog.Theor.Phys.Suppl.}
  {\bf 177} (2009)  143}, \href{http://arxiv.org/abs/0801.3278}{{\tt
  arXiv:0801.3278 [hep-ph]}}.

\bibitem{Buican:2008ws}
  M.~Buican, P.~Meade, N.~Seiberg, D.~Shih,
  ``Exploring General Gauge Mediation,''
  \href{http://dx.doi.org/10.1088/1126-6708/2009/03/016}{{\em JHEP} {\bf 0903}, 016 (2009)},
  \href{http://arxiv.org/abs/0812.3668}{{\tt arXiv:0812.3668 [hep-ph]}}.

\bibitem{Carpenter:2008wi}
  L.~M.~Carpenter, M.~Dine, G.~Festuccia, J.~D.~Mason,
  ``Implementing General Gauge Mediation,''
  Phys.\ Rev.\  {\bf D79}, 035002 (2009).
  [arXiv:0805.2944 [hep-ph]].


\bibitem{Meade:2009qv}
P.~Meade, M.~Reece, and D.~Shih, ``{Prompt Decays of General Neutralino NLSPs
  at the Tevatron},'' \href{http://dx.doi.org/10.1007/JHEP05(2010)105}{{\em
  JHEP} {\bf 1005} (2010)  105}, \href{http://arxiv.org/abs/0911.4130}{{\tt
  arXiv:0911.4130 [hep-ph]}}.

\bibitem{Meade:2010ji}
  P.~Meade, M.~Reece, D.~Shih,
  ``Long-Lived Neutralino NLSPs,''
  \href{http://dx.doi.org/10.1007/JHEP10(2010)067}{{\em JHEP} {\bf 1010}, 067 (2010)},
  \href{http://arxiv.org/abs/1006.4575}{{\tt arXiv:1006.4575 [hep-ph]}}.

\bibitem{Ruderman:2010kj}
J.~T. Ruderman and D.~Shih, ``{Slepton co-NLSPs at the Tevatron},''
  \href{http://dx.doi.org/10.1007/JHEP11(2010)046}{{\em JHEP} {\bf 1011} (2010)
   046}, \href{http://arxiv.org/abs/1009.1665}{{\tt arXiv:1009.1665 [hep-ph]}}.

\bibitem{Ruderman:2011vv}
J.~T. Ruderman and D.~Shih, ``{General Neutralino NLSPs at the Early LHC},''
  \href{http://arxiv.org/abs/1103.6083}{{\tt arXiv:1103.6083 [hep-ph]}}.

\bibitem{Kats:2011it}
Y.~Kats and D.~Shih, ``{Light Stop NLSPs at the Tevatron and LHC},''
  \href{http://dx.doi.org/10.1007/JHEP08(2011)049}{{\em JHEP} {\bf 1108} (2011)
   049}, \href{http://arxiv.org/abs/1106.0030}{{\tt arXiv:1106.0030 [hep-ph]}}.

\bibitem{Carpenter:2008he}
  L.~M.~Carpenter,
  ``Surveying the Phenomenology of General Gauge Mediation,''
  \href{http://arxiv.org/abs/0812.2051}{\tt arXiv:0812.2051 [hep-ph]}.

\bibitem{Rajaraman:2009ga}
  A.~Rajaraman, Y.~Shirman, J.~Smidt and F.~Yu,
  ``Parameter Space of General Gauge Mediation,''
  \href{http://dx.doi.org/10.1016/j.physletb.2009.06.047}{Phys.\ Lett.\  B {\bf 678}, 367 (2009)},
  \href{http://arxiv.org/abs/0903.0668}{\tt arXiv:0903.0668 [hep-ph]}.

\bibitem{DeSimone:2009ws}
  A.~De Simone, J.~Fan, V.~Sanz, W.~Skiba,
  ``Leptogenic Supersymmetry,''
  \href{http://dx.doi.org/10.1103/PhysRevD.80.035010}{Phys.\ Rev.\  {\bf D80}, 035010 (2009)}.
  \href{http://arxiv.org/abs/0903.5305}{\tt arXiv:0903.5305 [hep-ph]}.

\bibitem{Mason:2009qh}
  J.~D.~Mason, D.~E.~Morrissey, D.~Poland,
  ``Higgs Boson Decays to Neutralinos in Low-Scale Gauge Mediation,''
  \href{http://dx.doi.org/10.1103/PhysRevD.80.115015}{Phys.\ Rev.\  {\bf D80}, 115015 (2009)}.
  \href{http://arxiv.org/abs/0909.3523}{\tt arXiv:0909.3523 [hep-ph]}.

\bibitem{Katz:2009qx}
  A.~Katz and B.~Tweedie,
  ``Signals of a Sneutrino (N)LSP at the LHC,''
  \href{http://dx.doi.org/10.1103/PhysRevD.81.035012}{Phys.\ Rev.\  D {\bf 81}, 035012 (2010)},
  \href{http://arxiv.org/abs/0911.4132}{\tt arXiv:0911.4132 [hep-ph]}.

\bibitem{Abel:2009ve}
  S.~Abel, M.~J.~Dolan, J.~Jaeckel and V.~V.~Khoze,
  ``Phenomenology of Pure General Gauge Mediation,''
  \href{http://dx.doi.org/10.1088/1126-6708/2009/12/001}{{\em JHEP} {\bf 0912}, 001 (2009)},
  \href{http://arxiv.org/abs/0910.2674}{\tt arXiv:0910.2674 [hep-ph]}.

\bibitem{Kribs:2009yh}
  G.~D.~Kribs, A.~Martin, T.~S.~Roy, M.~Spannowsky,
  ``Discovering the Higgs Boson in New Physics Events using Jet Substructure,''
  \href{http://dx.doi.org/10.1103/PhysRevD.81.111501}{Phys.\ Rev.\  {\bf D81}, 111501 (2010)}.
  \href{http://arxiv.org/abs/0912.4731}{\tt arXiv:0912.4731 [hep-ph]}.

\bibitem{Katz:2010xg}
  A.~Katz and B.~Tweedie,
  ``Leptophilic Signals of a Sneutrino (N)LSP and Flavor Biases from
  Flavor-Blind SUSY,''
  \href{http://dx.doi.org/10.1103/PhysRevD.81.115003}{Phys.\ Rev.\  D {\bf 81}, 115003 (2010)},
  \href{http://arxiv.org/abs/1003.5664}{\tt arXiv:1003.5664 [hep-ph]}.

\bibitem{Abel:2010vb}
  S.~Abel, M.~J.~Dolan, J.~Jaeckel and V.~V.~Khoze,
  ``Pure General Gauge Mediation for Early LHC Searches,''
  \href{http://arxiv.org/abs/1009.1164}{\tt arXiv:1009.1164 [hep-ph]}.

\bibitem{Thalapillil:2010ek}
  A.~M.~Thalapillil,
  ``Low-energy Observables and General Gauge Mediation in the MSSM and NMSSM,''
  \href{http://arxiv.org/abs/1012.4829}{\tt arXiv:1012.4829 [hep-ph]}.

\bibitem{Jaeckel:2011ma}
  J.~Jaeckel, V.~V.~Khoze and C.~Wymant,
  ``Mass Sum Rules and the Role of the Messenger Scale in General Gauge
  Mediation,''
  \href{http://arxiv.org/abs/1102.1589}{\tt arXiv:1102.1589 [hep-ph]};
  J.~Jaeckel, V.~V.~Khoze and C.~Wymant,
  ``RG Invariants, Unification and the Role of the Messenger Scale in General
  Gauge Mediation,''
  \href{http://arxiv.org/abs/1103.1843}{\tt arXiv:1103.1843 [hep-ph]}.

\bibitem{Dimopoulos:1996vz}
  S.~Dimopoulos, M.~Dine, S.~Raby and S.~D.~Thomas,
  ``Experimental Signatures of Low Energy Gauge Mediated Supersymmetry
  Breaking,''
  \href{http://dx.doi.org/10.1103/PhysRevLett.76.3494}{Phys.\ Rev.\ Lett.\  {\bf 76}, 3494 (1996)}
  \href{http://arxiv.org/abs/hep-ph/9601367}{\tt arXiv:hep-ph/9601367};
  S.~Dimopoulos, M.~Dine, S.~Raby, S.~D.~Thomas and J.~D.~Wells,
  ``Phenomenological implications of low energy supersymmetry breaking,''
  Nucl.\ Phys.\ Proc.\ Suppl.\  {\bf 52A}, 38 (1997)
  \href{http://arxiv.org/abs/hep-ph/9607450}{\tt arXiv:hep-ph/9607450}.

\bibitem{Dimopoulos:1996yq}
  S.~Dimopoulos, S.~D.~Thomas and J.~D.~Wells,
  ``Sparticle spectroscopy and electroweak symmetry breaking with
  gauge-mediated supersymmetry breaking,''
  \href{http://dx.doi.org/10.1016/S0550-3213(97)00030-8}{Nucl.\ Phys.\  B {\bf 488}, 39 (1997)}
  \href{http://arxiv.org/abs/hep-ph/9609434}{\tt arXiv:hep-ph/9609434}.

\bibitem{Ambrosanio:1997rv}
  S.~Ambrosanio, G.~D.~Kribs and S.~P.~Martin,
  ``Signals for gauge-mediated supersymmetry breaking models at the CERN  LEP2
  collider,''
  \href{http://dx.doi.org/10.1103/PhysRevD.56.1761}{Phys.\ Rev.\  D {\bf 56}, 1761 (1997)}
  \href{http://arxiv.org/abs/hep-ph/9703211}{\tt arXiv:hep-ph/9703211}.

\bibitem{Culbertson:2000am}
  R.~L.~Culbertson {\it et al.}  [SUSY Working Group Collaboration],
  ``Low scale and gauge mediated supersymmetry breaking at the Fermilab
  Tevatron Run II,''
  \href{http://arxiv.org/abs/hep-ph/0008070}{\tt arXiv:hep-ph/0008070}.

  \bibitem{MatchevThomas}
K.~T. Matchev and S.~D. Thomas, ``{Higgs and $Z$ boson signatures of
  supersymmetry},'' \href{http://dx.doi.org/10.1103/PhysRevD.62.077702}{{\em
  Phys. Rev.} {\bf D62} (2000)  077702},
\href{http://arxiv.org/abs/hep-ph/9908482}{{\tt arXiv:hep-ph/9908482}}.

\bibitem{BMTW}
H.~Baer, P.~G. Mercadante, X.~Tata, and Y.-l. Wang, ``{The Reach of Tevatron
  upgrades in gauge mediated supersymmetry breaking models},''
  \href{http://dx.doi.org/10.1103/PhysRevD.60.055001}{{\em Phys. Rev.} {\bf
  D60} (1999)  055001},
\href{http://arxiv.org/abs/hep-ph/9903333}{{\tt arXiv:hep-ph/9903333}};
  H.~Baer, P.~G.~Mercadante, X.~Tata, Y.~-l.~Wang,
  ``The Reach of the CERN large hadron collider for gauge mediated supersymmetry breaking models,''
  \href{http://dx.doi.org/10.1103/PhysRevD.62.095007}{Phys.\ Rev.\  {\bf D62}, 095007 (2000)}.
  \href{http://arxiv.org/abs/hep-ph/0004001}{\tt hep-ph/0004001}.

\bibitem{prospino}
  W.~Beenakker, R.~Hopker, and M.~Spira, ``{PROSPINO: A Program for the
  production of supersymmetric particles in next-to-leading order QCD},''
  \href{http://arxiv.org/abs/hep-ph/9611232}{{\tt arXiv:hep-ph/9611232
  [hep-ph]}}, \url{http://www.thphys.uni-heidelberg.de/~plehn/}.

\bibitem{Meade:2006dw}
  P.~Meade, M.~Reece,
  ``Top partners at the LHC: Spin and mass measurement,''
  \href{http://dx.doi.org/10.1103/PhysRevD.74.015010}{Phys.\ Rev.\  {\bf D74}, 015010 (2006)},
  \href{http://arxiv.org/abs/hep-ph/0601124}{\tt hep-ph/0601124}.

\bibitem{ArkaniHamed:2007fw}
  N.~Arkani-Hamed, P.~Schuster, N.~Toro, J.~Thaler, L.~-T.~Wang, B.~Knuteson, S.~Mrenna,
  ``MARMOSET: The Path from LHC Data to the New Standard Model via On-Shell Effective Theories,''
  \href{http://arxiv.org/abs/hep-ph/0703088}{\tt hep-ph/0703088}.

\bibitem{Alwall:2008ve}
  J.~Alwall, M.~-P.~Le, M.~Lisanti, J.~G.~Wacker,
  ``Searching for Directly Decaying Gluinos at the Tevatron,''
  \href{http://dx.doi.org/10.1016/j.physletb.2008.06.065}{Phys.\ Lett.\  {\bf B666}, 34 (2008)},
  \href{http://arxiv.org/abs/0803.0019}{\tt arXiv:0803.0019 [hep-ph]}.

\bibitem{Dube:2008kf}
  S.~Dube, J.~Glatzer, S.~Somalwar, A.~Sood, S.~Thomas,
  ``Addressing the Multi-Channel Inverse Problem at High Energy Coliders: A Model Independent Approach to the Search for New Physics with Trileptons,''
  \href{http://arxiv.org/abs/0808.1605}{\tt arXiv:0808.1605 [hep-ph]}.

\bibitem{Alwall:2008ag}
  J.~Alwall, P.~Schuster, N.~Toro,
  ``Simplified Models for a First Characterization of New Physics at the LHC,''
  \href{http://dx.doi.org/10.1103/PhysRevD.79.075020}{Phys.\ Rev.\  {\bf D79}, 075020 (2009)},
  \href{http://arxiv.org/abs/0810.3921}{\tt arXiv:0810.3921 [hep-ph]}.

\bibitem{Alves:2011wf}
  D.~Alves, N.~Arkani-Hamed, S.~Arora, Y.~Bai, M.~Baumgart, J.~Berger, M.~Buckley, B.~Butler {\it et al.},
  ``Simplified Models for LHC New Physics Searches,''
  \href{http://arxiv.org/abs/1105.2838}{\tt arXiv:1105.2838 [hep-ph]}.

\bibitem{chickenfoot}
R.~Essig, E.~Izaguirre, J.~Kaplan and J.~G.~Wacker,
``Heavy Flavor Simplified Models at the LHC,''
\href{http://arxiv.org/abs/1110.6443}{\tt arXiv:1110.6443 [hep-ph]};
C.~Brust, A.~Katz, S.~Lawrence, R.~Sundrum,
``SUSY, the Third Generation and the LHC,''
\href{http://arxiv.org/abs/1110.6670}{\tt arXiv:1110.6670 [hep-ph]};
M.~Papucci, J.~T.~Ruderman, A.~Weiler, ``Natural SUSY Endures,'' \href{http://arxiv.org/abs/1110.6926}{\tt arXiv:1110.6926 [hep-ph]}.

\bibitem{Abazov:2010us}
{\bf D0} Collaboration, V.~M.~Abazov {\it et al.},
  ``Search for diphoton events with large missing transverse energy in
    6.3~fb$^{-1}$ of $p\bar{p}$ collisions at $\sqrt{s}=1.96$~TeV,''
  \href{http://dx.doi.org/10.1103/PhysRevLett.105.221802}{Phys.\ Rev.\ Lett.\  {\bf 105}, 221802 (2010)},
  \href{http://arxiv.org/abs/1008.2133}{\tt arXiv:1008.2133 [hep-ex]}.

\bibitem{CMSsecretperson}
CMS Collaboration (private communication).

\bibitem{ATLASsecretperson}
ATLAS Collaboration (private communication).

\bibitem{Martin:1993ft}
  S.~P.~Martin and P.~Ramond,
  ``Sparticle Spectrum Constraints,''
  \href{http://dx.doi.org/10.1103/PhysRevD.48.5365}{Phys.\ Rev.\  D {\bf 48}, 5365 (1993)}
  \href{http://arxiv.org/abs/hep-ph/9306314}{\tt arXiv:hep-ph/9306314}.

\bibitem{Kribs:2008hq}
G.~D. Kribs, A.~Martin, and T.~S. Roy, ``{Supersymmetry with a Chargino NLSP
  and Gravitino LSP},''
  \href{http://dx.doi.org/10.1088/1126-6708/2009/01/023}{{\em JHEP} {\bf 0901}
  (2009)  023}, \href{http://arxiv.org/abs/0807.4936}{{\tt arXiv:0807.4936
  [hep-ph]}}.

\bibitem{ArkaniHamed:1997fq}
  N.~Arkani-Hamed, M.~A.~Luty, J.~Terning,
  ``Composite quarks and leptons from dynamical supersymmetry breaking without messengers,''
  \href{http://dx.doi.org/10.1103/PhysRevD.58.015004}{Phys.\ Rev.\  {\bf D58}, 015004 (1998)},
  \href{http://arxiv.org/abs/hep-ph/9712389}{\tt hep-ph/9712389}.

\bibitem{Franco:2009wf}
  S.~Franco, S.~Kachru,
  ``Single-Sector Supersymmetry Breaking in Supersymmetric QCD,''
  \href{http://dx.doi.org/10.1103/PhysRevD.81.095020}{Phys.\ Rev.\  {\bf D81}, 095020 (2010)},
  \href{http://arxiv.org/abs/0907.2689}{\tt arXiv:0907.2689 [hep-th]}.

\bibitem{Craig:2011yk}
  N.~Craig, D.~Green, A.~Katz,
  ``(De)Constructing a Natural and Flavorful Supersymmetric Standard Model,''
  \href{http://dx.doi.org/10.1007/JHEP07(2011)045}{{\em JHEP} {\bf 1107}, 045 (2011)},
  \href{http://arxiv.org/abs/1103.3708}{\tt arXiv:1103.3708 [hep-ph]}.

 \bibitem{Gabella:2007cp}
 M.~Gabella, T.~Gherghetta and J.~Giedt,
 ``A gravity dual and LHC study of single-sector supersymmetry breaking,''
 \href{http://dx.doi.org/10.1103/PhysRevD.76.055001}{Phys.\ Rev.\  D {\bf 76}, 055001 (2007)}
 \href{http://arxiv.org/abs/0704.3571}{\tt arXiv:0704.3571 [hep-ph]}.


 \bibitem{Abazov:2010wq}
{\bf D0} Collaboration, V.~M.~Abazov {\it et al.},
  ``Search for scalar bottom quarks and third-generation leptoquarks in $p\bar p$ collisions at $\sqrt s = 1.96$~TeV,''
  \href{http://dx.doi.org/10.1016/j.physletb.2010.08.028}{Phys.\ Lett.\  {\bf B693}, 95 (2010)},
  \href{http://arxiv.org/abs/1005.2222}{\tt arXiv:1005.2222 [hep-ex]}.

\bibitem{Beenakker:2010nq}
W.~Beenakker, S.~Brensing, M.~{Kr\"{a}mer}, A.~Kulesza, E.~Laenen, {\em et
  al.}, ``{Supersymmetric top and bottom squark production at hadron
  colliders},'' \href{http://dx.doi.org/10.1007/JHEP08(2010)098}{{\em JHEP}
  {\bf 1008} (2010)  098}, \href{http://arxiv.org/abs/1006.4771}{{\tt
  arXiv:1006.4771 [hep-ph]}}.

\bibitem{ATLAS-CONF-2011-100}
{\bf ATLAS} Collaboration, ``{Measurement of the top-quark pair production
  cross-section in $pp$ collisions at $\sqrt s = 7$~TeV in dilepton final
  states with ATLAS},'' ATLAS-CONF-2011-100,
  \url{http://cdsweb.cern.ch/record/1369215}.


\bibitem{Aaltonen:2009sf}
{\bf CDF} Collaboration, ``{Search for pair production of supersymmetric top
  quarks in dilepton events from $p\bar{p}$ collisions at $\sqrt{s} =
  1.96$~TeV},'' \href{http://dx.doi.org/10.1103/PhysRevLett.104.251801}{{\em
  Phys. Rev. Lett.} {\bf 104} (2010)  251801},
  \href{http://arxiv.org/abs/0912.1308}{{\tt arXiv:0912.1308 [hep-ex]}}.

\bibitem{Aaltonen:2011na}
{\bf CDF} Collaboration, T.~Aaltonen {\it et al.},
  ``Search for new physics in $t\bar t + \slash E_T \to b\bar b q \bar q q \bar q + \slash E_T$ final state in $p \bar p$ collisions at $\sqrt{s}=1.96$~TeV,''
  \href{http://arxiv.org/abs/1107.3574}{{\tt arXiv:1107.3574 [hep-ex]}}.

\bibitem{Lester:1999tx}
  C.~G.~Lester and D.~J.~Summers,
  ``Measuring masses of semiinvisibly decaying particles pair produced at
  hadron colliders,''
  \href{http://dx.doi.org/10.1016/S0370-2693(99)00945-4}{Phys.\ Lett.\  B {\bf 463}, 99 (1999)},
  \href{http://arxiv.org/abs/hep-ph/9906349}{\tt arXiv:hep-ph/9906349}.

\bibitem{Lester:2011nj}
  C.~G.~Lester,
  ``The stransverse mass, $M_{T2}$, in special cases,''
  \href{http://dx.doi.org/10.1007/JHEP05(2011)076}{{\em JHEP} {\bf 1105}, 076 (2011)},
  \href{http://arxiv.org/abs/1103.5682}{\tt arXiv:1103.5682 [hep-ph]}.

\bibitem{Barr:2009wu}
  A.~J.~Barr, C.~Gwenlan,
  ``The Race for supersymmetry: Using $m_{T2}$ for discovery,''
  \href{http://dx.doi.org/10.1103/PhysRevD.80.074007}{Phys.\ Rev.\  {\bf D80}, 074007 (2009)}.
  \href{http://arxiv.org/abs/arXiv:0907.2713}{\tt arXiv:0907.2713 [hep-ph]}.

  \bibitem{lepmt2paper}
  Y.~Kats, P.~Meade, M.~Reece and D.~Shih, in preparation.

\bibitem{CMS-PAS-EXO-11-020}
{\bf CMS} Collaboration, ``{Search for Stopped Heavy Stable Charged Particles
  in $pp$ collisions at $\sqrt{s} = 7$~TeV},'' CMS PAS EXO-11-020,
  \url{http://cdsweb.cern.ch/record/1369210}.

\bibitem{CMS-PAS-EXO-11-022}
{\bf CMS} Collaboration, ``{Search for Heavy Stable Charged Particles in $pp$
  collisions at $\sqrt{s} = 7$~TeV},'' CMS PAS EXO-11-022,
  \url{http://cdsweb.cern.ch/record/1370057}.

\bibitem{CMS-PAS-EXO-11-004}
{\bf CMS} Collaboration, ``{Search for Heavy Resonances Decaying to Long-Lived
  Neutral Particles in the Displaced Lepton Channel},'' CMS PAS EXO-11-004,
  \url{http://cdsweb.cern.ch/record/1380311}.

\bibitem{CMSdisplacedphotons}
H.~Liu, ``SUSY Searches with Displaced Vertices in CMS,"
\url{https://indico.cern.ch/materialDisplay.py?contribId=44&sessionId=2&materialId=slides&confId=149404}


\bibitem{ATLASdisappearingtrack}
Y.~Azuma, ``SUSY Searches with Displaced Vertices (Disappearing Tracks) in ATLAS,"
\url{https://indico.cern.ch/materialDisplay.py?contribId=43&sessionId=2&materialId=slides&confId=149404}


\bibitem{Aad:2011zb}
{\bf ATLAS} Collaboration, G.~Aad {\em et al.}, ``{Search for displaced
  vertices arising from decays of new heavy particles in 7 TeV $pp$ collisions
  at ATLAS},'' \href{http://arxiv.org/abs/1109.2242}{{\tt arXiv:1109.2242
  [hep-ex]}}.

\bibitem{Aad:2011yf}
{\bf ATLAS} Collaboration, G.~Aad {\it et al.},
  ``Search for stable hadronising squarks and gluinos with the ATLAS experiment at the LHC,''
  \href{http://dx.doi.org/10.1016/j.physletb.2011.05.010}{Phys.\ Lett.\  {\bf B701 } (2011) 1},
  \href{http://arxiv.org/abs/1103.1984}{\tt arXiv:1103.1984 [hep-ex]}.

\bibitem{Abazov:2011pf}
{\bf D0} Collaboration, V.~M.~Abazov {\it et al.},
  ``A search for charged massive long-lived particles,''
  \href{http://arxiv.org/abs/1110.3302}{\tt arXiv:1110.3302 [hep-ex]}.

\bibitem{Kawagoe:2003jv}
  K.~Kawagoe, T.~Kobayashi, M.~M.~Nojiri and A.~Ochi,
  ``Study of the gauge mediation signal with nonpointing photons at the CERN
  LHC,''
  \href{http://dx.doi.org/10.1103/PhysRevD.69.035003}{Phys.\ Rev.\  D {\bf 69}, 035003 (2004)},
  \href{http://arxiv.org/abs/hep-ph/0309031}{\tt arXiv:hep-ph/0309031}.

\bibitem{Park:2011vw}
  M.~Park, Y.~Zhao,
  ``Recovering Particle Masses from Missing Energy Signatures with Displaced Tracks,''
  \href{http://arxiv.org/abs/1110.1403}{arXiv:1110.1403 [hep-ph]}.

\bibitem{Meade:2011du}
  P.~Meade, M.~Papucci, T.~Volansky,
  ``Odd Tracks at Hadron Colliders,''
  \href{http://arxiv.org/abs/1103.3016}{arXiv:1103.3016 [hep-ph]}.

\bibitem{Djouadi:2002ze}
A.~Djouadi, J.-L. Kneur, and G.~Moultaka, ``{SuSpect: A Fortran code for the
  supersymmetric and Higgs particle spectrum in the MSSM},''
  \href{http://dx.doi.org/10.1016/j.cpc.2006.11.009}{{\em Comput.Phys.Commun.}
  {\bf 176} (2007)  426}, \href{http://arxiv.org/abs/hep-ph/0211331}{{\tt
  arXiv:hep-ph/0211331 [hep-ph]}}.

\bibitem{Muhlleitner:2003vg}
M.~Muhlleitner, A.~Djouadi, and Y.~Mambrini, ``{SDECAY: A Fortran code for the
  decays of the supersymmetric particles in the MSSM},''
  \href{http://dx.doi.org/10.1016/j.cpc.2005.01.012}{{\em Comput.Phys.Commun.}
  {\bf 168} (2005)  46}, \href{http://arxiv.org/abs/hep-ph/0311167}{{\tt
  arXiv:hep-ph/0311167 [hep-ph]}}.

\bibitem{Sjostrand:2007gs}
T.~{Sj\"{o}strand}, S.~Mrenna, and P.~Skands, ``{A Brief Introduction to
  \textsc{Pythia} 8.1},''
  \href{http://dx.doi.org/10.1016/j.cpc.2008.01.036}{{\em Comput. Phys.
  Commun.} {\bf 178} (2008)  852}, \href{http://arxiv.org/abs/0710.3820}{{\tt
  arXiv:0710.3820 [hep-ph]}}.
  See~also~\url{http://home.thep.lu.se/~torbjorn/Pythia.html}.

\bibitem{Cacciari:2005hq}
M.~Cacciari and G.~P. Salam, ``{Dispelling the $N^3$ myth for the $k_t$
  jet-finder},'' \href{http://dx.doi.org/10.1016/j.physletb.2006.08.037}{{\em
  Phys.Lett.} {\bf B641} (2006)  57},
  \href{http://arxiv.org/abs/hep-ph/0512210}{{\tt arXiv:hep-ph/0512210
  [hep-ph]}}. See also M.~Cacciari, G.P.~Salam and
  G.~Soyez,~\url{http://fastjet.fr/}.

\bibitem{Sjostrand:2006za}
T.~{Sj\"{o}strand}, S.~Mrenna, and P.~Skands, ``{\textsc{Pythia} 6.4 Physics
  and Manual},'' \href{http://dx.doi.org/10.1088/1126-6708/2006/05/026}{{\em
  JHEP} {\bf 05} (2006)  026}, \href{http://arxiv.org/abs/hep-ph/0603175}{{\tt
  arXiv:hep-ph/0603175}}.

\bibitem{Pumplin:2002vw}
J.~Pumplin, D.~Stump, J.~Huston, H.~Lai, P.~M. Nadolsky, {\em et al.}, ``{New
  generation of parton distributions with uncertainties from global QCD
  analysis},'' \href{http://jhep.sissa.it/stdsearch?paper=07(2002)012}{{\em JHEP} {\bf 0207} (2002)  012},
  \href{http://arxiv.org/abs/hep-ph/0201195}{{\tt arXiv:hep-ph/0201195
  [hep-ph]}}.

\bibitem{Whalley:2005nh}
M.~R. Whalley, D.~Bourilkov, and R.~C. Group, ``{The Les Houches Accord PDFs
  (LHAPDF) and LHAGLUE},'' \href{http://arxiv.org/abs/hep-ph/0508110}{{\tt
  arXiv:hep-ph/0508110}}. See~also~\url{http://hepforge.cedar.ac.uk/lhapdf/}.

\bibitem{Cacciari:2008gp}
M.~Cacciari, G.~P. Salam, and G.~Soyez, ``{The anti-$k_t$ jet clustering
  algorithm},'' \href{http://dx.doi.org/10.1088/1126-6708/2008/04/063}{{\em
  JHEP} {\bf 0804} (2008)  063}, \href{http://arxiv.org/abs/0802.1189}{{\tt
  arXiv:0802.1189 [hep-ph]}}.

\bibitem{PGS}
J.~Conway {\em et al.}, ``{PGS 4 -- Pretty Good Simulation of high energy
  collisions},''
  \url{http://physics.ucdavis.edu/~conway/research/software/pgs/pgs4-general.htm}.

\bibitem{CLsrefs}
A.~L.~Read, ``Modified frequentist analysis of search results (the $CL_s$ method),'' CERN-OPEN-2000-205, \url{http://cdsweb.cern.ch/record/451614}.

\bibitem{Cranmer:2010hk}
  K.~Cranmer, I.~Yavin,
  ``RECAST: Extending the Impact of Existing Analyses,''
  \href{http://dx.doi.org/10.1007/JHEP04(2011)038}{{\em JHEP} {\bf 1104}, 038 (2011)},
  \href{http://arxiv.org/abs/1010.2506}{\tt arXiv:1010.2506 [hep-ex]}.


\end{thebibliography}
\end{document}